\documentclass[english,numberedappendix,letter,revtex4]{emulateapj}

\usepackage[T1]{fontenc}
\usepackage{amssymb}
\usepackage{amsmath}
\usepackage{epsfig}
\usepackage{pdflscape}

\def\aj{\rm{AJ}}
\def\apj{\rm{ApJ}}
\def\apjl{\rm{ApJ}}
\def\apjs{\rm{ApJS}}
\def\mnras{\rm{MNRAS}}

\shorttitle{Evolution of the Star-forming Main Sequence}
\shortauthors{Speagle et al.}

\newcommand{\sn}[2]{\ensuremath{#1 \times 10^{#2}}}
\newcommand{\qq}{\symbol{34}}
\newcommand{\rom}[2]{\ensuremath{#1_{\scriptsize\textrm{#2}}}}
\usepackage{babel}

\begin{document}

\title{A Highly Consistent Framework for the Evolution of the Star-Forming ``Main Sequence'' from \protect{$z \sim 0-6$}}
\author{J.~S. Speagle \altaffilmark{1,2}, C.~L. Steinhardt\altaffilmark{3,4,2}, P.~L. Capak\altaffilmark{3,4}, J.~D. Silverman\altaffilmark{2}}

\altaffiltext{1}{Harvard University Department of Astronomy, 60 Garden St., MS 46, Cambridge, MA 02138, USA; \href{mailto:jspeagle@cfa.harvard.edu}{jspeagle@cfa.harvard.edu}}
\altaffiltext{2}{Kavli IPMU, University of Tokyo, Kashiwanoha 5-1-5, Kashiwa-shi, Chiba 277-8583, Japan}
\altaffiltext{3}{California Institute of Technology, MC 105-24, 1200 East California Blvd., Pasadena, CA 91125, USA}
\altaffiltext{4}{Infrared Processing and Analysis Center, California Institute of Technology, MC 100-22, 770 South Wilson Ave., Pasadena, CA 91125, USA}

\begin{abstract}
Using a compilation of 25 studies from the literature, we investigate the evolution of the star-forming galaxy (SFG) Main Sequence (MS) in stellar mass and star formation rate (SFR) out to $z \sim 6$. After converting all observations to a common set of calibrations, we find a remarkable consensus among MS observations ($\sim 0.1$ dex 1$\sigma$ interpublication scatter). By fitting for time evolution of the MS in bins of constant mass, we deconvolve the observed scatter about the MS within each observed redshift bins. After accounting for observed scatter between different SFR indicators, we find the width of the MS distribution is $\sim 0.2$ dex and remains constant over cosmic time. Our best fits indicate the slope of the MS is likely time-dependent, with our best fit $\log\textrm{SFR}(M_*,t) = \left(0.84 \pm 0.02 - 0.026 \pm 0.003 \times t\right) \log M_* - \left(6.51 \pm 0.24 - 0.11 \pm 0.03 \times t\right)$, with $t$ the age of the Universe in Gyr. We use our fits to create empirical evolutionary tracks in order to constrain MS galaxy star formation histories (SFHs), finding that (1) the most accurate representations of MS SFHs are given by delayed-$\tau$ models, (2) the decline in fractional stellar mass growth for a ``typical'' MS galaxy today is approximately linear for most of its lifetime, and (3) scatter about the MS can be generated by galaxies evolving along identical evolutionary tracks assuming an initial $1\sigma$ spread in formation times of $\sim 1.4$ Gyr. %In addition, we observe evolution in radio SFRs as compared to other SFR indicators $\propto (1+z)^{\sim 0.8}$ (i.e. over-luminous radio sources), and find that $sBzK$-selection generates systematically smaller scatter relative to other selection methods.
\end{abstract}

\keywords{galaxies: evolution -- galaxies: star formation -- radio continuum: galaxies -- surveys}

\section{Introduction}
\label{sec:intro}
Wide-field and deep multi-wavelength surveys have allowed us to study statistically large samples of galaxies at a wide range of redshifts with unprecedented detail. Substantial progress in stellar population synthesis (SPS) modeling \citep{fiocrocca-volmerange97,bruzualcharlot03,maraston05,percival+09,conroy+09,conroygunn10} and improved global diagnostics of galactic star formation (\citealt{murphy+11}; \citealt{hao+11}; \citealt{kennicuttevans12} (KE12), and references within) have enabled the determination of key physical quantities of galaxies from these data: photometric redshifts, star formation rates (SFRs; $\psi$), stellar masses ($M_*$), dust attenuation, and stellar ages \citep{arnouts+99,benitez+00,bolzonella+00,collisterlahav04,ilbert+06,feldmann+06,brammer+08,hildebrandt+10,abdalla+11,acquaviva+11,pirzkal+12,johnson+13,moustakas+13,dahlen+13}.

These advances in redshift estimation have allowed the determination of accurate rest frame colors for many of these objects, and indicate that galaxies out to high redshifts fall into two  distinct groups in color-color space: ``star-forming'' (SF) and ``quiescent'' \citep{labbe+05,wuyts+07,williams+09,ilbert+10,brammer+11,ilbert+13}. New studies of physical quantities have revealed key differences between these groups, such as a strong correlation at fixed redshift between $M_*$ and $\psi$ among star-forming galaxies (SFGs). This SF ``Main Sequence'' (MS) generally takes the form
\begin{equation}\label{eq:ms_1}
\log \psi = \alpha \, \log M_* + \beta,
\end{equation}
with $\alpha$ and $\beta$ free parameters of the fit. $\alpha$ is usually measured to be between $0$ and $1$ (\citealt{chen+09}; \citealt{reddy+12} (R12a)), with values of $\sim 0.6$\,--\,$1$ preferred \citep{rodighiero+11}, and both $\alpha$ (MS slope, i.e. power-law index) and $\beta$ (MS normalization) likely functions of time, $\alpha(t)$ and $\beta(t)$. This relationship has been shown to hold for over 4\,--\,5 orders of magnitude in mass \citep{santini+09} and from $z = 0$ to $z \sim 6$ (\citealt{brinchmann+04} (B04); \citealt{salim+07} (S07); \citealt{noeske+07} (N07); \citealt{elbaz+07} (E07); \citealt{daddi+07} (D07); \citealt{chen+09} (C09); \citealt{pannella+09} (P09); \citealt{santini+09} (S09); \citealt{oliver+10} (O10); \citealt{magdis+10} (M10); \citealt{lee+11} (L11); \citealt{rodighiero+11} (R11); \citealt{elbaz+11} (E11); \citealt{karim+11} (K11); \citealt{shim+11} (S11); \citealt{bouwens+12} (B12); \citealt{whitaker+12} (W12); \citealt{zahid+12} (Z12); \citealt{lee+12} (L12); \citealt{reddy+12b} (R12); \citealt{salmi+12} (S12); \citealt{moustakas+13} (M13); \citealt{kashino+13} (K13); \citealt{sobral+14} (So14); Steinhardt et al. 2014, subm. (St14); Coil et al. 2014, in prep. (C14)). This relation is quite tight, with only $\sim 0.20$\,--\,$0.35$\,dex of observed scatter\footnote{Throughout this paper, we use the term ``scatter'' to refer to the 1$\sigma$ dispersion of galaxies around the best fit MS parameters, rather than the uncertainties in the fitted parameters themselves.} (D07; M10; W12). From this point onwards we will refer to each of these studies by their abbreviation (see also Tables~\ref{tab:msfr_info} and~\ref{tab:msfr}).

These studies typically find that galaxies on this SF MS formed stars at much higher rates in the distant universe than they do today: the average SFR at fixed stellar mass has decreased at a steady rate by a factor of $\sim 20$ from $z \sim 2$ to $z=0$ (D07; E07; W12; So14).
%\footnote{So14's self-consistent analysis finds that the typical SFR in H$\alpha$ decreases by a factor of $\sim 13$ from $z\sim 2.2$\,--\,0.4, which is $\sim 20$ when extrapolated to $z=0$.}
This has been linked to the rapid quenching of star formation (\citealt{bell+07}; \citealt{brammer+11}; \citealt{ilbert+13}; M13) and the ``downsizing paradigm''\footnote{``Downsizing'', as originally defined in \citet{cowie+88}, is the movement of star formation from more massive to less massive systems with time. Coupled with observed evolution in the cosmic star formation history (cSFH; \citealt{lilly+96}; \citealt{madau+96}; \citealt{hopkinsbeacom06}), ``downsizing'' has instead been taken to be an evolutionary scenario where more massive objects evolve more quickly. We use the phrase ``downsizing''  and ``downsizing paradigm'' to refer to the former and latter, respectively.}  for galaxy evolution \citep{cowie+88}. In addition, SFGs in clusters, groups, and the field display similar MS relations up to $z \sim 2.2$ (although with differing quiescent fractions and overall mass distributions), indicating that the underlying physics governing MS evolution are relatively insensitive to environment \citep{peng+10,koyama+13,lin+14}.

Although there have been a host of studies of the MS in the past decade, quantitative comparisons between them have been difficult, as studies have not standardized their calibrations and methodology. Differences in, e.g., assumed stellar initial mass function (IMF), luminosity-to-SFR ($L$\,--\,$\psi$) conversions, SPS models, dust attenuation, and emission line contributions can lead to differences in derived stellar masses and SFRs as high as a factor of 2\,--\,3 (M10; KE12; Z12; R12a; \citealt{stark+13}). These effects have not yet been systematically calibrated against each other, which has made it difficult to determine actual MS evolution, especially if both the normalization \textit{and} slope of the MS are changing over time. For instance, while some studies have found significant evolution in MS slope as high as $\alpha(z)=0.70-0.13z$ from $z \sim 0$\,--\,$2.5$ (W12), others seem to indicate little to no evolution over the same redshift range (D09; K11; So14).

Additionally, variation between MS slopes from various studies at a given redshift is also significant, reaching as high as $\gtrsim 0.6$ (E07; O10; \citealt{mitchell+14}), twice as large as the \textit{total} evolution observed by W12. As the slope and normalization are highly degenerate, samples that have similar overall distributions of masses and SFRs but have been selected differently can have large differences in their MS fits, leading to changes in the derived slopes by up to $\sim 0.4$ (K11; W12). The magnitude of these effects precludes robust interpretations of derived MS properties.

The inability to directly compare observations has also made it difficult to quantify how the scatter about the MS has evolved with time. While observations out to $z \sim 2.5$ find scatter to be roughly constant around $\sim 0.3$\,dex (N07; W12), the scatter observed at each median redshift has been convolved with evolution of the MS within its redshift bin, as well as with additional scatter resulting from uncertainties in stellar mass and SFR (N07). S12 are the first to attempt to account for this effect by simultaneously fitting a power-law correction as a function of redshift to their derived MS fits. This method, however, is limited by the redshift range spanned by their data ($0.5 < z < 1.3$) and somewhat dependent on the chosen functional form. As a result, the evolution of the ``true'' scatter about the MS across a wide range of redshifts has not yet been thoroughly investigated.

To overcome these limitations, interpublication comparisons have used average SFRs (either across the whole sample or at a specific mass) after simple IMF offsets to determine the approximate evolution of the average MS galaxy's SFR, rather than the derived MS's themselves (M10; Z12). This method has been useful in estimating the evolution of the cosmic star formation rate density (i.e. per cubic Mpc) (cSFR) to first order \citep{madau+98,hopkinsbeacom06}. However, it averages over the observed $M_*-\psi$ relations, and so does not take into account much of the information surrounding the mass dependencies that govern the MS.

In order to directly compare MS observations against each other and so constrain MS evolution and systematic errors, we have compiled 64 MS observations from 25 studies published since 2007, spanning $z \sim 0$\,--\,$6$, and converted them to the same absolute calibrations. These have been taken from a variety of fields, selected using different methodologies, include both stacked and non-stacked data, and have SFRs determined from all methods currently available. By taking into account the different mass ranges in each study consistently, we not only accurately determine MS evolution, but also quantify the extent to which selection can affect observed MS determinations. These results allow us to determine the evolution of both the MS and the ``true'' scatter about it as a function of cosmic time.

This paper is organized as follows. In \S\,\ref{sec:dataset}, we describe the data included in this work.  In \S\,\ref{sec:technicalities} we discuss some of the technical differences between different views of the MS and how we deal with them when converting MS observations to a common metric.  In \S\,\ref{sec:methodology}, we describe our mass-dependent method of fitting this inter-publication dataset. Our best fits and their corresponding evolutionary tracks are listed in \S\,\ref{sec:results}. We discuss some of their implications in \S\,\ref{sec:disc}. We summarize our results and offer some concluding remarks in \S\,\ref{sec:conc}.

Throughout this work, we standardize to a $(h,\Omega_M,\Omega_\Lambda)=(0.7,0.3,0.7)$ Wilkinson Microwave Anisotropy Probe (WMAP) concordance cosmology \citep{spergel+03}, AB magnitudes \citep{okegunn83}, a Kroupa \citep{kroupa01,kroupaweidner03} IMF (integrated from $0.1$\,--\,$100$\,$M_\odot$), KE12 $L$\,--\,$\psi$ relations\footnote{Although we refer to them as KE12 relations, these are taken from \citet{hao+11} and \citet{murphy+11}. KE12 has compiled them in one place for convenience.}, and \citet{bruzualcharlot03} (BC03) SPS models. Throughout the paper, $t$ will be used to refer to the age of the Universe (in Gyr), $M_*$ is measured in $M_\odot$, and $\psi$ is measured in $M_\odot$\,yr$^{-1}$. All masses discussed below are stellar masses unless stated otherwise.

\section{Observations of the Main Sequence}
\label{sec:dataset}

In order to get a robust selection of MS observations, we include papers which meet the following criteria: 
\begin{enumerate}
\item \textit{Includes a published $M_*$\,--\,$\psi$ or $M_*$\,--\,$\phi$ ($\phi \equiv \psi/M_*$) relation, or else numbers from which such a fit can be derived.} In order to accurately compare MS observations against each other, we require published values of $\alpha$ (slopes) and $\beta$ (normalizations) or otherwise analogous quantities.
\item \textit{Fit(s) include more than two data points (if stacked) or 50 galaxies (if directly observed)}. This requirement is mainly to avoid biases resulting from small number statistics and to enable the determination of a $\chi^2$ value to check the goodness of fit and thus possible variance and/or errors.
\item \textit{Includes the specifics of their fits, list references where such specifics may be obtained, or else provide data from which such specifics can be easily estimated.} In order to attempt to properly calibrate MS observations against each other, we must know what specific calibrations were used for each observation. 
\item \textit{Published no earlier than 2007.} We wish to limit ourselves to more recent observations with larger statistics, better estimates of physical parameters, and improved selection criteria. This is also when the idea of a ``Main Sequence'' was first coined by N07, and when observations of star-forming galaxies began to become more systematized.
\end{enumerate}

The papers which meet this criteria are listed in Table~\ref{tab:msfr_info} along with their calibrations and data types. The best-fit MS parameters for each of the individual studies are listed in Table~\ref{tab:msfr}. Our common set of calibrations are listed in Table~\ref{tab:calibration}, the corresponding offsets for each study in Table~\ref{tab:corr}, and the final set of relationships calibrated to a common basis in Table~\ref{tab:msfr_corr}. More details about each of the studies included here, as well as the rationale behind the respective offsets applied to each one, can be found in Appendix~\ref{app:data}. Note that these studies are not all independent; several listed here have analyzed the same set(s) of data (see Table~\ref{tab:msfr}).

In brief, we include data from 25 papers (64 MS relations), which can be broadly subdivided\footnote{Note that studies that use multiple datasets are double-counted.} as follows:
\begin{itemize}
\item 12 (26), 11 (35), and 2 (3) studies (MS relations) are derived assuming Salpeter, Chabrier, and Kroupa IMFs, respectively.
\item 13 (15), 9 (36), and 3 (13) utilize ``bluer'', ``mixed'', and ``non-selective'' selection methods (see \S\,\ref{subsubsec:selection}), respectively. These include 8 (9), 15 (43), and 3 (12) whose parent samples were selected based on their restframe UV, optical/NIR, and FIR emission, as well as 5 (6), 2 (3), 4 (4), 1 (7), 2 (3), 1 (1), 2 (14), 1 (4), and 8 (22) whose subsamples (used in the analysis) were selected via Lyman-break criteria, blue color, $sBzK$ criteria, bimodalities in the $M_*$\,--\,$\psi$ plane, emission lines, LIRG criteria, $NUVrJ$ or $UVJ$ color, a 2$\sigma$-clipping procedure (for the reported fit), or no substantive cut.
\item 6 (12) derive SFRs based on emission/absorption lines, 8 (9) from dust-corrected UV, 4 (11) from combined UV+IR data, 2 (7) from IR alone, 3 (16) from 1.4\,GHz radio observations, and 3 (9) from SED fitting alone. Of the emission/absorption line studies, 4 (7) utilize H$\alpha$ emission.
\item 19 (39) and 6 (25) derive masses and SFRs using non-stacked and stacked data, respectively.
\end{itemize}
In addition, masses, SFRs, and other physical parameters are derived using a range of model parameters, which include:
\begin{itemize}
\item 7 different SPS models/template sets, along with 2 analytical $M_*$/$L$ relations
\item 5 different parametrizations of SFHs
\item 7 different extinction curves, along with 3 independent observational estimates from IRX observations/correlations (M99; R12a; B12)
\item Assumed metallicites ranging from $Z = 0.005$\,--\,$2.5\,Z_\odot$.
\end{itemize}

We adjust each relation onto a common scaled based on the calibrations discussed in \S\,\ref{sec:technicalities}, which are briefly summarized here. The assumed stellar IMF is converted to a Kroupa IMF using the conversion factors taken from Z12 %\footnote{While most data use standard integration limits for the IMF from $0.1$\,--\,$100$\,$M_\odot$, data from O10 instead integrates from $0.15$\,--\,$120$\,$M_\odot$. This leads to slightly different conversion factors, as detailed in Appendix~\ref{app:data}.}
and the $L$\,--\,$\psi$ relation to those taken from KE12. Differences between SPS models (e.g., BC03 and CB07) are accounted for using the conversion factors from M10 and So14. IRX values (i.e. ``extinction'' corrections) are taken from either R12a ($z < 4$) or B12 ($z > 4$). Radio SFRs have been adjusted based on the $\psi_{1.4}/\psi{\textrm{other}} \propto (1+z)^{\sim 0.8}$ evolution observed here using the median redshifts of each redshift bin. When necessary, we include emission line effects on the masses using the conversion factors from \citet{stark+13} and adjust for differences in cosmology using our assumed $(h,\Omega_M,\Omega_\Lambda)=(0.7,0.3,0.7)$ WMAP concordance cosmology \citep{spergel+03} and first-order volume corrections (see \S\,\ref{subsubsec:cosmology}). Differences between selection methods and their effects on derived MS parameters are accounted for by subdividing them using our ``bluer'', ``mixed'', and ``non-selective'' classifications. To reduce the impact systematic uncertainties and selection effects have in our sample, we exclude data in the first and last 2\,Gyr of the Universe where the two are most important. Any other possible differences are not accounted for in this work. The calibrations and the areas they impact are briefly noted in Table~\ref{tab:calibration}, while their effects on the interpublication scatter and fitted MS parameters are shown in Table~\ref{tab:error_budget}. Based on these results, we take our ``best'' sample as the combination of our applied calibration offsets and ``time edge'' cuts restricted to mixed observations only.

These data encompass a wide range of assumed inputs and observations in the literature and are a census of most of the methods available today utilized to derive MS relations. The calibrations likewise incorporate many of the most up-to-date observational evidence as well as recent advances in modelling. By combining the two, we present what we hope is the broadest and most accurate census of MS observations to date.

\section{Calibrating the Main Sequence}
\label{sec:technicalities}

Differences in the assumptions and techniques used to derive the MS can lead to major offsets in the final derived $M_*$\,--\,$\psi$ relations\footnote{For a more in-depth discussion of many of the points discussed below, see \citet{bastian+10}, \citet{kroupa+13}, KE12, \citet{walcher+11}, and \citet{conroy13}.}. As outlined in Table~\ref{tab:msfr_info}, every one of these has been interpreted differently by various studies, leading to substantial difficulties in comparing different MS observations.

In order to properly compare these studies, in each case an offset is developed to produce a set of calibrations and assumptions, thereby putting all studies on a common basis. We denote all calibration offsets for the MS relation outlined in this section with the form $C_j$,
%\footnote{Note that this has no relation to spherical harmonics and/or multipoles.}
where $j$ denotes the particular attribute being adjusted for, and $C_j$ is in dex. This common basis is described in Table~\ref{tab:calibration}, while the impact it has on scatter between MS observations (i.e. interpublication scatter) is shown in Table~\ref{tab:error_budget}. The corresponding calibration offsets applied to each sample are listed in Table~\ref{tab:corr}. All non-reference acronyms used both here and throughout the rest of the paper are listed in Appendix~\ref{app:acronyms}.

Because studies have generally not released data tables containing individual objects, it is often impossible to perfectly adjust results to the common basis in Table~\ref{tab:calibration}.  Adjusting each study requires individual tuning, often in consultation with the authors.  In many cases, it is only possible to estimate an average adjustment to this common basis, expecting that it will produce a better result than making no adjustment.  For some adjustments (described later in this section) the situation is too ambiguous to find even an average value.  As a general principle, we choose to adjust data in every case where such an adjustment is unambiguously better or supported by results from the literature, but otherwise prefer to leave data unaltered rather than implement adjustments that may prove erroneous (although see \S\,\ref{subsec:radio}).

We find that the largest offsets arise due to differences in assumed $L$\,--\,$\psi$ conversion ($C_\psi$) and stellar IMF ($C_M$), which can lead to differences of several tenths of a dex.  This is fortunate, because both allow an unambiguous recalibration to a common standard.  Choices of SPS model ($C_S$) also play a significant role, with different treatments of short-lived but extremely luminous stellar phases (e.g., the thermally pulsating asymptotic giant branch) leading to differences of $\sim 0.1$\,--\,$0.2$\,dex. In addition, we find that adjusting radio/IR SFR studies for missing UV light (``extinction'' corrections; $C_E$) boosts SFRs upwards by $\sim 0.1$\,dex. This effect is offset, however, by the -0.1\,dex adjustment used to account for bias present in radio studies between the mean (derived through median stacking) and median (used by most other studies) of a lognormal distribution.

After applying these calibrations, we find that stacked radio SFRs display systematic deviations from other SFR indicators $\propto (1+z)^{\sim 0.8}$ (i.e. the IR-to-1.4\,GHz conversion decreases as $(1+z)^{-0.8}$). We note that evolution is expected, and apply an empirical correction using the median redshifts of each radio MS observation, which leads to radio SFR calibration offsets ($C_R$) as high as $\sim -0.6$\,dex. Outside of these main calibrations, different cosmologies ($C_C$) or emission line effects ($C_L$) have relatively negligible ($< 0.05$\,dex) effects for most redshifts included here. Based on previous results in the literature (discussed below), we do not choose to adjust our results for differences in assumed star formation history (SFH), different dust attenuation \textit{curves} (we correct for dust as a whole when it has not been applied), possible photo-z biases, differences in SED fitting procedures, or other possible observational biases. Lastly, we find that differing selection methods (i.e. ``bluer'' vs. ``mixed'' vs. ``non-selective''; see \S\,\ref{subsubsec:selection}) can lead to substantially different MS slopes, with bluer (non-selective) MS slopes biased towards values closer to unity (zero) relative to mixed slopes (see Figure~\ref{fig:slope_selection}). These are shown in Tables~\ref{tab:calibration} and~\ref{tab:error_budget}.

Each of these effects are discussed in more detail below. In \S\,\ref{subsec:big_problems}, we discuss seven calibration issues that could result in large offsets ($\gtrsim 25\%$) between different MS studies. These include: stellar IMF (\S\,\ref{subsubsec:imf}), $L$\,--\,$\psi$ conversion (\S\,\ref{subsubsec:l_psi}), SPS model (\S\,\ref{subsubsec:sps_model}), SFH (\S\,\ref{subsubsec:sfh}), dust attenuation (\S\,\ref{subsubsec:ext}), dust attenuation curve (\S\,\ref{subsubsec:ext_law}), and emission line effects (\S\,\ref{subsubsec:emission}). In \S\,\ref{subsec:small_problems}, we discuss four other calibration issues that likely only have minor impacts ($\lesssim 25\%$) on MS normalizations. These include: cosmology (\S\,\ref{subsubsec:cosmology}), use of photometric redshifts\footnote{The full impacts of the widespread use of photometric redshifts are not well-quantified outside of direct comparisons with spectroscopic redshifts, which significantly limits the conclusions reported here.} (\S\,\ref{subsubsec:photoz}), SED fitting procedures (\S\,\ref{subsubsec:sed_fitting}), and metallicity ($Z$; \S\,\ref{subsubsec:metallicity}). In \S\,\ref{subsec:biases}, we discuss the effects various observational biases have on MS parameters. These include the bias between the derived mean and median of a log-normal distribution (\S\,\ref{subsubsec:lognormbias}), the effects different selection methods have on derived MS parameters (\S\,\ref{subsubsec:selection}), systematic disagreements of $sBzK$-selected data relative to the other data included here (\S\,\ref{subsubsec:bzk}), and the impact of various other observational biases (incompleteness, Eddington bias, and Malmquist bias) on the MS (\S\,\ref{subsubsec:biases}). In \S\,\ref{subsec:radio}, we discuss the observed disagreements between radio SFR observations compared to other SFR indicators.

\begin{deluxetable}{l c c}[!ht]
\tabletypesize{\scriptsize}
\tablewidth{0pt}
\tablecaption{Main Sequence Calibrations \label{tab:calibration}}
\tablehead{
\colhead{Parameter} &
\colhead{Impact} &
\colhead{Calibration}
}
\startdata
Radio SFRs & SFR & $(1+z)^{-0.8}$ (see \S\,\ref{subsec:radio}) \\
Selection Effects & MS slope & ``mixed'' (see \S\,\ref{subsubsec:selection}) \\
$L$\,--$\psi$ relation & SFR & \citet{kennicuttevans12}\,\tablenotemark{a} \\
Assumed IMF & M/SFR & \citet{zahid+12} \\
SPS Model & M & \citet{magdis+10}\,\tablenotemark{b} \\
Extinction ($z < 4$) & SFR & \citet{reddy+12} \\
Extinction ($z > 4$) & SFR & \citet{bouwens+12} \\
Emission Lines & M & \citet{stark+13} \\
Cosmology & SFR & \citet{spergel+03} \\
Assumed SFH & M/SFR & None \\
Extinction Curve & SFR\,\tablenotemark{c} & None \\
Metallicity & M/SFR & None \\
Photo-z's & M/SFR & None \\
SED Fitting & M/SFR & None
\enddata
\tablecomments{A list of the assumptions, the areas they impact, and the calibrations we have choosen to establish (or not) to account for varying assumptions, listed in order from largest to smallest. Assumptions without a corresponding calibration have not been accounted for in this work. Note that M ($M_*$) = stellar mass, SFR ($\psi$) = star formation rate, and MS = Main Sequence. See \S\,\ref{sec:technicalities} for more details. \\
\tablenotemark{a}\,Taken from \citet{hao+11} and \citet{murphy+11}.
\tablenotemark{b}\,Calibrations for data from \citet{sobral+14} are instead taken from D. Sobral (priv. comm.).
\tablenotemark{c}\,Might also affect masses (see, e.g., \citealt{kriekconroy13}).
}
\end{deluxetable}

\subsection{Major Influences}
\label{subsec:big_problems}

\subsubsection{Initial Mass Function}
\label{subsubsec:imf}

At present, several different stellar IMFs are used to derive MS properties. These are usually presumed to be universal -- i.e., unchanging with respect to time, current and/or past SFH, metalicity, etc. Current evidence is conflicting: \citet{bastian+10} claim that the IMF is likely universal, while \citet{kroupa+13} argue that the IMF becomes more top-heavy (i.e. forming higher fractions of more massive stars) with increasing SFRs. Possible ramifications of this for MS evolution are discussed in \citet{dave08}, but at present the issue remains unresolved. In this work, we assume a universal IMF. Evolution in the IMF as a function of the SFR could change the derived MS slope, and evolution as a function of redshift could affect our evolutionary fits.

The most common of these IMFs are those of \citet{salpeter55}, \citet{chabrier03}, and \citet{kroupa01}, most commonly (but not universally) integrated from $0.1$\,--\,$100\,M_\odot$. These will be referred to as Salpeter, Chabrier, and Kroupa IMFs, respectively. The assumed IMF impacts both the derived masses and SFRs, leading to variations of up to $\sim 40\%$ (KE12). At present, there are several different factors used to convert between these different IMFs (E07; S07; C09; K11; Z12; \citealt{papovich+11}). We choose the IMF offsets taken from Z12 (also seen in S07 and E07) because they have been calculated recently and assume the same SPS model (BC03) that we standardize to here. These take the form
\begin{equation}\label{eq:imf}
M_{*,K} = 1.06\,M_{*,C} = 0.62\,M_{*,S},
\end{equation}
with the subscripts referring to Kroupa, Chabrier, and Salpeter IMFs, respectively. These correspond to mass offsets of $C_{M_*,C} = +0.03$ and $C_{M_*,S} = -0.21$\,dex. These agree well with the SFR offsets used to convert from \citet{kennicutt98} (K98) to KE12 (which assume Salpeter and Kroupa IMFs, respectively) for SFRs derived from the FUV and NUV. In all cases, the shift between a Chabrier and Kroupa IMF is essentially negligible. Although all adjustments have been applied for completeness, we note that our results are unchanged if the Chabrier IMF-derived masses are left as they are.

This mass adjustment is functionally equivalent to shifting the MS left or right (i.e., increasing/decreasing the SFR at a given mass) with the observed mass ranges adjusted accordingly (see Tables~\ref{tab:msfr} and~\ref{tab:msfr_corr}). These lead to calibration offsets in the normalization, $C_M$, of $-\alpha \times C_{M_*}$. For a slope of unity, these changes merely result in a shift of the observed range of the MS relation rather than the actual MS relation itself. However, as the majority of data compiled here have slopes of \textit{less} than unity (see Figure~\ref{fig:slope_selection}), and the SFR offsets are \textit{not} equivalent to the mass offsets in some cases (see KE12), the majority of these changes \textit{do} impact the observed normalizations significantly. For these reasons, we choose to only apply explicit IMF adjustments to the derived masses, as the $L$\,--\,$\psi$ relations outlined in the next section implicitly include such adjustments.

\subsubsection{SFR Indicators and the \protect{$L$\,--$\psi$ Relation}}
\label{subsubsec:l_psi}
SFRs are calculated based on observed galaxy luminosities over spectral ranges that correlate with active star formation in the past 10\,--\,100\,Myr. These most commonly are the UV continuum (from $\sim 1500$\,--\,$2800$\,{\AA}), H$\alpha$ emission, and the total IR (TIR) continuum (from $\sim 3$\,--\,$1100$\,$\mu$m). In addition, other SFR indicators, such as 1.4\,GHz emission, have further been developed by exploiting the tight observed radio\,--\,IR correlation \citep{condon92,yun+01,bell03}, as well as from SED fitting to individual bands (cf. S12) or multiband photometry (cf. M13). These indicators are sensitive to the SFR on different timescales: while H$\alpha$ probes SFRs on $< 10$\,Myr timescales, UV and TIR (and by extension 1.4\,GHz) probe SFRs on $\sim 100$\,Myr timescales\footnote{This might affect correlations with mass, especially in star formation is ``bursty''.}.  For additional discussion on the nature of SFR indicators and the assumptions used to derive them, see \citet{hao+11}, \citet{murphy+11}, \citet{murphy+12}, KE12.

Most notably, the studies included here calculate integrated luminosities over the entire wavelength range of interest by fitting specific templates to observed bands. In the IR, these templates most often are taken from \citet{charyelbaz01} (CE01), \citet{draineli01} (DL01), \citet{dalehelou02} (DH02), and \citet{draineli07} (DL07). In the UV, the most commonly used templates are taken from BC03, although \citet{brammer+08} (B08) and \citet{brammer+11} (B11)\footnote{B08 models are calculated based on PEGASE.2 and BC03 models, but the scheme by which this is done is non-trivial (see their Section~2 for more info). B11 models are modified B08 models that take emission line contributions into account.} are also used. To account for additional strong emission lines,  \citet{charlotlonghetti01} (CL01) templates are also sometimes used.

Each of these SFR indicators traces $\psi$ in different ways, over different timescales, and with different calibration issues (see, e.g., Table~1 and Section~3 of KE12), with different $L$\,--\,$\psi$ conversions differing by up to $\sim 50\%$ (see, e.g., the radio SFR calibrations from \citealt{yun+01} and \citealt{bell03}). The standard calibration for most calculated SFRs today is K98, based on a single power-law Salpeter IMF. While K98 gave reasonable SFR calibrations between SFR indicators, for many other wavelengths often studied, the relative calibrations are sensitive to the precise form of the IMF.

KE12 have taken advantage of major improvements in stellar evolution and atmospheric models over the last decade to update the $L$\,--\,$\psi$ relations presented in K98 to a Kroupa IMF, a broken 2-part IMF with a turnover below $\sim 1\,M_\odot$. A Chabrier IMF, which has a log normal distribution from $0.1$\,--\,$1\,M_\odot$, yields nearly identical results to those of KE12 \citep{chomiukpovich11}. As KE12 provide a \textit{self-consistent} set of $L$\,--\,$\psi$ relations for a more realistic IMF (see their Table~1), we opt to convert all previously derived SFRs to this new metric. The ratio of the $L$\,--\,$\psi$ relationships used in individual papers relative to those of K98 and KE12 are listed in Table~\ref{tab:msfr_info}. For SFRs derived from a combination of IR and UV data, we weigh $\psi_{\textrm{UV}}$ and $\psi_{\textrm{IR}}$ according to the calibration presented in \S\,\ref{subsubsec:ext}. For more information on these conversions, see KE12, \citet{murphy+11}, and \citet{hao+11}. See \citet{ranalli+03}, \citet{rieke+09}, and \citet{calzetti+10} for $L$\,--\,$\psi$ conversions at 2\,--\,10\,keV, 24\,$\mu$m, and 70\,$\mu$m, respectively, and \citet{murphy+12} for an empirical comparison of the radio SFR calibration presented here. Additional composite $L$\,--\,$\psi$ relations (i.e. multi-wavelength dust corrections) can be found in \citet{kennicutt+09} and \citet{hao+11}. See \citet{calzetti+07} and \citet{calzetti+10} for more discussion on many of the issues presented here. For convenience, we include a short description of the SFR calibrations used in this work below.

Assuming a solar metallicity and a constant SFR, \citet{murphy+11} find that Starburst99 stellar population models yield a relation between the SFR and the production rate of ionizing photons, $Q(H^{0})$, of
\begin{equation}
\log \psi = \log Q(H^{0}) - 53.14,
\end{equation}
for $Q(H^{0})$ measured in s$^{-1}$ and a starburst age of $\sim 100$\,Myr. Assuming Case B recombination and an electron temperature $T_{\textrm{e}} = 10^4$\,K, the H$\alpha$ recombination line strength is then related to the SFR via
\begin{equation}
\log \psi_{\textrm{H}\alpha} = \log L_{\textrm{H}\alpha} - 41.27,
\end{equation}
for $L_{\textrm{H}\alpha}$ measured in ergs\,s$^{-1}$. This is a factor of 0.68 that of the corresponding calibration from K98 and probes (0-3-10)\,Myr (min-mean-90\%) timescales. Note that the two coefficients are nearly independent of starburst age for ages $\gtrsim 10$\,Myr.

As the integrated UV spectrum is dominated by young stars (K98; S07; \citealt{calzetti+05}), it is a sensitive probe of recent star formation activity. By convolving the output Starburst99 spectrum with the Galaxy Evolution Explorer (GALEX; \citealt{martin+05}) FUV transmission curve, \citet{murphy+11} find
\begin{equation}
\log \psi_{\textrm{FUV}} = \log L_{\textrm{FUV}} - 43.35,
\end{equation}
for $L_{\textrm{FUV}}$ measured in ergs\,s$^{-1}$. This is a factor of 0.63 that of the corresponding calibration from K98 and probes (0-10-100)\,Myr timescales. Likewise, for the NUV, they find
\begin{equation}
\log \psi_{\textrm{NUV}} = \log L_{\textrm{NUV}} - 43.17,
\end{equation}
for $L_{\textrm{NUV}}$ measured in ergs\,s$^{-1}$. This is a factor of 0.64 that of the corresponding calibration from K98 and probes (0-10-200)\,Myr timescales.

\begin{deluxetable}{l c c c c c}[!ht]
\tabletypesize{\scriptsize}
\tablewidth{0pt}
\tablecaption{Impact of Calibrations on MS Parameters \label{tab:error_budget}}
\tablehead{
\colhead{Calibrations} &
\colhead{$\sigma_{i,o}$} &
\colhead{$\sigma_{i,e}$} &
\colhead{$\sigma_{i,f}$} &
\colhead{$d\psi/dt$} &
\colhead{$\psi(0)$} 
}
\startdata
Before (B/M/N) & 0.20 & 0.17 & 0.15 & -0.18 & 2.38 \\
Before (B/M) & 0.19 & 0.13 & 0.1 & -0.20 & 2.48 \\
Before (M) & 0.14 & 0.14 & 0.11 & -0.20 & 2.48 \\
All (B/M/N) & 0.17 & 0.14 & 0.13 & -0.15 & 2.25 \\
All (B/M) & 0.15 & 0.11 & 0.09 & -0.16 & 2.31 \\
All (M) & 0.09 & 0.09 & 0.09 & -0.16 & 2.30
\enddata
\tablecomments{The impact of our calibrations (detailed in \S\,\ref{sec:technicalities}) on interpublication scatters ($\sigma_i$, in dex) before ($\sigma_{i,o}$) and after ($\sigma_{i,e}$) data from the last $2$\,Gyr of the Universe are excluded from our sample, as well as after the first \textit{and} last $2$\,Gyr have been removed ($\sigma_{i,f}$; see \S\,\ref{sec:methodology}), along with the fitted linear evolution of $\psi(t) = (d\psi/dt)\,t + \psi(0)$, for $t$ measured in Gyr and $\psi$ in dex. Both are listed at fixed $\log M_* = 10.5$. The classification of ``bluer'' (B), ``mixed'' (M), and ``non-selective'' (N) studies is detailed in \S\,\ref{subsubsec:selection}.
}
\end{deluxetable}

Due to the presence of dust, much of the light emitted by young stars in the UV is absorbed and re-emitted in the IR. In order to derive a calibration for the TIR, \citet{murphy+11} assume that the entire Balmer continuum is absorbed and re-radiated by dust and that the dust emission is optically thin. After integrating the output Starburst99 spectrum from 912\,--\,3646\,{\AA}, they find
\begin{equation}
\log \psi_{\textrm{TIR}} = \log L_{\textrm{TIR}} - 43.41,
\end{equation}
for $L_{\textrm{TIR}}$ measured in ergs\,s$^{-1}$. This is a factor of 0.86 that of the corresponding calibration from K98 and probes (0-5-100)\,Myr timescales. Note that the exact timescales are sensitive to SFH (see, e.g., \citealt{hayward+14}).

To derive radio SFRs, most studies use the tight, empirical IR\,--\,radio correlation \citep{dejong+85,helou+85,yun+01,bell03}. This relation is most often expressed in terms of $q_{\textrm{IR}}$, where
\begin{equation}
q_{\textrm{IR}} \equiv \log\left(\frac{L_{\textrm{IR}}}{\sn{3.75}{12}\,L_{1.4}}\right),
\end{equation}
and
\begin{equation}
L_{1.4} = \sn{9.52}{18}\,S_{1.4}\,d_L^2\,4\pi(1 + z)^{-0.2},
\end{equation}
where $d_L$ is the luminosity distance of the galaxy in Mpc, $S_{1.4}$ is the 1.4\,GHz flux density in Jy, and a radio spectral index ($S_\nu \propto \nu^{\alpha_S}$) of $\alpha_S = -0.8$ is assumed (e.g., D09; K11). For $L_{\textrm{IR}} \equiv L_{\textrm{TIR}}$, $q_{\textrm{IR}} = 2.64 \pm 0.26$\,dex for SFGs in the local Universe \citep{bell03}; for $L_{\textrm{IR}} \equiv L_{\textrm{FIR}}$, $q_{\textrm{IR}}$ is instead $2.34 \pm 0.26$\,dex \citep{yun+01}. Using the \citet{bell03} $q_{\textrm{IR}}$ value, \citet{murphy+11} find
\begin{equation}
\log \psi_{1.4} = \log L_{\textrm{1.4}} - 28.20,
\end{equation}
for $L_{1.4}$ measured in ergs\,s$^{-1}$\,Hz$^{-1}$. This probes star formation activity in the last $\sim 100$\,Myr.

\subsubsection{Stellar Population Synthesis Model}
\label{subsubsec:sps_model}
In order to derive masses and SFRs, studies need to assume a specific SPS model. The basic ingredients needed to generate an SPS model are relatively straightforward, and are discussed extensively in \citet{conroy13}. Unfortunately, systematic uncertainties in calculating particular phases of stellar evolution, inadequacies in current stellar libraries, and other simplifying assumptions can lead to significant errors that are frequently not taken into account \citep{maraston05,conroy+09,percivalsalaris09,behroozi+10,conroygunn10,conroy13}. For example, uncertainties in modelling the little-understood evolution of thermally pulsating asymptotic giant branch (TP-AGB) stars, blue stragglers (BS), and horizontal branch (HB) stars, all of which are relatively luminous, are significant and can have major impacts on the integrated stellar spectrum \citep{maraston05,melbourne+12} ranging from $\sim 0.1$\,--\,$0.3$ dex  depending on SPS model (\citealt{salimbeni+09}; \citealt{conroy+09}; M10) in a way that is likely mass-dependent \citep{salimbeni+09}. SPS calculations also implicitly assume a well-sampled (i.e., fully populated) and unchanging IMF, which may not always be satisfied \citep{kroupa+13}.

Multiple SPS models are used when fitting for masses and deriving photometric redshifts (photo-z's; see \S\,\ref{subsubsec:photoz}). The models used in the compilation presented here\footnote{An extensive list can be found at {\href{http://www.sedfitting.org}{http://www.sedfitting.org}} and in \citet{walcher+11}.} are taken from \cite{fiocrocca-volmerange97,fiocrocca-volmerange99} (PEGASE.2), \citet{bruzualcharlot03} (BC03), \citet{maraston05} (M05), Charlot \& Bruzual (2007, 2011) (CB07, CB11)\footnote{Although used in the literature, these models have never been formally published.}, \citet{polletta+07} (P07), \citet{rowan-robinson+08} (R08), and \citet{gruppioni+10} (G10). In this study, all masses are calibrated as best possible to BC03 models, as described in Appendix \ref{app:data}. PEGASE.2 models are assumed to be similar to BC03 since they use similar stellar evolution tracks (i.e. the Padova 1994 stellar evolution tracks)\footnote{Technically, BC03 supplements the Padova 1994 tracks \citep{alongi+93,bressan+93,fagotto+94a,fagotto+94b,fagotto+94c,girardi+96} with tracks from the Padova 2000 \citep{girardi+00} and the Geneva \citep{schaller+92,charbonnel+96,charbonnel+99} libraries, as well as a couple others (see their Section~2), but for the most part are dominated by the Padova 1994 tracks.}, and so their derived masses are left unchanged. RR08 models, although empirically-grounded, are regenerated to higher-resolution (and given physical parameters) based upon the SPS models \citet{poggianti+01}. These again use similar stellar evolutionary tracks as BC03 models, and so are assumed to be similar. The models of P07 and G10 are fit only in addition to BC03 models in the studies listed here. As the relative rate of their fitting procedure relative to their BC03 counterparts is not detailed in any of the studies provided, possible differences are not accounted for here. Our assumption that these models lead to broadly similar physical parameters (at least for masses) are also supported by M13 at low redshift, who find that using several different SPS models (e.g., BC03, PEGASE) for the same set of priors results in almost identical stellar mass functions.

M05, CB07, and CB11 models, however, utilize different prescriptions to treat the TP-AGB phase that substantially differ from BC03 models. As the TP-AGB phase tends to dominate much of the starlight at certain wavelengths, the revised prescriptions tend to revise masses downward. We treat these models as identical because they implement similar TP-AGB prescriptions (R12), and implement an adjustment upwards of $C_{M_*,S} = +0.15$\,dex here based on the results of M10 (also an approximate average between the results of \citet{salimbeni+09} and \citet{conroy+09}). Most of the adjustments implemented in this way have the fortunate coincidence of being at similar redshifts ($z \gtrsim 2$) and being selected via Lyman-break criteria (see \S\,\ref{subsubsec:selection}). The exception is So14, for which the offset is closer to $\sim 1.6$ ($C_{M_*,S} = +0.20$\,dex; D. Sobral, priv. comm.). This gives SPS calibration offsets of $C_S = -\alpha \times C_{M_*,S}$.

Some studies choose to eschew using SPS models and SED fitting altogether in favor of analytical $M_*/L$ relations (calibrated on SPS models; e.g., \citealt{mccracken+10} and \citealt{gonzalez+11}) which can applied to a wider selection of data to get ``cheap'' masses (as in P09 and B12). In principle, since these relationships are derived from given SPS models, they should yield good masses on average for similar samples. In addition, many of these $M_*/L$ relations include a built-in color dependence that accounts for variation across the population (i.e. SF vs. quiescent; \citealt{bell+03}). Extending these relationships to larger samples and a wide range of masses, however, might lead to systematic effects in the derived masses relative to those derived directly through SED fitting. For instance, \citet{ilbert+10} find that using the analytical $M_*/L$ relationships of \citet{arnouts+07} for galaxies in the COSMOS \citep{scoville+07} field overpredict masses by an average of 0.2\,--\,0.4\,dex at fixed luminosity. Based on these findings, we adjust P09's masses by an additional $-0.2$\,dex (as they are derived from COSMOS field galaxies, albeit using a slightly different K-band $M_*/L$ conversion), but not those of B12 (which have not been investigated in a similar fashion and also are applied to similar data at similar redshifts). %We conduct a separate check using the new SPLASH data presented in Steinhardt et al. (2014, in prep.), and find that on in general the discrepancies between the \citet{arnouts+07} masses and the new SPLASH masses correlate well for the majority of the high-$z$ galaxies included in our study. Based on these findings and the fact that all the relations used in the works included here are calibrated on BC03 models, we opt not to correct for possible systematic biases these analytical methods might introduce.

\subsubsection{Star Formation History}
\label{subsubsec:sfh}
Different SFHs are needed as inputs to generate the SEDs used to derived galaxy physical properties. The most commonly used of these are declining (D) SFHs, taken from an exponentially-decaying burst with $\psi(t) = \psi_0\,e^{-t/\tau}$, where $\psi_0$ is the SFR at the onset of the burst (and also the scale-factor used in SED fitting procedures), $t$ is the time since the onset of the burst, and $\tau$ is SFR e-folding time. D-SFHs are usually modeled with a wide grid of values ranging from tens of Myr to several Gyr \citep{maraston+10}. Some fitting procedures modify typical D-SFHs by superimposing random starbursts (DRB-SFHs), usually modeled using a tophat function with a constant SFR and a range of intensities and timescales (S07; M13; So14).

Recent studies, however, motivated by the unphysicality of the extremely short ages often derived with D-SFHs -- plus the implied functional form of the MS for galaxies undergoing significant mass assembly -- have advocated \textit{rising} SFHs as better functional fits to the MS than D-SFHs. These have taken several forms: that of exponentially-rising (R) SFHs, with $\psi(t) = \psi_0\,e^{t/\tau}$ \citep{maraston+10,gonzalez+12}; power-law-rising (RP) SFHs, with $\psi(t) = \psi_0\,t^\alpha$ (\citealt{papovich+11}; but see \citealt{smit+12}); and linearly-rising (RL) SFHs, with $\psi(t) = \psi_0 + \frac{d\psi}{dt} t$ (L11). Frequently, constant (C) SFHs are also used as a go-between for the two options, with $\psi(t) = \psi_0$ (L12). Studies may also include ``delayed-$\tau$'' (DT) models, with $\psi = \frac{A}{\tau^2}te^{-t/\tau}$, with $A$ a normalization constant and the rest of the variables defined as above (St14; see also M13). These models allow the construction of D-SFHs ($t/\tau \gg 1$) and RL-SFHs ($t/\tau \ll 1$), as well as several that serve as intermediates between the two. %Finally, in an attempt to forego the use of parameterized SFHs and the dependencies they introduce altogether, some studies (e.g., \citealt{dye+08}) attempt to construct the SFH of a galaxy using separate age blocks, where the SFR in each block is assumed to be constant. While these constant ``blocked'' (CB) SFHs do have some appeal, they're utility is highly limited due to the coarse nature of broadband photometry. While versatile spectral analysis (VESPA; \citealt{tojeiro+07}), which can recover 2\,--\,5 stellar populations from a galaxy spectra and can have quite good age resolution \citet{leitner12}, CB-SFHs tend to become highly degenerate and are computationally intensive.

Given all these current different parametrizations of SFHs, however, it seems that the derived masses are largely independent of the chosen SFH assuming reasonable physical constraints on $\tau$  (\citealt{maraston+10}; R12; So14). The SFRs from R-SFHs/DRB-SFHs in this scenario, by contrast, can be $\sim 0.3$\,--\,$0.4$\,dex (i.e., around a factor of 2) higher than those from simple D-SFHs (\citealt{maraston+10}; So14). Often, however, $\tau$ values do not choose ``physical'' values, an artifact of the ``outshining'' problem -- i.e., that the youngest stars tend to dominate the SED in the UV, where the SFR is best constrained, and thus provide poor constraints on stellar ages without extensive multiband photometry and possibly more complex and physically-motivated SFHs. Instead, fits frequently choose incredibly small, unphysical values of $\tau$ that simply provide the best formal fits to the SED. When $\tau$ is free to choose these small values, \citet{maraston+10} find that the R-SFHs tend to derive masses of $\sim 0.2$\,--\,$0.3$\,dex less than those of D-SFHs, and SFRs $\sim 0.5$\,--\,$0.6$ greater than those of D-SFHs. 

This problem is not fully rectified by using slightly more complex, 2-component SFHs (S11; So14), and frequently requires ad-hoc limits on $\tau$. Indeed, findings from So14 and \citet{behroozi+13} seem to indicate that the uncertainty in SFHs leads almost \textit{all} fitted parameters other than the mass to be extremely unreliable. Given that the SFH of a typical MS galaxy, which likely includes both a rising and declining SFH component of varying degrees as a function of observed mass (and hence formation time; see \S\,\ref{subsec:ms_tracks}), as well as the small sample sizes of both studies, we do not utilize any possible SFH-based SFR adjustments here.

\subsubsection{Dust Attenuation}
\label{subsubsec:ext}
Most studies have measured extinction/attenuation photometrically\footnote{This excludes extinctions derived via, e.g., emission line ratios (e.g., \citealt{garnbest10}; \citealt{sobral+12}; \citealt{stott+13}) or H$\alpha$ vs. IR measurements \citep{ibar+13}.} by dust using $E$($B$\,--\,$V$), either derived through SED fitting ($E_S$) or using the IR-to-bolometric luminosity ratio (IRX) via the IRX-$\beta$ (where $\beta$ is the UV slope) relation ($E_\beta$) of, e.g., \citet{meurer+99} (M99). In the literature, $E_S$ values tend to be used cautiously because of the degeneracies between age and reddening (and hence the assumed metallicity and SFH), the parametrization of the extinction curve, and the very limited grid space, leading $E_\beta$ values to be preferred. However, while observational methods such as the IRX\,--\,$\beta$ relation have for a long time been found to give accurate UV-corrected SFRs compared to those derived from UV+IR observations (B12), it exhibits a significant amount (up to an order of magnitude in some cases) of scatter \citep{boquien+12}. In addition, results from \citet{wuyts+11b,wuyts+11,price+13} imply that simple extinction corrections are insufficient to accurately correct for dust, and that more complex geometrical (i.e., patchy) dust models are needed.

Regardless, by observing UV versus UV+IR emission from an ensemble of SFGs, one can apply average extinction corrections that should be sufficient to convert from the observed UV-derived SFR to the bolometric SFR. As observations imply that average IRX of galaxies evolves strongly at higher redshifts (e.g., B12), we will approximate the average IRX observations using results presented by R12a ($\psi_{\textrm{bol}} \sim (5.2 \pm 0.6)\,\psi_{\textrm{UV}}$) at low-$z$ ($z < 4$), and B12 ($\psi_{\textrm{bol}} \sim (2.5 \pm 0.5)\,\psi_{\textrm{UV}}$ at $z \sim 4$\,--\,$5$; see their Table~6 for the full list of corrections) at high-$z$ ($z > 4$). We apply these IRX values to weight the corresponding $\psi_{\textrm{UV}}$ and $\psi_{\textrm{IR}}$ components from $\psi_{\textrm{UV+IR}}$ data accordingly when adjusting SFR values using the $L$\,--\,$\psi$ relations from KE12 as well as to correct for dust attenuation in data which only reports observed UV luminosities.

\subsubsection{Extinction Curve}
\label{subsubsec:ext_law}
Multiple extinction curves have been used in the literature to account for the effects of dust on the observed SEDs in SPS-generated spectra. The ones used in the papers presented here are taken from: \citet{prevot+84} (P84), from observations of the Small Magellanic Cloud (SMC); \citet{cardelli+89} (C89), from various sources in the optical and NIR; \citet{calzetti+00} (C00), from observations of SFGs; \citet{madau95} (M95) and \citet{charlotfall00} (CF00), from observations of nebular attenuation; and \citet{charyelbaz01} (CE01), from observations of local galaxies. A hybird C00 model with a bump at 2175\,{{\AA}} (C00b) to account for graphite and polycyclic aromatic hydrocarbon (PAH) features is also used in some cases \citep{ilbert+09}. Although the impact of extinction can be as high as a factor of $\sim 5$ (R12a), the impact of using \textit{different} extinction curves appears negligible\footnote{This result, however, may be dependent on both the wavelength probed and the level of dust present (D. Kashino, priv. comm.).} \citep{papovich+01,dickinson+03}. This will not be accounted for here. %\footnote{As a brief qualitative check, we use the new SPLASH data (St14; Capak et al. 2014, in prep.) to examine the effects of varying extinction curves and  BC03 models. While variations tended to affect the general shape of the MS derived at high redshifts, as a whole the they had relatively effects on the slope/normalization considering the systematics involved (O. Ilbert, priv. comm.).}.

However, while the use of different extinction curves might produce similar results, using \textit{uniform} extinction curves might still produce subtle biases in derived physical results. In particular, if the general shape/amount of PAH emission is correlated with the fitted amount of extinction, then the use of models with constant (or a lack of) 2175\,{\AA} features will produce notable biases in SED-derived galaxy properties. Using a flexible parametrization of dust attenuation, \citet{kriekconroy13} report a negative correlation between the slope of the attenuation curve and the strength of the 2175\,{\AA} bump (i.e., SED types with steeper attenuation curves have stronger bumps.). They find this leads to biases in derived dust attenuation (large) as well as masses (small) and specific SFRs (sSFRs, $\phi \equiv \psi/M_*$; also small). In addition, they find edge-on and/or low-sSFR galaxies tend to have steeper attenuation curves, while face-on and/or high sSFR galaxies tend to have shallower attenuation curves, implying possible dependencies on orientation. Taking these findings into account may better improve future SED-fitting procedures.

While these findings imply that current SED-fitted physical parameters might display parameter-dependent systematic biases (but see \citealt{garnbest10}),
% no code used in this compilation is able to correct for this\footnote{The exception here is Le PHARE, which fits a bump with varying strengths superimposed on the usual C00 dust attenuation curve if it provides a better fit to the data. However, this is only used when deriving photometric redshifts, not physical parameters.}. We thus 
we do not attempt to account for this effect here.

\subsubsection{Emission Lines}
\label{subsubsec:emission}
Strong emission lines, such as Ly$\alpha$, H$\alpha$, H$\beta$, [OII], and [OIII] can significantly alter the SED by contaminating observed band photometry. These lines decrease (i.e., make more luminous) the observed magnitude in a given band by up to several tenths of a mag at higher redshifts (\citealt{ilbert+09}; S11; \citealt{stark+13}). %\footnote{See also Figure~3 from the Le PHARE documentation, available at {\href{http://www.cfht.hawaii.edu/~arnouts/LEPHARE/lephare.html}{http://www.cfht.hawaii.edu/~arnouts/LEPHARE/lephare.html}}, and Figure~3 from S11.} 
These differences, not accounted for (correctly) by most SPS models \citep{ilbert+09}, can significantly affect the derived physical parameters taken from the SED fitting process and impact the quality of derived photo-z's. The relative impact depends on the number of bands included in the fit, their respective width, and the redshift of the source: for surveys with a large number of bands (e.g., COSMOS), this effect will be somewhat washed out; however, for surveys with only a handful of bands, this effect can make a big difference \citep{kriekconroy13}.

\citet{stark+13} show the effects that emission line contributions can have on the observed $M_{\textrm{UV,1500}}$\,--\,$M_*$ (i.e. $M_*$\,--\,$\psi$) relationship at high redshifts (where emission line contamination is most severe), and demonstrate that while on average the slope of the relation remains the same, the overall fitted masses decreases substantially. We choose to implement high-$z$ corrections from their Figure~7 (taken from \citealt{robertson+13}), which lead to mass corrections of $C_{M_*,E} \sim (0,-0.03,-0.18,-0.40)$\,dex for galaxies at $z \sim (4,5,6,7)$, respectively. Like \citet{robertson+13}, we have chosen to apply the correction without the hypothesized redshift evolution of the H$\alpha$ equivalent width (EW) due to the age dependence it would introduce. If we had taken these into account, the corrections listed above would be even larger (e.g., up to an order of magnitude at $z \sim 7$). This leads to MS calibration offsets of $C_E = -\alpha \times C_{M_*,E}$.

\subsection{Minor/Unknown Influences}
\label{subsec:small_problems}

\subsubsection{Cosmology}
\label{subsubsec:cosmology}
The effects of differing cosmologies are accounted for by calculating the ratios between luminosity distance, $d_L(z)$, derived from two different cosmologies, and, given the observed redshift range of a sample, applying a $d_L^2$ correction ($C_{d_L}$) at the expected median $z$ of galaxies in the sample after weighting for first-order volume effects. This volume-weighting assumes an approximately constant number density and slowly changing mass distribution of MS galaxies within the redshift bin in question. The effect is negligible in these cases, only changing the derived $d_L$ corrections by less than a percent, but are more significant when we use it later to deconvolve the scatter about the MS (see \S\,\ref{subsec:scatter}). Relative to the possible impacts listed in \S\,\ref{subsec:big_problems}, we find this effect is small, in all cases $< 0.05$\,dex. As they are straightforward to derive, however, we choose to apply them out of completeness (see Table~\ref{tab:corr}). Because they boost the luminosity of the entire spectrum, cosmology differences should lead to both increased masses and SFRs. Our calibration offset is then $C_C = (1-\alpha) \times C_{d_L}$.

\subsubsection{Photometric Redshifts}
\label{subsubsec:photoz}
For the majority of galaxies used in the studies included here, redshifts have been derived photometrically (photo-z's) via SED fitting rather than spectroscopically (spec-z's). SPS models are used to derive these photo-z's, which simultaneously provide the masses (and sometimes SFRs) used in these studies. Photo-z's have varying precision, ranging from $0.8\%$\,--\,$3\%$ scatter compared to their spec-z counterparts \citep{ilbert+13}, and can be subject to ``catastrophic failures'' where the photo-z's and spec-z's disagree by more than $15\%$ ($\eta \equiv |z_{phot} - z_{spec}| / (1+z_{spec}) > 0.15$). Note that these statistics are only available when spec-z's are available, and thus are often based on only the brightest galaxies (which are often targeted in I-band selected surveys).

Besides just misfits caused by bad photometry, a small number of bands, or a multi-peaked redshift probability distribution function (PDF), catastrophic errors can occur systematically by, e.g., confusing the Lyman break at $\sim 1220$\,{\AA} and the Balmer/$4000$\,{\AA} break (St14). Although the errors from the average scatter are small, the effect of catastrophic failures on the $M_*$\,--\,$\psi$ relation relative to that of confirmed spectroscopic samples has yet to be fully investigated. We conduct a simple experiment to qualitatively assess the effects of catastrophic errors $M_*$\,--\,$\psi$ relationship, and find that their effect on the overall distribution appears small, even for a large fraction of catastrophic errors (see Appendix~\ref{app:eta_test}).

In many cases, photo-z's have not been compared with spec-z's across the full mass and redshift ranges to which they have been applied; this serves to both check their accuracy and are often necessary for calibration purposes \citep{hildebrandt+10,abdalla+11,dahlen+13}. Existing spec-z or narrow-band selected studies, however, seem to agree well with photo-z derived distributions (e.g., S12; C14). In addition, simulated errors and catastrophic failure rates agree with the measured spectroscopic samples and are accounted for in some works (e.g., \citealt{ilbert+13}). Finally, the quality of photo-z's have been checked with pair statistics and cross correlations, which seems to confirm errors derived from spec-z's \citep{benjamin+10}.

On the whole, photo-z methods do not seem to display large redshift biases relative to spec-z's, and the likely induced scatter is small relative to scatter about the MS and other systematic errors \citep{hildebrandt+10,abdalla+11,dahlen+13}. Any possible systematic offsets they have relative to spec-z's are not accounted for in this study.

\subsubsection{SED Fitting Procedure}
\label{subsubsec:sed_fitting}

Besides the variations in generating SEDs that have been detailed above, the SED fitting procedure used to derive photometric redshifts differs for different codes\footnote{See \citet{hildebrandt+10}, \citet{abdalla+11}, and \citet{dahlen+13} for a good sampling of current codes.}. %Codes such as HyperZ \citep{bolzonella+00} or Le PHARE \citep{arnouts+99,ilbert+06} follow a grid-based template fitting method, and achieve best fits using a standard $\chi^2$-minimization technique. Other codes, including ZEBRA \citep{feldmann+06} and iSEDfit (M13), implement varying Bayesian/hybrid approaches to the SED fitting procedure. Still others, including GalMC \citep{acquaviva+11}, $\pi$MC$^2$ \citep{pirzkal+12}, and SATMC \citep{johnson+13}, use Monte Carlo Markov Chain (MCMC) fitting techniques in order to attempt to determine best-fit physical parameters and better constrain the underlying posterior parameter distribution. Several photo-z codes, including ANNz \citep{collisterlahav04,abdalla+11} and the Sloan Digital Sky Survey (SDSS; \citealt{york+00}) pipeline, even opt out of the standard template-fitting procedure altogether and utilize neural networks, deriving a parametrization of photo-z's via a training set of multiband photometry and spec-z's, which they then apply to redshift ranges outside the training sample. 
Each of these fitting procedures, besides contamination from catastrophic errors, might exhibit biases in the determined photo-z's relative to the true spec-z's and/or each other. In particular, the best fit parameters derived from SED fitting tend to be sensitive to small changes in parameter space and errors on the photometry. This can be reduced by incorporating a wider range of parameter space into the final mass, such as by taking the median mass across all solutions in the entire multi-dimensional parameter space for each fit that lies within 1$\sigma$ of the best fit (So14; see Appendix~\ref{app:mass_disc}). At the moment, however, since such procedures are not widely used, we will not attempt to account for these effects here.

In order to directly test different photo-z codes/fitting procedures against each other, \citet{hildebrandt+10}, \citet{abdalla+11}, and \citet{dahlen+13} compare photo-z code performance against each other using identical samples. Their results indicate that, in general, all codes produce reasonable photo-z estimates in both an absolute and relative sense, although using a training set of spec-z priors reduces both the scatter and the fraction of catastrophic errors. Their findings also indicate that using a training set from a small region of the sky does not seem to produce biases when applied to larger survey areas, and that the median of all codes seems to do better than any individual code at matching spec-z's.

Most crucially, \citet{dahlen+13} find that photo-z errors and the fraction of catastrophic errors are the largest for data at higher magnitudes (i.e., are fainter with larger error bars), which implies the majority of photo-z errors should happen preferentially to low-mass, low-SFR galaxies observed within any given sample (precisely where spec-z's are lacking). While these results provide areas for photo-z codes to improves and that should be investigated, based on these overall positive results, we do not opt to attempt to account for possible differences among SED fitting procedures.

\subsubsection{Metallicity}
\label{subsubsec:metallicity}
Stellar evolutionary tracks (i.e. isochrones) used by SPS models can be strong functions of metallicity. Currently, SPS models do not model metallicity evolution self-consistently, which would involve tracing the evolving metallicity content of stellar populations over time from supernovae injections, mixing, elemental abundance patterns, etc., and their subsequent impact of star formation and evolution. Instead, many resort to using simple stellar populations (SSPs), which follow the evolution in time of the SED of a single, coeval stellar population at a single fixed metallicity and abundance pattern. The effects of using SSPs relative to populations where metallicity evolution is taken into account is not fully understood. As SSPs are utilized in all SPS models considered here and a fundamental assumption in the derivation of physical parameters from SEDs, we take this to be an unknown systematic that cannot be quantified and/or accounted for at this time. See \citet{conroy13} for further discussion.

While systematics from using fixed-metallicity SSPs are not accounted for, we can at least investigate a related assumption: the effects using \textit{different} metallicities in SSPs have on the derived physical parameters. At low redshifts ($z \lesssim 1$), results from M13 (see their Appendix~B) seem to indicate that assuming a fixed solar metallicity relative to a much wider metallicity distribution (from $\sim 0.2$\,--\,$1.5$\,$Z_\odot$) does not have a major impact on the resulting mass distribution both at fixed redshift and as a function of redshift (their impact on SFRs has yet to be thoroughly investigated). Based on these findings, and the fact that the majority of studies included here include sensible metallicity priors, we do not attempt to implement any adjustments due to  possible metallicity-induced effects.

%At much higher redshifts ($z \sim 7$), the results of \citet{schaererdebarros10} seem to imply much of the same. Namely, they find that assuming a fixed solar metallicity relative to lower metallicities (which might be more common at higher redshift) has much smaller impacts on the derived physical properties (most notably masses) than any of the other assumptions they investigate (SFH, inclusion of dust extinction, adopted extinction curve, inclusion of nebular emission and the assumptions made to do so), in line with many of the results discussed here.

\begin{figure*}[!ht]
\plotone{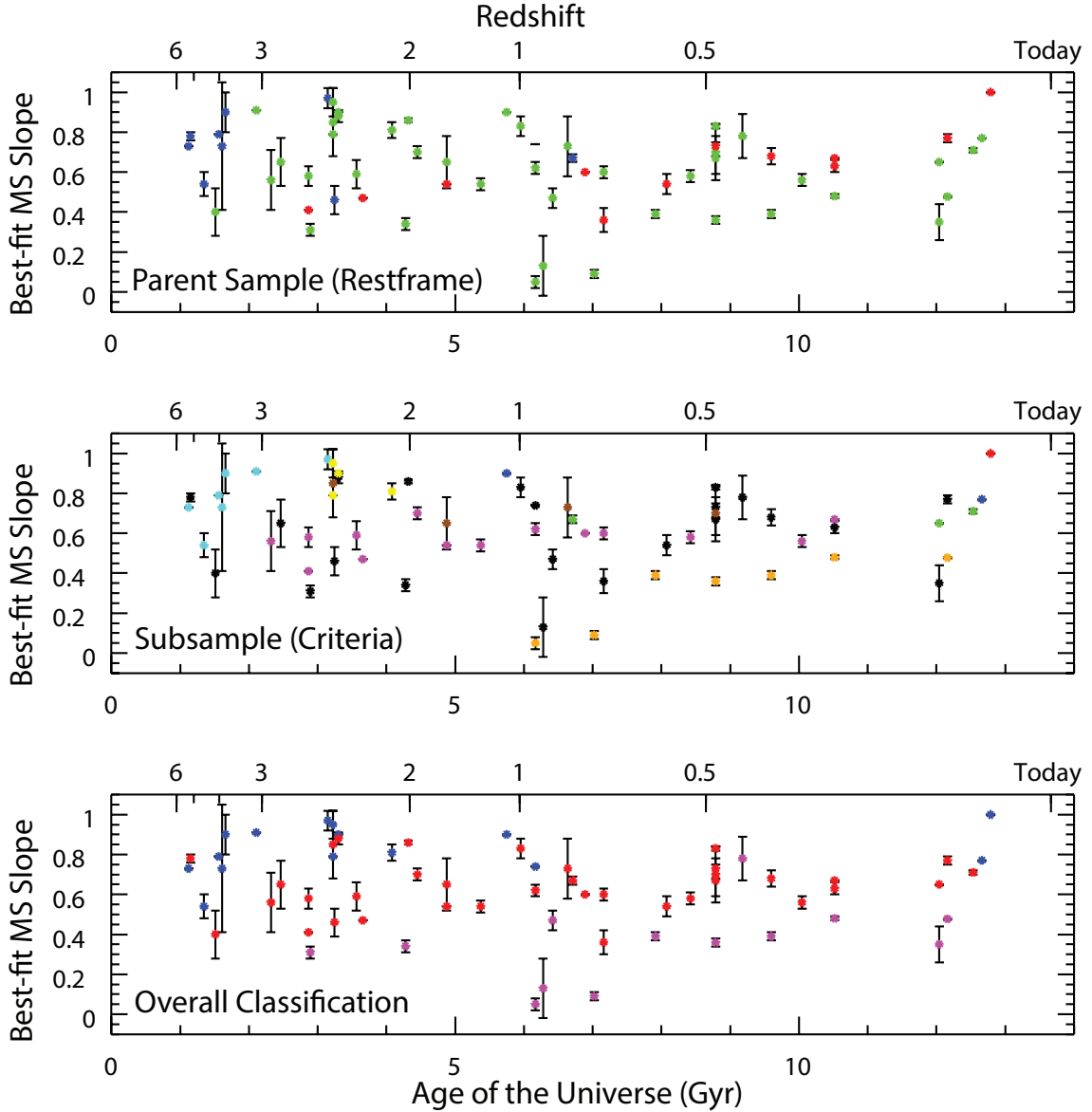}
\vspace{-10pt}
\caption{The slope of the MS as a function of time, color-coded according to varying classifications of selection type. Top: Color-coded based on restframe wavelength used to select the parent sample, with restframe UV-selected samples in blue, restframe optical/NIR-selected samples in green, and restframe FIR-selected samples in red. Middle: Color-coded based on the procedure used to select the star-forming subsample used in the reported fit. Those selected using Lyman-break criteria are in cyan, using blue color in blue, using $sBzK$ in yellow, using bimodalities in the $M_*$\,--\,$\psi$ plane in orange, using emission lines in green, using LIRG criteria in red, using NUVrJ or UVJ color criteria in purple, and using a 2$\sigma$-clipping procedure (for the reported fit) in brown. Subsamples established using simple mass/luminosity-completeness criteria and that otherwise have no substantive cuts are in black. Bottom: Color-coded according to likely biases, with blue points taken from studies which use ``bluer'' selection methods biased towards bluer, highly active, non-dusty galaxies, red points taken from studies which use more ``mixed'' selection methods not biased in the same way (see ~\S\,\ref{subsubsec:selection}), and purple points taken from ``non-selective'' studies that have not effectively separated star-forming and quiescent galaxies in their samples (see \S\,\ref{sec:methodology}). The readily apparent dichotomies in slope, with ``bluer'' data displaying slopes between 0.75\,--\,1, ``mixed'' data displaying slopes around $\sim 0.6$ (and almost always $< 0.8$), and ``non-selective'' data displaying slopes $\lesssim 0.4$, seems to support our classifications detailed in \S\,\ref{subsubsec:selection}. Errors in individual MS slope measurements have been taken from their respective studies when available.
}\label{fig:slope_selection}
\end{figure*}

\subsection{Observational Biases}
\label{subsec:biases}

\subsubsection{Bias between the Mean and Median of a Log-Normal Distribution}
\label{subsubsec:lognormbias}

While the mean and median of a log-normal distribution are approximately identical when calculated in log space, the expected mean of a log-normal distribution is skewed in linear space \citep{behroozi+13}. This offset depends on the scatter present in the distribution -- for a log-normal distribution with a median of 1 and scatter $\sigma$ (dex), the expected mean will instead be
\begin{equation}
\langle x \rangle = \exp\left[0.5\left(\sigma\ln 10\right)^2\right].
\end{equation}
This leads to an offset between the mean and median in log space of
\begin{equation}
\Delta x \approx 1.15\sigma^2.
\end{equation}
For an intrinsic scatter of $\sigma = (0.30,0.35)$\,dex, this corresponds to $\Delta x = (0.10,0.14)$\,dex. As all radio data included here (D09; P09; K11) have used median stacks to find the \textit{true mean} of the SFR for a given mass bin \citep{white+07}, this effect translates to a systematic overestimation of the SFR at a given mass by approximately 0.1\,dex compared to most other data included here. This effect has been included in the $C_\psi$ calibration offsets presented in Table~\ref{tab:corr}.

\subsubsection{Selection Effects}
\label{subsubsec:selection}
Selection effects within each study -- not to mention within the definition of the MS itself -- also can affect both the derived slopes and their evolution as a function of redshift (O10; K11; W12). While most of these (see Table~\ref{tab:msfr_info}) are efficient at selecting SFGs, they do not all select the same population. As K11 show in their Appendix~C, $B$\,--\,$z$ vs. $z$\,--\,$K$ ($sBzK$; \citealt{daddi+04b} (D04)) selection -- and a bluer selection criteria in general -- is biased towards more ``active'' SFGs (i.e. with higher (s)SFRs), excluding good portions of galaxies that are classified as SFGs via other selection mechanisms ($\textrm{NUV}-r$ vs. $r-J$; $\textrm{NUV}rJ$), and give steeper MS slopes (see also O10 and K11). W12 shows that this effect further translates into an inherent bias against redder, more dust-attenuated SFGs (see their Figure~3), which have lower slopes compared to their bluer, less dust-attenuated counterparts. See also \citet{sobral+11} for more discussion on this issue.

Because of these effects, selection methods that are inherently biased towards bluer, highly-active, non-dusty SFG populations will preferentially select a subset of the MS population with a higher slope relative to other selection mechanisms. These selection methods include: $sBzK$, used to select SFGs from $1.4 < z < 2.5$ (D07; R11; K13); the Lyman break (\citealt{steidel+99}; \citealt{stark+09}; \citealt{bouwens+11}; B12), used to select high-$z$ Lyman-break galaxies (LBGs); and $U - g$ vs. $M_{\textrm{bol}}$, or any other cut on the color-magnitude diagram (CMD) that explicitly selects based on (blue) color (E07). We therefore classify these methods as ``bluer'' selection mechanisms, along with luminous infrared galaxy (LIRG) selected samples such as that of E11 -- although LIRGs are definitely SFGs, nearby LIRGS tend to be highly active SFGs with large amounts of dust attenuation and extreme amounts of star formation, in contrast to ``regular'' MS galaxies that are more similar to the Milky Way at low redshifts from, e.g., B04 and S07.

As can be seen by comparing selection methods from, e.g. E07 (see their Figure~2) and \citet{ilbert+13} (see their Figure~3), non-``bluer'' selection methods (broadly classified as ``mixed'') seem to provide not only a ``cleaner'' cut between SFGs and quiescent galaxies, but a more diverse star-forming population. The total classification scheme of these two selection types is listed in Table~\ref{tab:msfr_info}. As mentioned earlier, while these different selection methods do not seem to affect the average observed SFRs across different publications, they do seem to influence the derived slopes and the intrinsic scatter (see \S\,\ref{subsubsec:bzk}).

Although differences between bluer and mixed selection criteria can lead to differences in the derived MS relations, all MS studies should ideally only include SFGs in their analysis. Several of the studies included here do not opt to impose a color-color cut of some sort to separate out SFG and quiescent galaxy populations\footnote{Although we have used the terms extensively, the actual definition of what constitutes a ``star-forming'' vs. ``quiescent'' galaxy remains somewhat arbitrary. While there appears to be a strong bimodality in color-color space (e.g., \citealt{ilbert+13}), it is much less pronouced in mass\,--\,SFR space (S09; M13; So14; C14).} (C09, So14, and C14). These ``non-selective'' studies consequently display prominent differences from SFG-only studies. As quiescent galaxies ``contaminate'' their highest mass bins at a wide range of redshifts, their lower SFRs significantly reduce the slope. In addition, their increased prevalence at lower masses at lower redshifts (as more and more galaxies ``quench'') leads to increasing offsets in normalizations for any flux-limited survey. This effect is accentuated by increases in survey sensitivity, which can drive the SFR floor lower at all massess.

As expected, we find that all non-selective studies agree with other data relatively well at lower masses (especially at higher redshifts), but disagree significantly at higher masses due to shallower MS relations\footnote{This is not completely true for C14's data -- see Appendix~\ref{app:data} for a more extensive discussion.}.

In Figure~\ref{fig:slope_selection}, we plot the derived slopes of each MS sample color-coded by selection method of the parent sample, the subsample used for analysis, and our groupings listed above. We find that, while some biases in MS selection might emerge from parent samples selected primarily on restframe UV, the majority of biases occur in the precise selection of the subsample. As expected, we also find that ``bluer'' MS observations display slopes between $\sim 0.8$\,--\,$1$ and are relatively similar over the majority of the age of the Universe, while ``mixed'' observations center around $\sim 0.6$ and display possible time-dependencies\footnote{We note that this bimodality in slope determination seems to break down at higher redshift, where LBG-selected samples display lower slopes of $\sim 0.7$. In addition, most of the high slopes at lower redshift for our ``mixed'' data are from D09; the bimodality is much sharper when their data is excluded.}. Based on these results, we decide to use our bluer/mixed classification scheme to account for different biases inherent in SFG/MS selection, preferring mixed selection methods to bluer ones since they give us a larger and more diverse SFG sample while still excluding most quiescent galaxies.

\subsubsection{Scatter and $sBzK$ Selection}
\label{subsubsec:bzk}

We find that the true and deconvolved scatters (see \S\,\ref{subsec:scatter}) reported in all $sBzK$-selected studies (D07; R11; K13) are systematically lower than reported in other papers, even for large sample sizes (R11). Furthermore, their resulting values are low enough to likely be unphysical, especially given the possible $\sim 0.1$\,dex of intrinsic scatter in determining mass (even when considering possible convariances; see Appendix~\ref{app:mass_disc}). This seems to indicate that the scatter observed in these papers is not representative of the redshift range that they encompass, and hence that $sBzK$ is substantially biased compared to other selection mechanisms (data taken from LBGs, for instance, show similar scatters as other ``mixed'' samples; M10; L12; R12)).

%At first glance, this result seems plausible. UV selection methods by nature are selecting towards high-active, non-dusty SFGs; $sBzK$ especially requires a high S/N cut to be effective, further skewing the distribution towards brighter, bluer objects. And with more massive objects on average redder and more extincted, we would expect that our distribution should be truncated and more tightly correlated than normal. In contrast, this reasoning should apply to all types of UV selection mechanisms, most notably LBG selection. 

Most likely, this difference is due to an inherent bias built into the $sBzK$ selection mechanism itself. As outlined in D04, the $B$\,--$z$/$z$\,--$K$ line used to select $sBzK$ galaxies was designed to be parallel to the reddening vector. However, due to the age-extinction degeneracy, this means that age runs perpendicular to the selection function, and implies that you will systematically be missing older (and hence likely more massive and dusty) galaxies because of their redder colors. As LBG selection mechanisms are not explicitly designed this way, although redder SFGs are still selected against, the selection bias is not as systematic or complete as using $sBzK$. Thus, while $sBzK$ is effective at selecting for SFGs for $1.4 < z < 2.5$, the distribution and scatter of the sample is biased ($\sigma_{t,BzK} \sim 0.1$ while $\sigma_{t,\textrm{med}} \sim 0.2$; see \S\,\ref{subsubsec:scatter_true}).

However, tests we have conducted on the \citet{ilbert+09} COSMOS catalog find that $sBzK$-selection is actually quite efficient at selecting out SFGs between $1.4 < z < 2.5$ (O. Ilbert, priv. comm.; although see D09, P09, and K11). While this does not rule out the possibility of intrinsically biased selection, it does point to another possibility. Instead, the narrower distribution might likely arise due to systematic biases in calculating bolometric SFRs. Extinction corrections in $sBzK$ samples are determined exclusively from $B$\,--$z$ color (D04), and so the same bands used to select the sample are also used to determine the dust attenuation. Using a large sample of LBGs, B12 finds that using the same passbands for both selection and dust attenuation measurements leads to large biases in the derived dust attenuations. This implies we might be witnessing a similar problem with $sBzK$-selected sample here, where the problem is not with inherent selection biases, but with substantially biased calculations of dust attenuation and hence a narrower distribution bolometric SFRs.

We note that apart from $sBzK$, the only other data points which display lower-than-average scatter are those with small sample sizes ($N \lesssim$ 250), as well as those from S09 (although S09 uses a 2-sigma-clipped fitting procedures that biases the scatters by default; see Table~\ref{tab:msfr}).

\subsubsection{Additional Biases}
\label{subsubsec:biases}
There are several main biases that characterize observations of the SFG MS: incompleteness, Malmquist bias, and Eddington bias. If a survey is not mass-complete, observations will be biased towards bluer galaxies both due to the flux-limited nature of most surveys as well as many SED fitting procedures (which are sensitive to the signal-to-noise ratio of the photometry; see \citealt{dahlen+13}), which will affect properties of the MS relation below the mass completeness limit. This is easily rectified by only using data where the survey is approximately mass complete, as is done in, e.g., W12 (see their Figure~1). For the studies collected here, we find that on average this effect on the reported MS fits is small compared to the other issues discussed above, and therefore do not correct for it here.

There are also other competing effects in most surveys that tend to become more prominent near mass limits, most notably a Malmquist bias of selecting galaxies with larger SFRs at a given stellar mass in a flux-limited sample (see R12's Appendix~B). As the (s)SFR of galaxies are a strong function of their mass, Malmquist bias will result in higher (s)SFRs derived on average for a given mass for masses where the flux limit approaches the hypothesized distribution. As shown in R12 (see their Figure~26), this bias can lead to derivations of sSFRs from their true values on the low mass end by a factor of $\sim 3$\,--\,4. In general, such a bias is strongest the more flux-limited a sample is (and as such is different from just strict mass completeness), and most prominently affects galaxies on the low-mass end of the MS. As this leads to higher average SFRs for these objects as compared to higher mass objects, this would lead to a shallower fitted slope\footnote{In R12, simulations of this effect lead to a change in slope from unity to $\sim 0.5$, which could imply that all MS observations really should have slopes of approximately unity. However, as relatively mass-complete surveys (e.g., K11; W12) find slopes much less than unity, such an argument is strongly disfavored.}.

Another possible impact of Malmquist bias would be a strong selection effect towards bursting low-mass SFGs near the detection limits, as they would be more likely to be detected over their non-bursting counterparts. This might lead to a strong bias in SED-fitting procedures towards very young ages for low-mass, low-SFR systems (bottom-left of the MS), which might in turn bias MS slopes. As such a trend is not seen in R12, such a strong systematic bias is likely not strong.

On the high-mass/high-SFR end, Eddington bias (i.e. that random scatter in a given mass/luminosity bin will preferentially scatter objects up into higher mass/luminosity bins because they have comparatively fewer objects) tends to be much more dominant. Such bias might lead to a flattening of the MS at high masses as lower mass (and hence lower SFR) objects are scattered up into higher mass bins. This effect, however, should also lead to objects with lower SFRs being upscattered into higher SFR bins (assuming, of course, that the SFR/mass derivations are somewhat independent of one another, as they are constrained by different portions of the SED), causing an upturn in the MS relation. These two effects should then combine to produce a more densely populated upper population in the derived MS relation, with a downturn for samples with well-constrained SFRs and less well-constrained masses (e.g., empirically-derived H$\alpha$ SFRs from high-quality spectra and SED-fitted masses from multiband photometry), an upturn for samples with well-constrained masses and poorly constrained SFRs (i.e. extintion-corrected UV SFRs with SED-fitted masses from extensive, high-quality multiband photometry), and a similar slope for samples with about equivalent constraints on both (i.e. both masses and SFRs derived through SED fitting).

All these scenarios are only relevant, however, assuming photo-z accuracy does not depend on other physical parameters and is relatively good for the majority objects included in the fit. This is not necessarily true -- like masses and SFRs, photo-z accuracy is sensitive to the overall shape of the SED. This can lead to complex covariances which have not been fully explored (but see Appendix~\ref{app:eta_test}). If the photo-z is incorrect, then masses may likely be overestimated and SFRs derived from the incorrect portion of the spectrum, which will lead to more complicated behavior. Note that not all studies are affected by photo-z biases: $sBzK$-, LBG-, and line-selected samples have precise redshift distributions that are applied to the mass fitting rather than using the photo-z for each individual source.

In all cases, however, there will be a bias towards higher numbers of mass/SFR objects. These effects should not have a large impact on MS relations derived directly from nonstacked data (due to the small number of objects at the high mass end), although for stacked data (especially using mean instead of median stacks), this effect is expected be more prominent. In many studies, however, the slope of the MS is computed from binned data, with mass bins often assigned equal weight regardless of the size of each bin. In these cases, using mass bins is essentially the same as stacking and implies likely contributions from Eddington bias on the high-mass end of the MS. Although we do not attempt to correct for it here, we thus cannot exclude a significant contribution from Eddington bias for massive galaxies.

\begin{figure*}[!ht]
%\vspace{-180pt}
\plotone{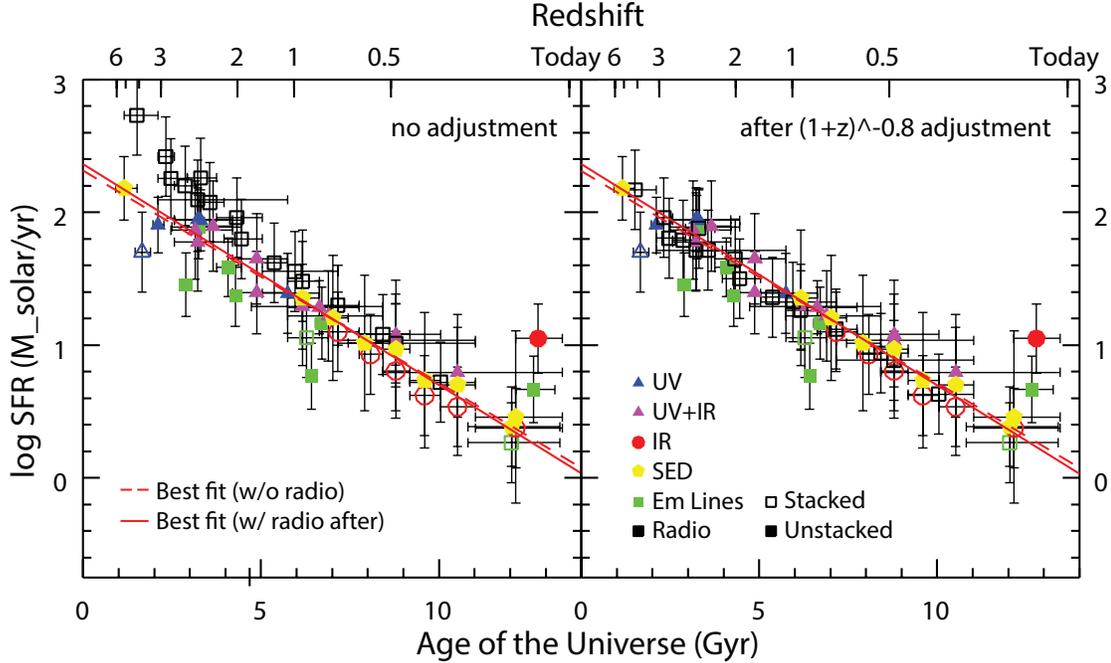}
\vspace{-10pt}
\caption{$\log\psi$ vs. $t$ for $\log M_* = 10.5$ before (left) and after (right) we introduce our -0.1\,dex stacking correction to convert between the mean and median in log space and the $(1+z)^{-0.8}$ calibration offset to the radio data, calculated using the best fit MS relations from each of the 25 studies listed in Table~\ref{tab:msfr_info}. Data points are colored based on the SFR indicators they primarily use: blue = UV, purple = UV+IR, red = IR, green = emission lines, yellow = SED fitting, black = radio. Stacked data in all panels are indicated by hollow points, while non-stacked data are filled. The horizontal errors indicate the redshift range spanned by each particular observation, while the vertical errors are the ``true'' scatter about each MS observation (see \S\,\ref{subsubsec:scatter_true}). For stacked data, the plotted true scatters are derived assuming a fiducial intrinsic scatter of $0.3$\,dex. The best fits excluding (pre-``correction''; dotted line) and including (post-``correction''; solid line) the radio data are overplotted in each panel. The marked difference in stacked radio SFR observations at higher redshifts, their excellent agreement following our corrections, and the small effect these new radio SFRs have on the fit all suggest we are observing a real systematic difference between radio SFRs as compared to other SFR indicators.
}\label{fig:radio_sfr_t}
\end{figure*}

\subsection{Disagreements in Radio SFR Data}
\label{subsec:radio}
In the left panel of Figure~\ref{fig:radio_sfr_t}, we plot SFRs at fixed $\log M_* = 10.5$ after applying the calibrations discussed above and listed in Table~\ref{tab:calibration}. As can be seen, SFRs derived from stacked radio observations are systematically larger than those derived via other methods, and also seem to display a steeper time dependence. This is consistent across all radio data included in this study and all mass ranges probed.

In order to characterize the possible time/redshift-dependent component of the observed offset, we fit our data using the methods described in \S\,\ref{sec:methodology} for the radio data alone as well as for all other data excluding the radio data. Parametrizing $\psi \propto (1+z)^\gamma$, we find that, at fixed $\log M_*=10.5$, $\gamma \sim 3.6$ for the radio-only fit and $\sim 2.8$ for the radio-excluded fit (i.e. larger radio SFRs relative to other SFR indicators). In the right panel of Figure~\ref{fig:radio_sfr_t}, we plot radio SFRs after accounting for this $(1+z)^{\sim 0.8}$ systematic offset (which we term $C_R$) using the median redshifts of each radio MS observation. As can be seen, these new radio data agree well with the rest of the observations included here and do not alter the fit substantially. Note that this is not entirely by design (although such a procedure by nature should induce overall agreement), as this redshift-dependent offset could just as easily have left a remaining constant offset between the radio SFRs and the other SFR indicators. Due our empirical methodology of accounting for these offsets, in our later series of fits (see Tables~\ref{tab:fits} and~\ref{tab:fits_z}) we fit the evolution of the MS with and without including radio SFR data as well as with and without this offset.

As this finding may have important implications interpreting studies heavily reliant on the precise redshift evolution of radio SFR data (e.g., \citealt{leitner12}), we investigate possible reasons for this disagreement in Appendix~\ref{app:bell}. Briefly, the systematic disagreements between radio SFR calibrations and other data that emerge after moving all data to a common set of calibrations mainly arises due to assumptions regarding $q_{\textrm{IR}}$ -- contrary to the straightfoward conversion presented in \citet{murphy+11}, the often-used \citet{bell+03} radio $L$\,--\,$\psi$ calibration uses a different $q_{IR}$ value than reported for the entire sample, and in addition attempts to account for light emitted by older stellar populations. Calibration assumptions themselves, however, cannot account for steeper redshift evolution observed here since they only adjust the normalizations -- the KE12 relations merely serve to highlight existing differences previously hidden in the data. Although scenarios involving radio suppression from the Cosmic Microwave Background (CMB) photons, redshift evolution of the radio spectral index $\alpha_S$ \citep{carilli+08,sargent+10}, or unknown biases present in the stacking procedures used here \citet{condon+12} appear to be the most reasonable explanations, they seem unlikely (at least, at lower redshift) based on existing data (A. Karim, priv. comm.; again, see Appendix~\ref{app:bell}).

Given the amount of data included here, the self-consistent nature of the $L$\,--\,$\psi$ conversions used in this work (see \S\,\ref{subsubsec:l_psi}), and the relatively straightforward way that both IR and radio SFRs are derived, we are fairly confident that this systematic $(1+z)^{\sim 0.8}$ disagreement is not a spurious effect. Although we are ultimately unsure of its origins, it is likely that some combination of the effects discussed above (and other possible issues likely not accounted for here) might serve as the underlying basis for the observed redshift evolution. Future studies should hopefully be able to clarify this issue.

\section{Fitting the Main Sequence}
\label{sec:methodology}

In order to fit a robust functional form for the MS that includes not only information on the slope and normalization as a function of time but also eliminates some of the degeneracies between $\alpha$ and $\beta$ between samples with similar observational properties, the observed mass ranges from each study (see Tables~\ref{tab:msfr} and~\ref{tab:msfr_corr}) are incorporated into the fit and considered the boundaries of that specific MS. Thus, only studies that contain objects at, e.g., $M_* = 10^{10.5} M_\odot$, are included when fitting for the SFR evolution of galaxies at that mass. These mass ranges have either been taken directly from the paper in question or estimated based on the data included in the relevant fits, rounded to the nearest $0.1$\,dex after excluding outlying points. For stacked data, the mass ranges have been taken from the medians of the lower and upper mass bins included in the fit, and thus the errors are approximately equivalent to the width of the bin, or $\sim 0.1$\,--\,$0.2$\,dex. When IMF and SPS model adjustments (among others) are significant (i.e. Salpeter to Kroupa or CB07 to BC03), the reported mass ranges have been adjusted accordingly in Table~\ref{tab:msfr_corr}.

Using this additional mass information, we proceed to fit the evolution of the SFR at fixed mass as a function of time,
\begin{equation}
\log\psi(t) = a_i t + b_i,
\end{equation}
with SFRs calculated from the reported MS fits in each individual paper. These are only included if the MS from the study in question \textit{is observed at that given mass} (i.e. is within the $\log M_*$ range observed). This allows us to account for observational limitations inherent in individual MS observations and different selection methods. We choose this form to parametrize the MS as am easy compromise between prior expectations and the observed data. Our decision to parametrize $\log\psi$ as a function of $t$ was motivated by the log-normal distribution of the MS in $M_*$\,--\,$\psi$ space and the likely dependence of MS evolution on the more physical time instead of redshift. A straightforward linear fit was then found to give the best fit to the data and was subsequently adopted.

This behavior was expected given the log-normal distribution of the MS (which implies logSFR), as well as the likely dependent variable governing evolution being time instead of redshift (linear t). We then simply chose a linear function as the simplest to fit the data and found it provided a good parametrization. This is now included in the paper.

By fitting $a_i$'s and $b_i$'s for a grid of masses, we then can derive a function of the form
\begin{equation}
\log\psi\left(t,\log M_*\right) = a(\log M_*)\,t + b(\log M_*),
\end{equation}
assuming a given parametrization for $a(\log M_*)$ and $b(\log M_*)$. In other words, instead of fitting the MS by simply averaging over all observed $\alpha$'s and $\beta$'s as a function of time, we average a subset of the observed slopes/normalizations for each mass bin and then fit the derived parameters within each mass bin as a function of mass. By doing this process for a grid of masses within a specified dynamical range, and specifying a minimum number of observations required to include a mass bin in the fit ($N_{\textrm{bin}}$), we are then able to derive a more robust, mass-dependent parametrization of the MS.

Before beginning our analysis, we wish to find a balance in the data between including all available observations and establishing a robust, self-consistent sample. In the hopes of reducing the impacts of systematic and observational biases on our parametrizations, we remove data in the first and last 2\,Gyrs (i.e., $\lesssim 2$\,Gyr and $\gtrsim 11.5$\,Gyr) from our analysis (as determined by the median redshifts of the respective samples). The rationale behind this is twofold. At the high redshift end, we get much higher uncertainties in masses and SFRs, which is to be expected: observations are more difficult, sample sizes are smaller, selection effects are worse, and hidden biases are more prominent. By removing these points, we restrict ourselves to observations where data are more tightly constrained.

\begin{figure*}[!ht]
\plotone{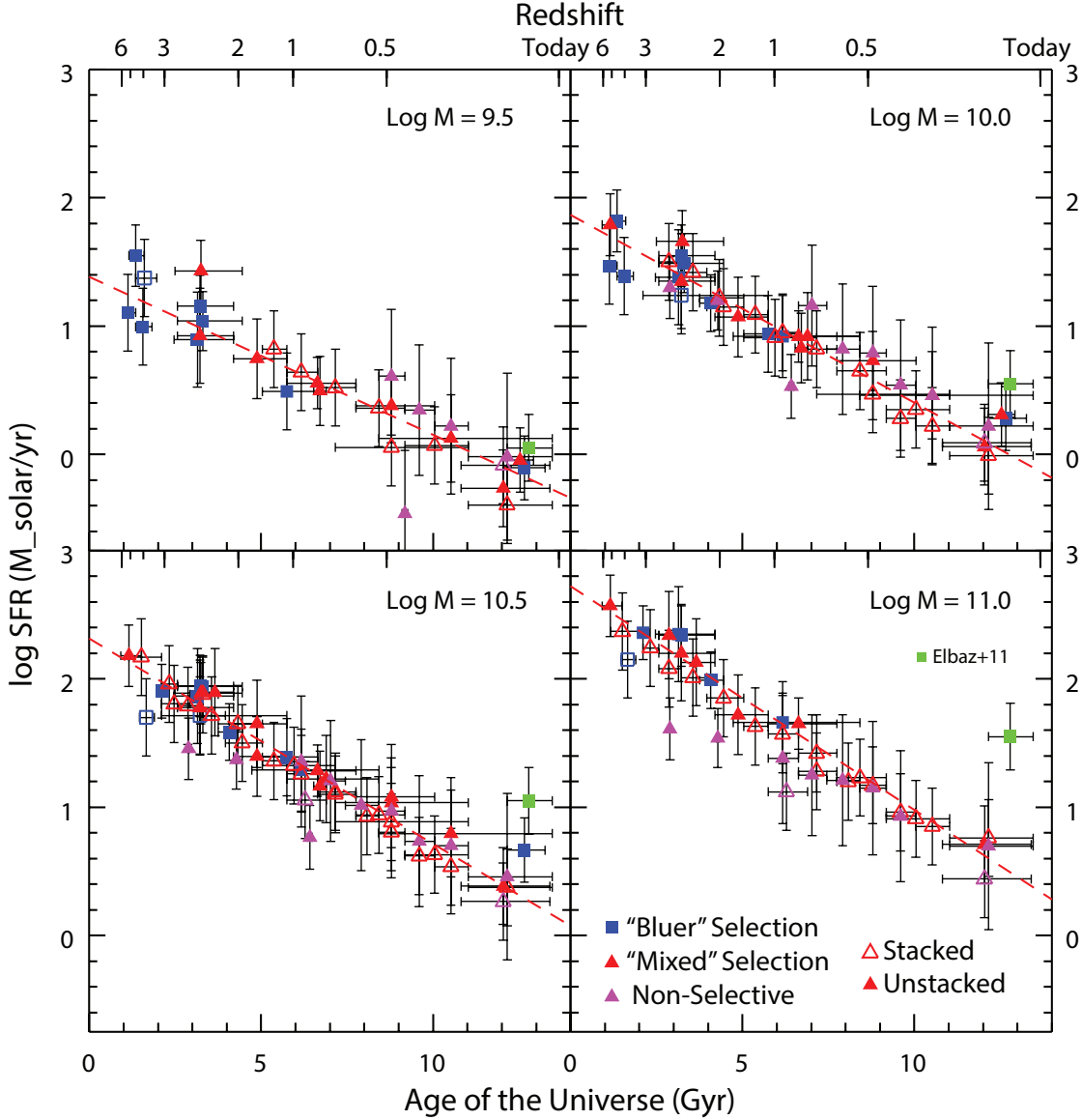}
\vspace{-10pt}
\caption{$\log\psi$ vs. $t$ for $\log M_* = (9.5, 10.0, 10.5, 11.0)$, calculated using the best fit MS relations from each of the 25 studies listed in Table~\ref{tab:msfr_info} and plotted using the same color/symbol scheme as Figure~\ref{fig:slope_selection}. Data from E11 (which is excluded from our fits) are plotted in green. The best fits (excluding non-selective data; see Figure~\ref{fig:slope_selection} and \S\,\ref{sec:methodology}) for a given mass are plotted as red dashed lines.
}\label{fig:sfr_t_relations_selection}
\end{figure*}

\begin{figure*}[!ht]
\plotone{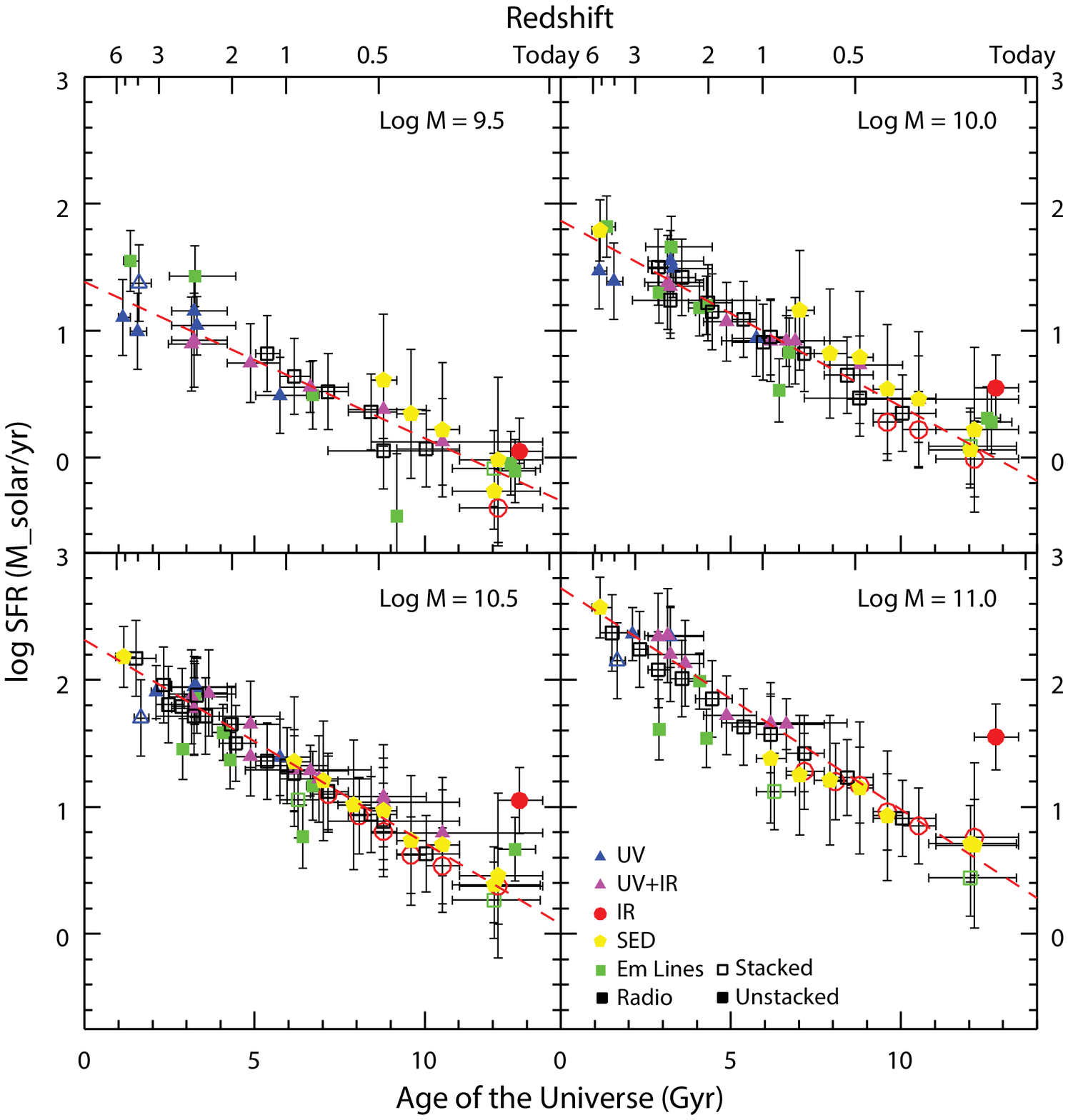}
\vspace{-10pt}
\caption{$\log\psi$ vs. $t$ for $\log M_* = (9.5, 10.0, 10.5, 11.0)$, calculated using the best fit MS relations from each of the 25 studies listed in Table~\ref{tab:msfr_info} and plotted using the same color/symbol scheme as Figure~\ref{fig:radio_sfr_t}. The best fits (excluding non-selective data; see Figure~\ref{fig:slope_selection} and \S\,\ref{sec:methodology}) for a given mass are plotted as red dashed lines.
}\label{fig:sfr_t_relations}
\end{figure*}

At low redshift, several studies have used SDSS data (S07; C09(1); Z12(1); C14(1)), and so have essentially just fit the same set of photometry in different ways with different selection criteria. In addition to almost identical datasets, these studies often need to utilize aperture corrections to account for missing light, which can lead to additional sources of uncertainty relative to higher-$z$ samples (B04; S07; C09; Z12). As we are not able to distinguish a ``best'' MS fit among them, we decide not to include them at all rather than unduly overweight the fit towards results in the local Universe (see Appendix~\ref{app:low_z}).

Both series of points, however, do provide useful information even though they are not included in our main sample. Most basically, their presence can test the validity of our parametrization of MS evolution at both early and late times. Good agreement at late times (low redshift) would be encouraging and indicate the fit is probably good, while good agreement at early times (high redshift) would be important in confirming that, contrary to some observations (M10; \citealt{weinmann+11}), the average sSFR for galaxies does not ``plateau'' at high redshift\footnote{As noted in \citet{behroozi+13}, the current data seem to be consistent with both a scenario where the sSFR plateaus at higher redshifts $\gtrsim 3$ and where it continues to increase with a relatively shallow slope.}. The differences between the SDSS observations on the low redshift end can give some sense of the different systematic biases present in the different assumptions that are made in the process of deriving a MS fit (since they are done on similar sets of photometry), while those at high redshift give more insight into the effects of selection biases, differing photometry, SED fitting procedures, and dust corrections. These will be discussed in \S\,\ref{sec:disc}.

As discussed in \S\,\ref{subsubsec:selection}, non-selective methods -- which do not probe the SF MS but rather the weighted average of star-forming and quiescent galaxies -- display prominent biases in both slope (as a function of redshift) and normalization (as a function of mass) compared to the other methods discussed above. The differences in slope as a function of redshift can most clearly be seen in Figure~\ref{fig:slope_selection}, where non-selective studies (purple) display slopes well below those of bluer- or mixed-selected studies. Differences in normalization as a function of mass can be seen by examining Figure~\ref{fig:slope_selection} and comparing normalizations of non-selective studies (purple) at $\log M_* = 10.0$ (where agreement with other data is fair) and $11.0$ (where non-selective SFRs are well below other studies).\footnote{While observational limits and the decreasing proportion of quiescent galaxies would imply non-selective studies should agree better at higher redshift, in the lower-right panel of Figure~\ref{fig:sfr_t_relations_selection} the opposite effect is observed. This might be due to the increasing influence \textit{any} quiescent contamination has in the sample when SFRs are on average higher (while quiescents are by definition low), or possible biases in C14's fitted $M_*$\,--\,$\psi$ distributions (where most of the non-selective sample is drawn from at high masses).} As a result, over a broad range of masses, MS observations (number) from C09 (2), So14 (4), and C14 (7) (all non-selective studies) tend to disagree by several tenths of a dex with other measurements from the literature. In order to develop a consistent picture of main sequence evolution, studies with large systematic differences should be considered in separate groups and comparisons drawn within a group. As a result, we exclude these non-selective studies from our main set of fits.

In addition to these non-selective studies, MS observations (number) from E11 (1) and Z12(3) (1) also display disagreements with the majority of the sample of up to several tenths of a dex. Here, however, we have the opposite problem: their samples tend to exhibit much higher SFRs at a given mass compared with other data in the literature. For Z12(3), this discrepancy is best highlighted at low masses in the top-right panel of Figure~\ref{fig:sfr_t_relations_selection}, where their data (green) lies substantially above the majority of other data at $z \sim 2$. For E11, on the other hand, this discrepancy mainly occurs at high masses, and best be seen in the lower-right panel of Figure~\ref{fig:sfr_t_relations_selection} where their data (also green) lies over an order of magnitude above most other low-$z$ data. As E11 restricts their sample to only include highly-active SFGs (i.e. LIRGs) and not ``typical'' MS galaxies (such as the large SDSS samples of B04, E07, S07, Z12, and C14, or the SWIRE sample of O10), their analysis is likely inherently biased; it is not surprising that their results disagree with all others at low redshifts. In the case of Z12(3), however, we are ultimately unsure where the root cause of the apparent disagreement arises, although the explanation might likely involve different choices of extinction corrections (especially at lower masses) or some aspect of their selection criteria. Although we opt to exclude these two results from all our MS fits, we note that including them has a negligible impact on the majority of our fits.

As mentioned above, SFRs at a given mass are calculated based on the reported best fit relations from the studies in question. As not all studies report errors on their MS fits, and the errors are symptomatic of the observed scatter about the MS, we instead incorporate errors into our fit by including the \textit{scatter} reported in the papers in question (see \S\,\ref{subsec:scatter}). As the scatter in almost all cases is larger than the reported errors, this process is a conservative estimate that will likely inflate our reported errors. However, as we are fitting observed SFRs as a function of time (which display scatter about the best MS fit) in bins of mass, injecting this amount of scatter is actually more physical. 

We plot the $\psi(t)$ relations derived for $\log M_* = 9.5$, $10.0$, $10.5$, and $11.0$ in Figures~\ref{fig:sfr_t_relations_selection} and~\ref{fig:sfr_t_relations}, color-coded by selection type and SFR indicator, respectively.

\begin{figure*}[!ht]
\plotone{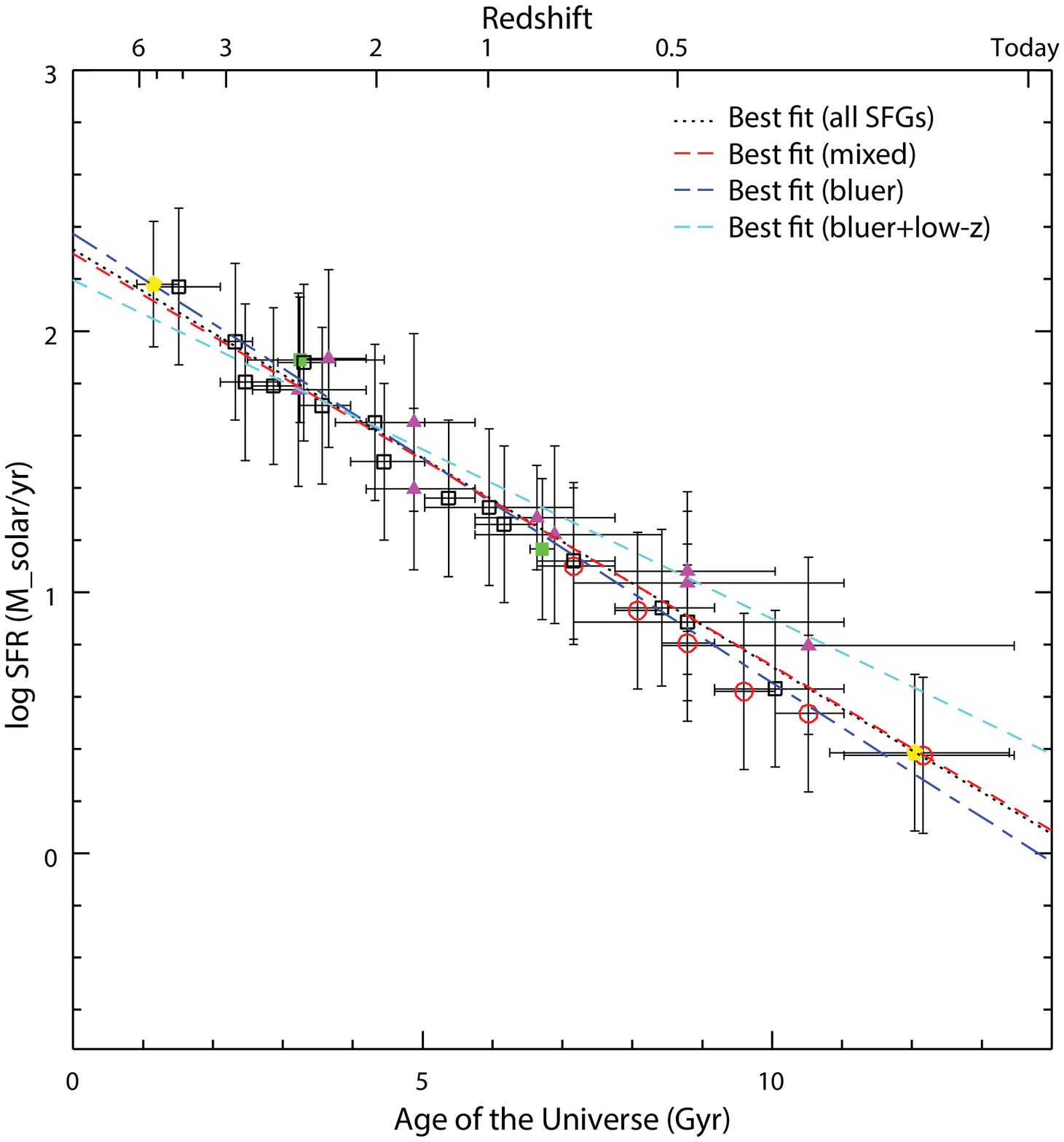}
\vspace{-10pt}
\caption{Similar to Figure~\ref{fig:sfr_t_relations}, but here redshift evolution (shown by the best-fit relations) is illustrated for various (sub)samples at $\log M_* = 10.5$. The dotted black, dashed red, and long dashed blue lines indicate the best-fit relations (excluding the first and last 2\,Gyrs of data) to all SFG observations, ``mixed'' observations only, and ``bluer'' observations only. The best-fit relation to bluer observations including low-$z$ SDSS data (dashed cyan) is shown for contrast. As can be seen in our bluer low-$z$ fit, the time evolution of the SFR is significantly reduced if we restrict ourselves to observations biased towards bluer studies only, in good agreement with \citet{karim+11}. There is good agreement among all three fits when low-$z$ (and high-$z$) data has been excluded.
}\label{fig:m105ri}
\end{figure*}

After compiling $a_i$'s and $b_i$'s for a grid of masses, we fit them as linear functions of $\log M_*$, with
\begin{align}
a(\log M_*) &= a_m \log M_* + a_0, \nonumber \\
b(\log M_*) &= b_m \log M_* + b_0,
\end{align}
which gives us
\begin{align}
\log\psi\left(t,\log M_*\right) &= \left(a_{m,t} \log M_* + a_{0,t}\right) \, t \nonumber \\
								&+ \left(b_{m,t} \log M_* + b_{0,t}\right).
\end{align}
Some quick rearranging allows us to write this in a more familiar form,
\begin{equation}
\log\psi\left(t,\log M_*\right) = \left({\alpha_t}t  + \alpha_c\right) \, \log M_* + \left({\beta_t}t + \beta_c\right),
\end{equation}
where we have redefined $\alpha_t \equiv a_{m,t}$, $\alpha_c \equiv b_{m,t}$, $\beta_t \equiv a_{0,t}$, and $\beta_c \equiv b_{0,t}$ for clarity/convenience. This leads us to a functional fit to the MS of the form as seen in equation~\ref{eq:ms_1}, with $\alpha(t) = {\alpha_t}t  + \alpha_c$ and $\beta(t) = {\beta_t}t + \beta_c$ now linear functions of time. We may also rewrite this new time-dependent equation, MS($t$), in terms of a mixed power-exponential model,
\begin{equation}
\psi(t) = \psi_{0,t} \, e^{t/\tau} \, M_*^{\alpha(t)},
\end{equation}
where $\psi_{0,t} = 10^{\beta_c}$ and $\tau = 1/(\beta_t \ln 10)$. As the errors on the fits are highly dependent on the number of observations available at a given mass, the errors from one mass bin to the next are highly correlated (as observations tend to cluster in the mass ranges they probe at a given redshift). However, as will be argued below, the extent to which these correlated errors (dependent on the estimates of the internal scatter for each observation as well as the scatter between observations) impact our results should be small and does not change our conclusions.

In order to compare our results to previous ones in the literature, we also consider a fit as a function of redshift,
\begin{align}
\log\psi\left(1+z,\log M_*\right) &= \left(a_{m,z} \log M_* + a_0\right) \, \log(1+z) \nonumber \\
								  &+ \left(b_m \log M_* + b_0\right),
\end{align}
which is fit in the same way as detailed above for the $\psi(t)$ case, the only difference being the substitution of $\log(1+z)$ for $t$. Here, we end up with a slightly different parametrization, with
\begin{equation}
\psi\left(z,M_*\right) = \psi_{0,z}(\log M_*) \times (1+z)^{a(\log M_*)},
\end{equation}
where $\psi_{0,z}(\log M_*) = 10^{b(\log M_*)}$. Unlike in the MS($t$) case, here we have fit the MS as a mass-dependent power law in $z$ and $M_*$. On average, values in the literature seem to report that $\psi \propto (1+z)^{\sim 3.5}$ out to $z \sim 2.5$ (O10; K11). As we have fit linear functions for $a(\log M_*)$ and $b(\log M_*)$, doing a simple linear fit to two fiducial $\log M_*$ values gives the predicted MS($z$) relation at any given redshift. Our best fit $\psi\left(z,M_*=10.5\right)$ evolution goes as $(1+z)^{\sim 2.8}$, in good agreement with the evolution assumed in \citet{sargent+12}.

It is also easy to recover $a(M,t)$ and $b(M,t)$ to compare to the $a(M,z)$ and $b(M,z)$ coefficients using our parametrization above. We include the values of both sets of coefficients at fixed mass ($\log M_* = 10.5$) in Tables~\ref{tab:fits} and~\ref{tab:fits_z}. Although we note that redshift-dependent parametrizations of MS evolution give decent fits to the data, we find that time-based parametrizations provide overall better fits (see Appendix~\ref{app:details}).

\begin{figure*}[!ht]
\vspace{-10pt}
\plotone{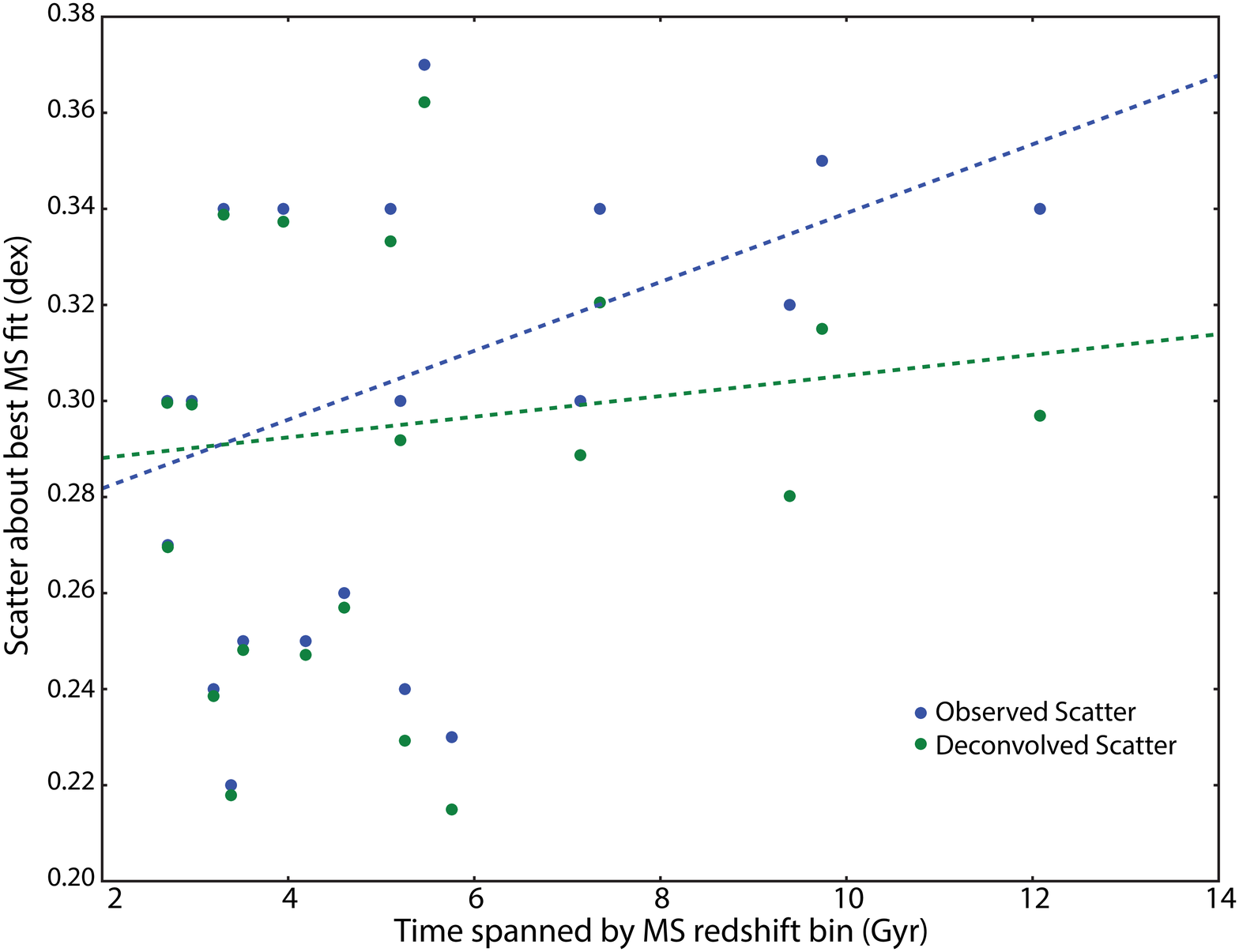}
\vspace{-10pt}
\caption{The scatter observed for the subset of non-stacked data before (blue) and after (green) our deconvolution process, plotted as a function of the width of the respective time bin it encompasses ($\Delta t$). The best-fit linear trends are overplotted (dashed lines). Studies with small sample sizes (N < 250) or that have been selected via $sBzK$ have been excluded from the fit, as well as the results of S09, So14, C14 (due to the former's sigma-clipped fit and the latter two's differing selection criteria, respectively). Our deconvolution process has the largest effects on data observed in the largest time bins, and removes any trends with time to within the fitted errors (slope of $0.002 \pm 0.005$\,dex per Gyr).
}\label{fig:scatter}
\end{figure*}

\subsection{Scatter about the Main Sequence}
\label{subsec:scatter}

\subsubsection{Deconvolving the Intrinsic Scatter about the Main Sequence}
\label{subsubsec:scatter_int}

Each MS observation we include has been measured within a predefined redshift window, and therefore has an associated time-window ($\Delta t$) that accompanies it. Because of this, the \textit{observed} scatter about the MS ($\sigma$) is actually the intrinsic scatter (i.e. ``deconvolved'' scatter, $\sigma_d$) about the MS convolved with the MS's evolution within the given time interval. They are therefore overestimates. In order to deconvolve the observed scatter with this effect, we use our best fits to approximate MS evolution within each time bin ($\Delta t$), and subtract this evolution from the observed scatter. In order to account for first-order volume effects, we assume that within each redshift bin in question, the number density and mass distribution of MS galaxies is approximately constant, and weight each result according to the changing comoving volume within each redshift bin. Accounting for this effect slightly decreases the total magnitude of the deconvolution (from that of a simple tophat function), and so is again a conservative estimate on the effects of MS evolution within each redshift bin. 

We then refit the MS using these new, smaller scatters until we get a convergent solution. We find that differential mass evolution within each time bin is small relative to the absolute changes in SFRs, and approximate the evolution of the MS as one entirely in normalization, with
\begin{align}
\alpha(t) &= \alpha \nonumber \\
\beta(t)  &= (-0.15)t,
\end{align}
in good agreement with results from So14 ($\beta(t) = -0.18 \pm 0.02$) after taking into account differences in selection\footnote{Restricting the sample to include just So14's data, we find $\beta(t) = -0.20 \pm 0.02$, fully consistent with the their results.}. We note, however, that this differential mass evolution is \textit{not} negligible (see \S\,\ref{sec:results}), and would imply that there should be a greater evolution at higher masses, and hence more scatter, relative to lower masses (see \S\,\ref{subsec:ms_tracks}). The actual change between $\sigma$ and $\sigma_d$ in most cases is negligible (see Table~\ref{tab:msfr}), but for the data with the largest $\Delta t$'s the correction leads to a reduction of $\sim 0.04$\,dex (see Figure~\ref{fig:scatter}). The results are not sensitive to the precise value chosen for $\beta(t)$ here, only the approximate magnitude, with shifts in $\beta(t)$ of $\pm 0.02$ leading to changes in $\sigma_d$ of at most $\sim 0.01$\,dex, and does not affect our results from \S\,\ref{subsubsec:scatter_true}.

\begin{figure*}[!ht]
\vspace{-10pt}
\plotone{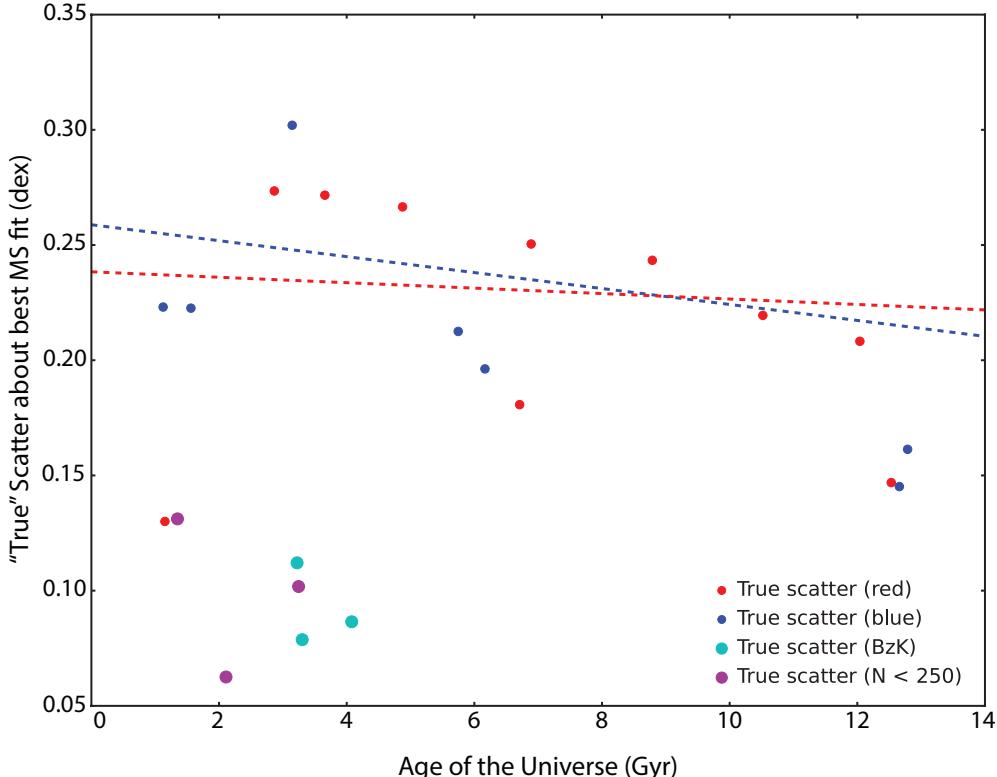}
\vspace{-10pt}
\caption{The ``true'' scatter about the MS as a function of time, with red and blue points representative for mixed and bluer selection methods, cyan points for $sBzK$ studies, and purple points indicating samples with less than 250 galaxies. The prevalence of $sBzK$ points well-below the majority of the data suggests that it is somehow ``missing'' a significant portion of SFGs (see \S\,\ref{subsubsec:bzk}), while the lower scatters among smaller samples indicates the prevalence of selection effects. The best fit relations for each respective selection method, excluding both of these categories, are plotted as dashed lines in their respective colors. Given the uncertainties, this seems to indicate on average a constant MS scatter over the majority of the age of the Universe.
}\label{fig:scatter_bzk_2}
\end{figure*}

\subsubsection{Calculating the ``True'' Scatter about the Main Sequence}
\label{subsubsec:scatter_true}

The deconvolved, ``intrinsic'' scatter, however, is still not the ``true'' scatter ($\sigma_t$) about the MS, as errors in observing/deriving relevant physical quantities (i.e. photo-z's, masses, and SFRs) will further tend to inflate the observed distribution. In order to account for this observation-induced scatter, we look for metrics by which to estimate the approximate variation present in any given indicator used to derive one of these quantities relative to another (hopefully more reliable) metric.

For photo-z's, this comparison is easy in principle, as we are able to compare them against their more accurate spec-z counterparts\footnote{We note again, however, that such a comparison is only available for the most massive/luminous galaxies.}. Based on our first-order volume-weighted calculations, the average scatter in the photo-z vs. spec-z relation -- ignoring contamination by catastrophic errors, interdependencies on masses and SFRs, and other observational biases -- will tend to increase the observed scatter by at most $0.02$\,dex (for average phot-z vs. spec-z errors of $2\%$). These are essentially negligible and are not accounted for here. Accounting for this effect would tend to further decrease the observed scatters, making the $M_*$\,--\,$\psi$ relation even more tightly correlated.

At the moment, masses are mainly derived through SED fitting and do not have a range of available cross-checks. Several studies (e.g., S12) have found other indicators that might work as ``empirical'' masses, any real vetting of such an indicator has yet to be undertaken. Comparisons between dynamical and stellar mass measurements, however, are available, and provide helpful estimates of the uncertainties in mass estimates. These generally display scatters of $\sim 0.2$\,dex, and in addition find that any evolution systematic difference with redshift is small ($\lesssim 0.3$\,dex) up to $z\sim 2$ \citep{forsterschreiber+09,taylor+10}. This seems to indicate that SED-derived masses on quite accurate, and any intrinsic scatter present in stellar masses is at the $\sim 0.15$\,dex level or lower, well below the $0.3$\,dex of scatter observed in the MS.

To further characterize uncertainties in the SED-fitting procedure, we try and compare the cross-correlation between two methods of determining masses from the same SED fitting procedure and their effects on the derived $M_*$\,--\,$\psi$ relations. While this cannot test some of the uncertainties that go into the SED fitting procedure itself (as the dynamical mass comparison above), it can test to see if there is some intrinsic scatter present in mass determinations that might inflate the $M_*$\,--\,$\psi$ distribution. Using the different mass determinations from So14 (D. Sobral, priv. comm.; see Appendix~ \ref{app:mass_disc}), we tentatively find that the intrinsic variation induced by using masses taken from the best-fitting SED relative to a more robust indicator is small, at around $\sim 0.1$\,dex. As this has not been rigorously tested, we do not opt to include this additional correction when calculating $\sigma_t$.

The variety of SFR indicators in use today allows us to establish some level of intrinsic observational-induced SFR scatter. In order to determine the scatter that might be introduced due to observational techniques and/or modeling assumptions, we search for previous instances where cross-calibrations have been established. Using results from N07, S07, \citet{noeske+07b}, S09, \citet{nordon+10}, \citet{wuyts+11}, R12, K13, \citet{price+13}, \citet{carollo+13}, \citet{utomo+14}, and Pforr et al. (2014, in prep), we find that on average all SFR indicators used today show quite good agreement\footnote{Although there are conflicting results concerning SED and UV-corrected SFRs relative to other SFR indicators, with SED/UV-corrected SFRs displaying good agreement with other SFR indicators in some cases (D07;  \citealt{carollo+13}) and up to 0.5\,dex scatter/offsets in others \citep{wuyts+11}, we will assume here that the scatter is consistent with other methods.}, albeit with $\sim$\,0.3 dex cross-calibration scatter ($\sigma_{cc}$), irrespective of extinction curve, sample size, and SPS model\footnote{Note that this does \textit{not} hold true for assumptions regarding $\tau$ in SFHs \citep{maraston+10,price+13}, as well as in recently quenched galaxies or more ``starbursty'' SFHs \citep{utomo+14,hayward+14}.}. $\sigma_{cc}$ in each case appears to be uncorrelated with other variables, which holds even for indicators that require extinction corrections such as H$\alpha$ \citep{wuyts+11}. Results are also consistent with no redshift evolution between the indicators themselves or their cross-correlations (N07; S07; S09; \citealt{wuyts+11}; R12).

The most reasonable interpretation is that the observed $\sim 0.3$\,dex scatter among cross-correlations is due to $\sim$\,.2 dex of ``intrinsic'' scatter for any given individual SFR indicator (see Appendix~\ref{app:sfr_disc}; see also Section~2 of N07). This can be easily subtracted in quadrature from the deconvolved scatters to yield approximate ``true'' scatters ($\sigma_t$) in the reported MS relations. The results are included in Table~\ref{tab:msfr}.

We use these new $\sigma_t$ values in our MS fit, and iteratively recalculate $\sigma_d$ and $\sigma_t$ until our fit converges.

\subsubsection{Success of the Deconvolution Procedure}
\label{subsubsec:scatter_success}

We investigate the ``success'' of our deconvolution procedure by examining trends in $\sigma$ against $\Delta t$ using non-stacked data. If we assume an unchanging MS scatter, we would expect that $\sigma(\Delta t)$ should increase for larger time bins due to the increasing effects of MS evolution and/or worse data. After removing the results of S09, So14, and C14 (the former due to their sigma-clipping fitting procedure, the latter two as they are non-selective) as well as several biased/outlying data points ($sBzK$-selected observations and those with a sample size of $< 250$; see \S\,\ref{subsubsec:bzk} and Figure~\ref{fig:scatter_bzk_2}), we find a best fit relation of
\begin{equation}
\sigma(\Delta t) = (0.28 \pm 0.01) + (0.015 \pm 0.005) \, \Delta t.
\end{equation}
This seems to indicate that data observed over a larger time range has been increasingly impacted by MS evolution.

After deconvolution, we find a new best-fit relationship of
\begin{equation}
\sigma_d(\Delta t) = (0.29 \pm 0.02) + (0.003 \pm 0.005) \, \Delta t.
\end{equation}
This $\sigma_d(t)$ relation has a consistent zero-point with the $\sigma(t)$ relation, yet exhibits a five-fold reduced dependency on $\Delta t$. Thus, our deconvolution procedure significantly reduces trends of larger scatters with $\Delta t$ (see Figure~\ref{fig:scatter}). 

We obtain a very similar relationship for $\sigma_t$, with
\begin{equation}
\sigma_t(\Delta t) = (0.20 \pm 0.02) + (0.007 \pm 0.006) \, \Delta t.
\end{equation} 
In order to check that our results are not the heavily influenced by selection effects, we also calculate the best fits for the mixed and bluer subsamples (excluding the same studies as above). We find that
\begin{align}
\sigma_{t,\textrm{mixed}}(\Delta t) &= (0.20 \pm 0.03) + (0.009 \pm 0.010) \, \Delta t \nonumber \\
\sigma_{t,\textrm{bluer}}(\Delta t) &= (0.21 \pm 0.06) - (0.001 \pm 0.043) \, \Delta t,
\end{align}
for the mixed and bluer subsamples, respectively. These seem to imply that our deconvolution procedure is relatively robust to selection effects.

We also wish to investigate possible evolution of $\sigma_t$ over time. Our best fits for $\sigma_t(t)$ are
\begin{align}
\sigma_{t,C}(t) &= (0.25 \pm 0.03) - (0.005 \pm 0.003) \, t \nonumber \\
\sigma_t(t,\textrm{mixed}) &= (0.24 \pm 0.05) - (0.003 \pm 0.006) \, t \nonumber \\
\sigma_t(t,\textrm{bluer}) &= (0.26 \pm 0.03) - (0.008 \pm 0.02) \, t,
\end{align}
for the combined, mixed, and bluer subsamples, respectively (see Figure~\ref{fig:scatter_bzk_2}). The lack of time evolution seen in the mixed (our preferred) subsample is in good agreement with the common assumption that scatter on the MS is approximately constant with time (N07; W12). The $\sim 2\sigma$ time evolution in the combined sample is mainly due to the lower scatters reported in E07 and E11 at low redshift, which drive the fit to lower values. Once these two data points are removed, the time evolution is consistent with being negligible. Given these results, we conclude that the scatter about the MS ($\sigma$, $\sigma_d$, and $\sigma_t$) is consistent with being constant in time.

Considering the range of $\Delta t$ and $t$ values which spanned by these studies and the variety of different assumptions that go into each MS observation included here, these results seem to indicate that the real scatter about the MS is
\begin{align}
\sigma_d &\sim 0.3\,\textrm{dex} \nonumber \\
\sigma_t &\sim 0.2\,\textrm{dex},
\end{align}
consistent with the analysis of \citet{munozpeeples14}.

\subsection{Accounting for Observational Limits and Selection Effects}
\label{subsec:cuts}

We account for observational biases and other effects by fitting the MS using a variety of cuts in selection mechanism (mixed, bluer, non-selective), detection type (stacked or not), SFR indicator (emission lines, UV, UV+IR, IR, radio, SED), and the number of points required per mass bin (ranging from 5\,--\,35). Each of these cuts is reported for two sets of data -- the uncalibrated (Table~\ref{tab:msfr}) and the calibrated data (Table~\ref{tab:msfr_corr}) -- and two sets of fits -- ``extrapolated'' fits, where the data fit without including any type of mass-weighting scheme (i.e. including all MS observations published for the given cuts and just averaging over their reported $\alpha$ and $\beta$), and mass-dependent fits, where the data is fit as described previously.

1$\sigma$ uncertainties on the fits are calculated two different ways. The first is from standard fitting procedures (using Scipy's \texttt{ODRpack}), which take into account the true scatter derived for each of the observations as well as the interpublication scatter ($\sigma_{i}$). The second is from bootstrapping (via resampling), where for each trial we randomly adjusting the upper and lower bounds of each observation by up to $\pm 0.2$\,dex (equivalent to the bin width for some of our stacked data, the true scatter, and a conservative overestimate of our rounding procedure), re-fit, and take the 1$\sigma$ deviation around the median after 100 runs. We find that our functional form for the MS is robust to possible errors on the reported $\log M_*$ ranges, with resampled errors only $\approx 50\%$ larger than the formal fitting uncertainties\footnote{Using more runs (e.g., 500) gave consistent error estimates, so this effect is not due to extra variation caused by too few trials.}. To be conservative, we report these higher errors.

\section{Results}
\label{sec:results}

\subsection{The Evolution of the Galaxy ``Main Sequence''}
\label{subsec:ms_evol}

Our results for both our time- and redshift-dependent fits given a variety of (sub)samples and fitting assumptions are listed in Tables~\ref{tab:fits} and~\ref{tab:fits_z}, respectively. We discuss the details behind the various fitting assumptions and how they impact our analysis in Appendix~\ref{app:details}. Our best fits and their comparisons to previous results in the literature are discussed below.

Based on arguments in \S\,\ref{subsubsec:selection} and Appendix~\ref{app:details} (also see Table~\ref{tab:error_budget}), we limit our ``best'' results to mixed data converted to our common calibration after we have removed data from our ``time edges''. We require our best MS fit to include a moderate threshold on the number of data points included in each mass bin ($N_{\textrm{bin}} = 15$), where we have included enough data points to avoid over-biasing towards individual studies (e.g., S09/O10 at lower/higher mass), but not so much that we eliminate a large portion of the available mass range and lose some of the flexibility of our mass-dependent parametrization. Our best MS fit is:
\begin{align}
\log\psi(M_*,t) &= \left(0.84 \pm 0.02 - 0.026 \pm 0.003 \times t\right) \log M_* \nonumber \\
&- \left(6.51 \pm 0.24 - 0.11 \pm 0.03 \times t\right),
\end{align}
where the listed errors are derived from resampling.

\begin{figure*}[!ht]
\plotone{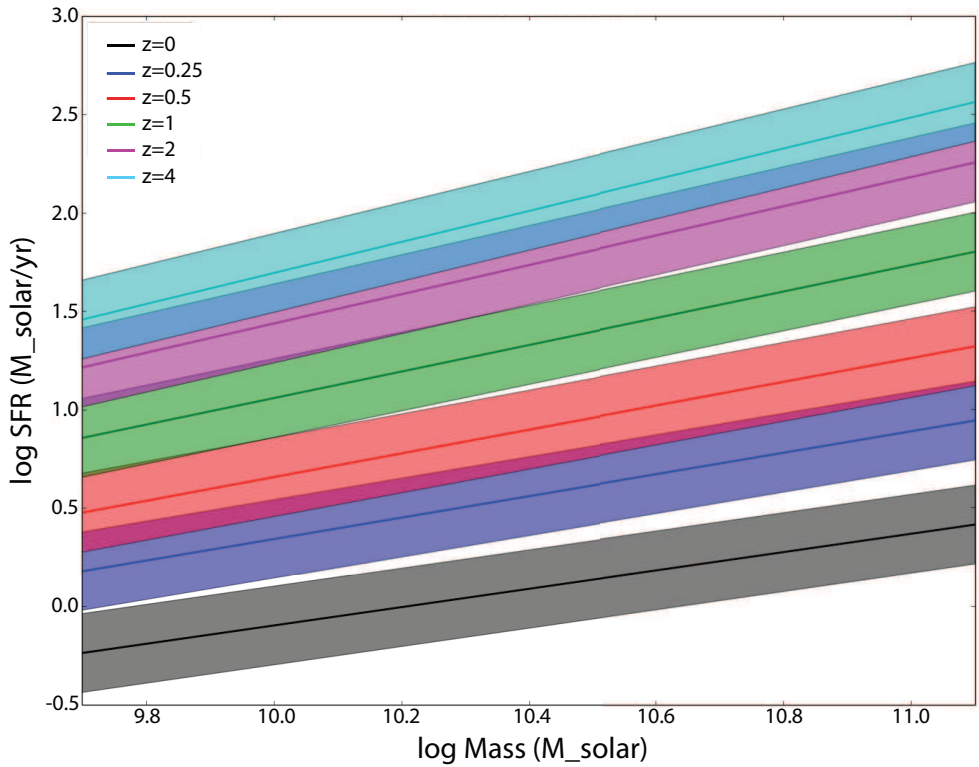}
\vspace{-10pt}
\caption{Several of our ``consensus'' MS relations taken from our best fit to observations from the literature (see \S\,\ref{subsec:ms_evol}) plotted at several given redshifts. The widths of the distributions are taken to be the ``true'' scatters ($\pm 0.2$\,dex) rather than the likely observed scatters ($\sim 0.3$\,dex) for improved clarity, and the mass bounds are taken directly from the fit. The changing MS slope and $\sim 2$ orders of magnitude evolution in SFR at fixed mass from $z=4$ to $0$ are easily visible. As the first and last 2\,Gyrs of data are not included in the fit, the $z=0$ and $z=4$ slopes should be viewed as predictions of high-/low-$z$ MS relations rather than simply best fits to data available at those redshifts (which would tend to fit well by default).
}\label{fig:consensus_ms}
\end{figure*}

The interpublication scatter around this fit is $\sigma_i = (0.08,0.09,0.11)$\,dex, for the minimum, median, and maximum values within the fitted mass range $\log M_* = 9.7$\,--\,$11.1$, respectively. This encompasses a majority of the age of the Universe ($z \sim 0.25$\,--\,$2.75$), and provides good fits to the observed SFRs all the way out $z \sim 5$ (see Figure~\ref{fig:sfr_t_relations}). For convenience, we also plot the related best-fit MS relations at several fixed redshifts in Figure~\ref{fig:consensus_ms}.

We note that our best fit provides good fits to the data out to $z \sim 5$ (St14); we might, however, opt to include all high-$z$ data to try and better constrain the fit. If we re-include data from the first 2\,Gyr of the Universe in our fit, we instead get
\begin{align}
\log\psi(M_*,t) &= \left(0.80 \pm 0.02 - 0.022 \pm 0.003 \times t\right) \log M_* \nonumber \\
&- \left(6.09 \pm 0.23 - 0.07 \pm 0.03 \times t\right),
\end{align}
consistent with our earlier fit. In addition to incorporating high-$z$ data, we might also choose to see how the varies if we do not include our empirically-derived (and more tentative) $C_R$ calibration offsets. Once these are removed, our best fit is instead
\begin{align}
\log\psi(M_*,t) &= \left(0.96 \pm 0.05 - 0.045 \pm 0.006 \times t\right) \log M_* \nonumber \\
&- \left(7.41 \pm 0.48 - 0.27 \pm 0.06 \times t\right),
\end{align},
with a slightly higher median interpublication scatter of $\sigma_i = 0.14$\,dex. As expected, the fit exhibits stronger time evolution, and has slightly larger errors due to the larger interpublication scatter. We list all our best fits for various subsets of the data in Table~\ref{tab:best_fits}.

In order to confirm the validity of our fits, we check to see whether our redshift fits can reproduce the observed dependencies ($\psi \propto (1+z)^{\gamma}$) reported in other works. For O10's data, we find $\gamma \sim 3.2 \pm 0.3$, in agreement with the $\gamma \sim 3.4 \pm 0.3$ derived there. For radio observations without our $(1+z)$ adjustments (i.e. D09, P09, and K11), we find a redshift dependence of $\gamma \sim 3.6 \pm 0.1$, in agreement with the values reported in K11.\footnote{If we only include K11 the fit, our best fit evolution remains unchanged.} As expected, the best fit to the $(1+z)^{\sim 0.8}$ corrected radio SFRs is $\gamma \sim 2.8 \pm 0.1$. In almost all cases, our observed redshift evolution is milder than the $\gamma \sim 3.5$ reported previously (O10; K11). At $\log M_* = 10.5$, we find that the power law index for the fit including all data or mixed data only is $\gamma \sim 2.8$ (2.8\,--\,2.9 if we exclude radio; 2.5 if we exclude all stacked observations), in good agreement with the evolution assumed in \citet{sargent+12}.

For bluer observations only, we find $\gamma \sim 2.4$, in excellent agreement with the result from K11 (this reduces even further to $\gamma \sim 1.4$ after high-$z$ data are included in the fit). As with the more rapid time-evolution of the MS slope reported for combined UV+IR/IR-selected observations only, we also find that the redshift dependence among these observations is the steepest ($\gamma \sim 3.1$). Taken together, these observations indicate that the parametrized dependence of the (s)SFR with redshift goes as $\gamma \sim 2.8$.

\subsection{Main Sequence Evolutionary Tracks}
\label{subsec:ms_tracks}
Using the MS as an empirical constraint, we can formulate evolutionary tracks for typical MS galaxies with the assumption that they must obey the MS at all times by simply integrating along the MS. This type of ``Main Sequence Integration'' (MSI; \citealt{renzini09}; \citealt{peng+10}; \citealt{leitner12}) can provide information on the typical SFH of a MS galaxy at some time in the absence of major mergers\footnote{The recent results of \citet{hayward+14} and \citet{utomo+14} indicate that IR SFRs can be significantly overestimated for quiescent galaxies (or galaxies with more ``starbursty'' SFHs). As post-starburst/starburst systems are a minority of MS galaxies at a given redshift (E11; \citealt{sargent+12}; Lackner et al. 2014, subm.), their effect on our results should be small.}. The evolution of stellar mass in a given galaxy can be described as
\begin{equation}
\frac{dM_*(t)}{dt} = (1-\eta(t))\,\psi(M_*,t),
\end{equation}
where $\frac{dM_*(t)}{dt}$ is the mass growth rate, $\psi(M_*,t)$ is the SFR as parametrized by our best MS fits, and $\eta(t)$ is the galaxy wide fractional mass-loss rate as a function of time. Thus, given an initial mass, time, and a parametrization of the fractional mass-loss rate, we can easily integrate to find the mass (and consequently SFR) at any future time. Conversely, we can also use this method to integrate backwards from some starting point. We opt here to integrate \textit{forward} in time, fine-tuning formation time for a given initial mass via trial and error to arrive at the desired final mass.

\begin{figure*}[!ht]
\vspace{-10pt}
\plotone{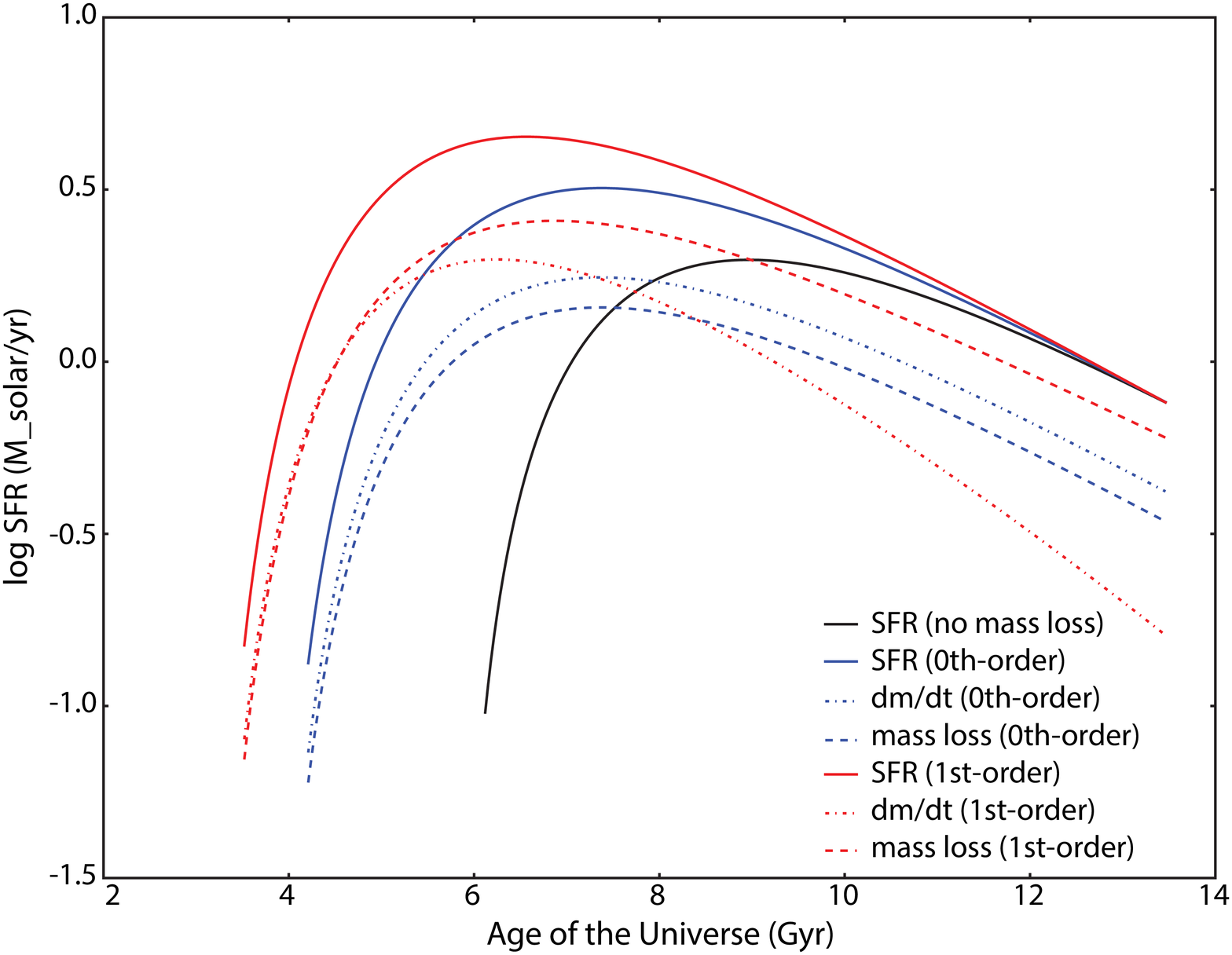}
\vspace{-10pt}
\caption{Calculated $\log \psi$ tracks for galaxies that strictly evolve along the MS with final stellar masses of $10^{10}\,M_\odot$ at the present day, assuming direct SFR-to-$M_*$ growth (black), a zeroth order ``return rate'' correction (from stellar mass to gas mass) to account for mass loss from young, massive stars (blue), and an additional first-order correction (accurate to within a few percent of the final mass) to account for mass loss from older stellar populations (red). Both stellar mass loss prescriptions are taken from \citet{leitnerkravtsov11} and \citet{leitner12} and assume a Chabrier IMF (functionally equivalent to a Kroupa IMF). Initial seed masses of $10^7 M_\odot$ are assumed. Dashed lines indicate the return rate for the given correction as a function of time, and dashed-dotted lines indicate the mass growth the galaxy over the same period.
}\label{fig:logsfr_track}
\end{figure*}

\begin{figure*}[!ht]
\vspace{-10pt}
\plotone{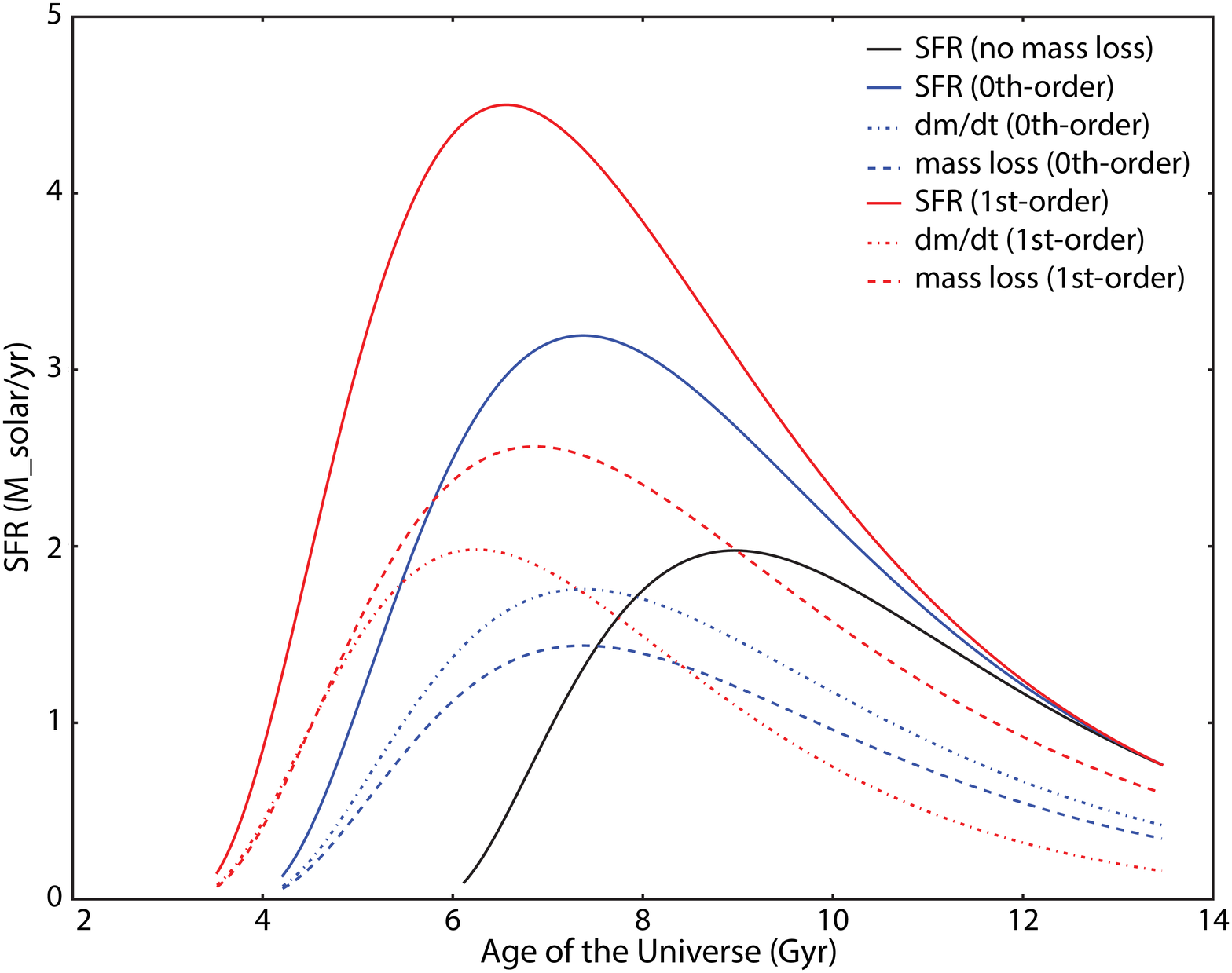}
\vspace{-10pt}
\caption{As Figure~\ref{fig:logsfr_track}, but for $\psi$. The magnitude of the peak for galaxies including first-order approximations for stellar mass loss can more easily be seen. In addition, a mix of ``fast growth'' (i.e. power-law rising), approximately constant, and exponentially declining SFH over time are clearly visible, implying that delayed-$\tau$ models (which naturally include all three phases) provide the most accurate parametrizations of typical MS SFHs.
}\label{fig:sfr_track}
\end{figure*}

\begin{figure*}[!ht]
\vspace{-10pt}
\plotone{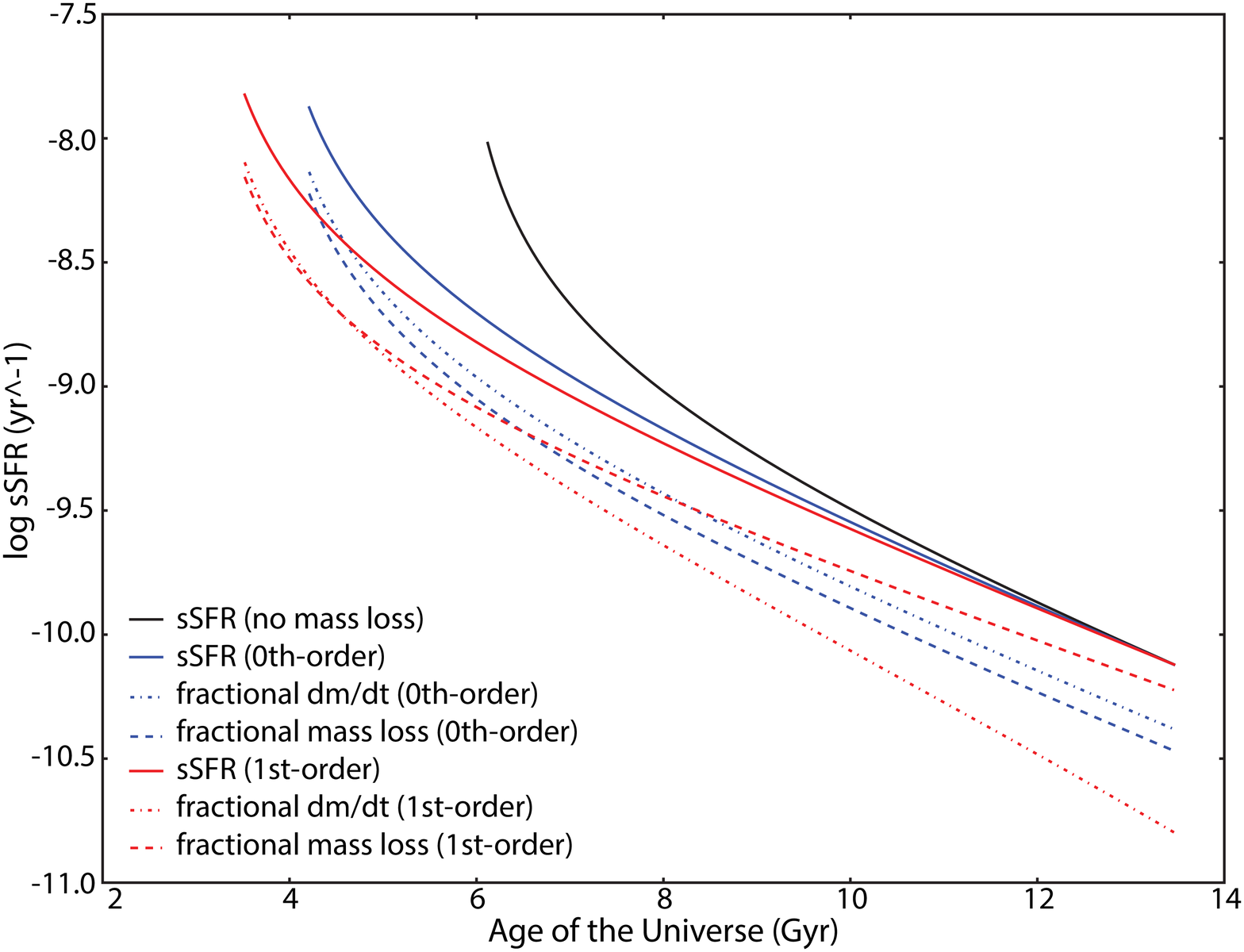}
\vspace{-10pt}
\caption{As Figure~\ref{fig:logsfr_track}, but for the sSFR. As can be seen, the sSFR (i.e. the inverse growth time scale of a galaxy) decreases over time to the present day values in an approximately linear fashion after an initially steep decline. Including stellar mass loss prescriptions leads to a more gradual decline in the sSFR (in log space) compared to the direct SFR-to-mass growth.
}\label{fig:logssfr_track}
\end{figure*}

In Figures~\ref{fig:logsfr_track},~\ref{fig:sfr_track}, and~\ref{fig:logssfr_track}, we plot our MSI tracks for $\log \psi$, $\psi$, and $\log\phi$, respectively, for a typical $\log M_*(t=t_0) = 10$ galaxy (in the sense that this is a typical galaxy mass) assuming a seed mass of $\log M_* = 7$. We choose this starting mass based on observations of globular clusters and ``super'' star clusters, which are $\sim 10^{6-7}\,M_\odot$ and come into being almost instantaneously ($< 5$\,Myr). We note that by assuming such a seed mass, we are extrapolating MS evolution to lower masses than observed at higher redshifts; however, as mass growth is extremely rapid at masses smaller than $M_* = 10^7 M_\odot$, the calculated formation times and ages are only weakly a function of mass and do not affect our conclusions. See  Appendix~\ref{app:leitner} for more discussion on some of the uncertainties present our methodology.

We include the effects of stellar mass loss using the zeroth-order and first-order approximations of the fractional galaxy-wide stellar mass-loss rate presented \citet{leitnerkravtsov11} and \citet{leitner12} assuming a Chabrier IMF (functionally equivalent to a Kroupa IMF for our purposes). These take into account mass lost through the death of massive, young stars (zeroth-order; $\eta(t) \sim 0.45$) as well as mass loss from the remaining ensemble of older stars (first-order; $d\eta/dt \sim 2/3\,\,\times$ galaxy age, normalized to a  $\log M_*(t=t_0) = 11$ galaxy). The latter expression is accurate to within a few percent of the final mass. As the use of these mass-loss prescriptions for a given final mass leads to less-efficient growth, these result in a shift to earlier formation times as well as earlier and larger peaks in SFRs. In the following discussion, we only discuss MSI tracks for $\log M_* (t=t_0) = 10$ galaxies, which serve as a soft upper limit to typical SFGs seen today (B04; S07). However, we have checked that the majority of our conclusions still hold for galaxies with smaller masses down to $\log M_* (t=t_0) \sim 8$. %(indeed, they look more like smaller subsets of the $10^{10} M_\odot$ tracks)

Two star-forming limits of the MS are clearly present in our tracks. At lower masses, we see that a typical MS galaxy experiences extremely rapid mass growth in a very short amount of time. This amounts to taking the limits $\left[\alpha(t),\beta(t),\zeta(t)\right] \approx \left[\alpha,\beta,\zeta\right]$ at some particular time, where $\zeta(t_{\textrm{gal}})$ is the mass growth efficiency ($\dot{M_*}/\psi$ or $1-\eta$) as a function of time, which is a function of the age of the galaxy ($t_{gal}$) and the past SFH. In this limit, the mass growth as a function of mass is a simple power law,
\begin{equation}
\frac{dM_*}{dt} \propto M^\alpha,
\end{equation} 
and thus
\begin{equation}
M_*(t) = (\alpha - 1) \times (C - t)^{1/(1-\alpha)},
\end{equation} 
where $C$ is a constant of integration. This ``fast growth'' behavior is clearly visible in Figure~\ref{fig:logsfr_track}, where $\log\psi$ grows by over an order of magnitude in an extremely short period of time.

At later times, especially in the most realistic case with the first-order approximation, we can see an approximate linear decline in $\log\psi$. This can be best understood in the ``slow growth'' mass limit, where $\log M_*(t) \approx M_*$, and growth is dominated by evolution in $\alpha$ and $\beta$. In this case, at a given mass, we end up with
\begin{equation}
\log\psi \sim (\alpha_t \, \log M_* + \beta_t) \, t,
\end{equation}
which leads to approximately linear decay in $\log\psi(t)$ (i.e., exponential decay). This behavior can again be seen in Figure~\ref{fig:logsfr_track}. By looking at $\psi$ rather than $\log\psi$, however, we actually see that the power-law rise can be easily approximated by a linear rise because it dominates such a small portion of a typical MS galaxy's SFH (see Figure~\ref{fig:sfr_track}). 

We examine the evolution in sSFR and find that MS galaxies seem to become progressively less efficient (i.e. decreasing fractional mass growth) at forming stars over the course of their lifetime (see Figure~\ref{fig:logssfr_track}). This is a natural extension of the sub-linear nature of the $\log M_*$\,--\,$\log\psi$ relation, and decreases in a log-linear manner for the majority of the MS galaxy's lifetime (i.e. during the ``slow growth'' phase). By extension, this leads to an approximately linear fractional mass growth rate for the majority of a galaxy's lifetime, minus initial upturns at early times (where gas is plentiful and the ``fast growth'' to ``slow growth'' transition is still taking place) and downturns at late times (where the gas supply is being exhausted and the mass loss from old stars is significant).

Using these MSI tracks, we are then able to better judge what comprises ``typical'' MS SFHs, which may help to improve future model fits. As the current variation in the ways the SFH is parametrized (e.g., exponentially declining, power-law rising, delayed-$\tau$) might influence SED-derived SFRs by up to several tenths of a dex \citep{maraston+10}, choosing the proper SFHs for SED fitting is important if we wish to use them to eventually reliably derive physical properties of SFGs \textit{beyond} just masses.

Our findings indicate that at young ages/early times/low masses, MS galaxy SFHs seem to be well-characterized by a rising SFHs, which can be well-modeled by any of the current linear, exponential, or power-law parametrizations, provided the ages are young. At middle ages/times/masses, MS galaxies are forming stars at around their peak SFRs, and are best fit by constant SFHs or very slowly rising/declining SFHs, depending how far around the peak they are currently being observed. Finally, at later times/older ages/higher masses, we see that MS galaxies are best characterized by traditional exponentially declining SFHs.

Based on our MSI-derived SFHs, we would expect there to be a changing distribution of SFHs with masses (from linearly rising to constant to exponentially declining as the fitted mass increases) since, at a given redshift, mass is a proxy for age. As of now, no paper has done a detailed study of the distribution of best-fit SFHs within MS relations at fixed redshifts. A study along these lines should help us better understand the parameter space explored when fitting MS galaxies (both physically and computationally), and possibly lead to development of priors and other procedures that can cut down on the size of the parameter space that is often explored when fitting photometric data.

A number of studies have shown that masses are relatively robust to variations in the SFH (e.g., \citealt{maraston+10}; R12; So14). However, if one hopes to use SED-derived SFRs (which can be quite sensitive to the SFH) in addition to masses, these results indicate that typical MS SFHs (for galaxies that follows the MS at all times and do not undergo any disruptive major mergers) include a combination of all three types of SFHs -- rising, constant, and exponentially declining -- as a function of mass/age/time. We find that this type of behavior is naturally reproduced using delayed-$\tau$ SFHs (see \S\,\ref{subsubsec:sfh}), and that no study included here except St14 has fit their samples using these SFHs. Following recent results from \citet{leitner12} and M13, we further note that models including random burst components or multiple component SFHs will likely not significantly improve fits to the SED (S11; So14), and likely do not need to be included.

\subsection{Generating Scatter about the Main Sequence}\label{subsec:t_syn}

While the MSI tracks are useful at exploring average MS evolution, they often fail to incorporate and/or generate \textit{scatter} about the MS (L11; \citealt{leitner12}; see also \citealt{munozpeeples14}). We propose a simple model to generate scatter about the MS and briefly explore its implications (see also Steinhardt \& Speagle 2014, subm.). We first assume that MS galaxies all follow the deterministic evolutionary track set by our best-fit MS evolution included here, and that their evolution is only governed by their initial formation time. For a galaxy born at, e.g., $t = 2$\,Gyr, we calculate its evolution in $M_*$\,--\,$\psi$ space using the MSI tracks outlined above that incorporate our first-order stellar mass loss approximation. In addition, galaxies at a given mass form at a variety of different times \citep{pressschechter74}. In order to accommodate this initial spread in formation times, we further assume that, at any given time, an ensemble of MS galaxies with a spread in formation times all follow the same evolutionary tracks. Thus, a galaxy born at, e.g., $t = 2$\,Gyr would follow the same evolutionary track as one born $\pm 0.5$\,Gyr before or after it. By extension, the galaxies born earlier/later would be $\pm 0.5$\,Gyr ``ahead'' or ``behind'' at every possible time.

Due to both stellar mass loss and a MS slope of less than unity (which prevents runaway exponential growth), galaxies evolving along the same set of evolutionary tracks at different times end up at slightly different locations in the $M_*$\,--\,$\psi$ plane. If such an ensemble of ``coeval'' MS galaxies are observed at a common time, this translates into the upper and lower $\pm 1\sigma$ bounds of MS-like distribution. The only difference between an observed MS relation and our ``coeval'' MS relation is we have assumed that instead of a spread in SFR at a given mass (or vice versa) there is instead a completely fixed $M_*$\,--\,$\psi$ relation with an initial spread in \textit{time}. As such an ensemble of galaxies follows the same evolutionary path (by construction), they can be thought of as evolving ``in sync''. Consequently, we term this $1\sigma$ spread in (formation) time a \textit{synchronization timescale}, $\tau_s$.

This could alternately be interpreted as determining the accuracy of a MS inversion, where we have turned our $\log\psi(t)$ fits (at fixed mass) into $t(\log\psi)$ ones. Thus, $\tau_s$ is the accuracy to which we can determine the cosmic epoch when an observed MS galaxy is active given its mass and SFR: we have just transformed our ``vertical'' scatter (in $\log\psi$) to a ``horizontal'' one (in $t$). $\tau_s$ is thus a physically meaningful quantity.

The behavior of $\tau_s$ as a function of time also has important implications for galaxy evolutionary models. At one extreme, galaxy evolution can be thought of as being driven by a variety of disruptive stochastic processes such as major mergers. At the other, galaxy evolution is instead driven by more deterministic ones such as ``cold mode'' accretion \citep{keres+05,dave08}. If galaxy evolution is driven by disruptive, stochastic processes, then over a given interval an ensemble galaxies at higher redshifts should have smaller $\tau_s$ values compared to galaxies at low redshift as galaxies slowly ``evolve apart''. However, if galaxies simply follow a common, deterministic track (as we have assumed here using the MS) or follow an attractor solution\footnote{While these deterministic solutions may be smooth, they may also be approached through, e.g., a large number of random star formation episodes via the central limit theorem, and thus intrinsically remain stochastic processes. While we thus might be able to rule out large, disruptive, and/or rare stochastic processes, our model is fully consistent with many smaller, more common ones.} (set by the MS), then $\tau_s$ should remain nearly redshift-independent across a wide range of redshifts.

As the sets of tracks offset by our choice of synchronization timescale serve as upper and lower $1\sigma$ bounds for the MS observed at any particular time, we can use $\tau_s$ to derive get the corresponding true scatter ($\sigma_t$) in SFR at a given mass\footnote{Since the slope of the MS is not constant in our fits, the scatter varies as a function of mass. In this section, we assume $\log M_* = 10.5$.}. The true scatter at a given mass is actually quite sensitive to the synchronization timescale that we pick, with changes of just $\sim 0.3$\,--\,$0.4$\,Gyr altering $\sigma_t$ by $\sim 0.05$\,dex. Our best-fit constant value of $\sigma_t \sim 0.2$\,dex directly translates into a constant $\tau_s$ of $\sim 1.4$\,Gyr. A spread in formation times combined with smooth SFHs for individual galaxies thus can account for the observed scatter, and provides a consistent view of the evolution of the MS Mergers, by contrast, must likely either remove objects from the MS entirely (and must do so relatively quickly and/or infrequently) or have very little effect on the overall $M_*$\,--$\psi$ relation (see also S11). Although this model provides a consistent framework for interpreting the MS, it is important to note that this model is not unique: a model with zero age spread and a stochastic component in the SFH of individual galaxies could plausibly also reproduce the data (although possibly by construction; see, e.g.,  \citealt{munozpeeples14}).

In addition, such a small $\tau_s$ implies that we can determine the cosmic epoch a galaxy is active to within $\sim 10\%$ ($\sim 15\%$ including observational errors on the SFR) knowing only its mass and SFR. Combined with a good prior on the redshift, such accuracy could easily be turned into a strong prior on photo-z codes that could significantly constrain the parameter space that needs to be explored for any specific galaxy and would help to distinguish between competing solutions at markedly different redshifts. Using this information thus offers the possibly of markedly reducing systematic catastrophic errors. This could be improved further by combining information on correlations among other physical parameters, such as the mass\,--\,dust attenuation relation from \citet{garnbest10}, IRX$(z)$ results from R12a and B12, and the form of the MS presented here. We plan to investigate possible improvements in photo-z accuracy and computational efficiency that can arise from such priors in a future work.

\section{Discussion}
\label{sec:disc}

\subsection{Extrapolations to High Redshift}
\label{subsec:high_z}
The fits presented in this paper have purposely excluded data within the first and last 2\,Gyr of the Universe in order to avoid biases and strong selection effects. In this section, will discuss how well our fits do when compared to current published high-$z$ results. Comparisons of our MS fits to low-$z$ results and their implications for MS systematics are described in Appendix~\ref{app:low_z}.

As shown in Figure~\ref{fig:sfr_t_relations}, the time-based extrapolations are overall consistent with the MS observed in the distant Universe, with observations from D09 (radio stacking, K-band limited survey), S11 (spectroscopic follow-up of H$\alpha$-emitting LBGs), L11 (LBG survey), L12 (LBG survey), B12 (LBG survey), and St14 (IR-limited survey combined with COSMOS field data) following our best fits to within $\sim 0.5$\,dex. We find that at higher masses, the observations of D09, S11, and St14 seem to follow our fits most closely, while at slightly lower masses those of L11, L12 and B12 (all deep photometric LBG surveys) provide slightly better fits (see Figure~\ref{fig:sfr_t_relations} and~\ref{fig:sfr_t_relations_selection}). 

This dichotomy of results between flux-limited and LBG surveys might indicate two possibilities. The first would be that selection effects (e.g., a lower limit on the SFR, completeness) play a significant role in MS observations at these high redshifts, even when surveys are not near detection limits. The second would be that extinction corrections in some of these studies have been under/overestimated at higher/lower masses\footnote{For instance, L12's disagreement with our best fit at higher masses could be explained if B12's extinction corrections (i.e. IRX) have been underestimated by $\sim 0.1$\,dex in more massive galaxies.}. These in turn might point to systematic issues with IRX\,--\,$\beta$ and SED-fitted extinction corrections. We note that both issues would serve to increase the MS slopes at these redshifts towards unity, in agreement with our MS fits.

The good agreement at high redshift is somewhat surprising given that the $\log\psi$\,--\,$t$ relationship is likely not indefinitely linear in time. Assuming an Eddington-limited starburst \citep{capak+08,younger+08}, such an upper-limit to the SFR would be approximately
\begin{equation}
\psi_{\textrm{max}} = 560\,\sigma_{400}^2 D_{kpc} \kappa_{100}^{-1}\,M_\odot\,\textrm{yr}^{-1},
\end{equation}
where $D_{\textrm{kpc}}$ is the characteristic physical scale of the ``starburst'' in kpc, $\sigma_{400}$ is the line-of-sight velocity dispersion in units of 400\,km\,s$^{-1}$, and $\kappa_{100}$ is the dust opacity in units of 100\,cm$^{2}$\,g$^{-1}$. By assuming that the starbursting system is in virial equilibrium ($\sigma^2 \propto M_{\textrm{tot}}/D_{\textrm{kpc}}$, where $M_{\textrm{tot}}$ is the total mass contained within $D_{\textrm{kpc}}$) and that $M_* \propto M_{\textrm{tot}}^a$, where $a$ is an arbitrary constant, we arrive at $\psi_{\textrm{max}} \propto M_*^a$. For $a = 1$, such a system would display similar behavior to Eddington-limited accretion for black holes. The fact that we do not see any type of limit up to $z \sim 5$\,--\,$6$ (i.e. we still observe a continual increase in sSFR; St14) is intriguing.

On the whole, the agreement between our fits and current high-$z$ observations is quite good (St14), and justifies our fits as functions of time instead of redshift; other studies that fit MS evolution as function linear in redshift (Z12; W12) or as a power-law (K11) do not provide nearly as good fits to the data over the full redshift range that we span. While we are not able to definitively rule out a possible ``plateauing'' of the sSFR in the redshifts explored here, our results seem to favor a scenario where the sSFR continues increasing until at least $z \sim 5$ (and likely plateaus afterwards).

\subsection{Implications on Galactic Star Formation Processes}
\label{subsec:sfr_t}

Star formation in galaxies has been found to correlate well with several large-scale observables. The Kennicutt-Schmidt (KS) relation \citep{schmidt59,kennicutt98b} implies that the SFR surface density ($\Sigma_{\psi}$, $M_\odot$\,yr$^{-1}$\,kpc$^{-2}$) in a wide range of galaxies is related to the gas surface density ($\Sigma_{G}$, $M_\odot$\,pc$^{-2}$) via a simple scaling relation of the form $\Sigma_\psi \propto \Sigma_G^{1.4}$ \citep{schmidt59,kennicutt98b}. This implies that the physics governing star formation is only dependent on the amount of gas available, and is otherwise unrelated to other properties of the host galaxy or any past star formation.

According to the Elmegreen-Silk (ES) relation \citep{elmegreen97,silk97}, however, this does not capture all of the physics. Galaxies are instead expected to consume similar fractional amounts of gas during each orbit, and by including the dynamical timescales ($\tau_{d}$) of the galaxies in question, the SFR surface density can instead be parametrized as $\Sigma_{\psi} \propto \Sigma_{G}/\tau_{d}$ \citep{kennicutt98b}. By contrast, \citet{shi+11} propose that it might not be the kinematics of the host galaxy at all that determine $\Sigma_\psi$, but rather past star formation. This would indicate that feedback effects -- such as supernova and/or Active Galactic Nuclei (AGN) -- play a significant role in determining a galaxy's current SFR. By parameterizing by past star formation via the stellar mass surface density ($\Sigma_*$, $M_\odot$\,pc$^{-2}$), \citet{shi+11} propose an ``Extended'' KS (EKS) law of the form $\Sigma_{\psi} \propto \Sigma_{G}\Sigma_{*}^{0.5}$.

In addition to these scaling relations, hydrodynamical simulations of star-forming galaxies at high-$z$ (e.g., \citealt{finlator+06}) also predict a tight relationship between mass and SFR, with a stronger dependence ($M_* \propto \psi$) that evolves slowly with redshift \citep{dave08}. In this scenario, a slope slightly below unity occurs due to feedback, which leads to the growth of hot halos around higher mass galaxies and slows down gas accretion.

In this context, the marked \textit{linearity} of the decline in $\log \psi$ over time in these data is striking.  One possible implication is that the age of a galaxy, rather than more stochastic events such as major mergers, is the most important indicator and/or driver of gas availability and star formation (see also \S\,\ref{subsec:t_syn}). While other factors such as environment likely play a role in galaxy evolution by, e.g., influencing gas availability and/or delivery (Khabiboulline et al. 2014, subm.), this steady decline seems to favor a much steadier, environment-independent mode of star formation\footnote{This environmental-independence is only true at a given mass. Environment likely plays a role in determining the mass of a galaxy and possibly ``accelerating'' evolution along the MS, even if it does not change the form of the MS itself \citep{koyama+13,lin+14}.}.

By using observations of MS evolution included here, we can attempt to distinguish between each of the different scenarios outlined above. If we substitute a few scaling relations ($\psi \propto R^2\Sigma_{\psi}$, $M_* \propto R^2\Sigma_{*}$, and $\tau_d \propto R/V$, where $R$ is the characteristic radius of the system and $V$ is the corresponding characteristic velocity), assume the system is in virial equilibrium ($V^2 \propto M_{\textrm{tot}}/R$, where $M_{\textrm{tot}}$ is the total mass contained within R), assume that $M_* \propto M_{\textrm{tot}}$, and that the dark matter halo follows a simple spherical distribution ($M_{\textrm{tot}} \propto R^3$), we can transform the KS, ES, and EKS scaling relations listed above into more MS-like relations. Substituting and rearraging, we derive MS relations for $\log\psi \sim \alpha \, \log M_* + \beta$ with $\alpha = 2/3, 2/3$, and $5/6$ and $\beta = 1.4\log\Sigma_G, \log\Sigma_G$, and $\log\Sigma_G$ (up to a scaling constant) for the KS, ES, and EKS laws, respectively. 

For each of these parametrizations, evolution the normalization of the SFR could be due to evolution in the average gas surface densities, while evolution in the slope could be due to deviations from some of the assumptions assumed here. These findings seem in good agreement with results from \citet{magdis+12} that MS galaxies likely follow a single, tight $L_{\textrm{IR}}$\,--\,$M_{\textrm{gas}}$ (i.e. $\psi$\,--\,$\Sigma_G$) relation from redshift $z \sim 0$\,--\,$2$.

In order to compare the results above (involving SF laws in disks) with those of the MS (involving total masses), however, we are required to somehow implement a correction to account for possible bulge components (and bulge growth) of MS galaxies. Using the recent results of \citet{abramson+14} that show an increase in MS slope of $\sim 0.25$ after limiting their analysis to disk components only, our best MS fits seem to imply a $M_{\textrm{disk}}$\,--\,$\psi_{\textrm{disk}}$ relation that begins at approximately unity and decays to $\sim 0.65$ at $z = 0$. As the KS and ES relations both predict MS slopes of $\sim 0.65$, while the EKS law predicts a slope closer to $\sim 0.85$, our findings seem to support the former two views of star formation over the latter\footnote{Note that the slope for MS (disk) relation from \citet{abramson+14} at $z \sim 0$ ranges from $0.57$\,--\,$0.76$ ($0.80$\,--\,$1.0$).} (see also Appendix~\ref{app:mass_function}). Taken together with observations that the ES relation encompasses a broader range of galaxy classes than the KS relation while proving equally good fits \citep{daddi+10,magdis+12}, we interpret this as tentative evidence supporting the ES view of star formation.

In conclusion, if we assume current hydrodynamical simulations are representative of galaxy evolution at high-$z$, our results seem to indicate that MS galaxies transition from ``cold mode'' accretion scenarios (initial conditions) to more general ES-dominated modes of star formation (equilibrium conditions) over time.

\section{Conclusion}
\label{sec:conc}

64 measurements of the star-forming (SF) ``Main Sequence'' (MS) from 25 papers have been combined to determine MS evolution out to $z \sim 6$. They have been recalibrated to use a common set of assumptions, correcting for stellar IMF, $L$\,--\,$\psi$ conversion, cosmology, SPS model, dust attenuation, emission lines, and possible $q_{\textrm{IR}}$ evolution. Our main conclusions are as follows:

\begin{enumerate}
\item By taking into account mass information, we are able to derive a robust functional form for the MS that indicates strong differential mass evolution (i.e. a time-dependent MS slope), with a best fit of $\log\psi(M_*,t) = (0.84 \pm 0.02 - 0.026 \pm 0.003 \times t) \log M_* - (6.51 \pm 0.24 - 0.11 \pm 0.03 \times t)$.
This provides good fits from $z \sim 0$\,--\,$6$. Almost all of our fits show strong departures from unity.
\item Using our fits, we are able to deconvolve the scatter around the MS with the scatter due to evolution within any given time/redshift bin. After accounting for intrinsic scatter among SFR indicators, we find that the true scatter among the MS is likely $\sim 0.2$\,dex rather than the $\sim 0.3$\,dex often reported. Future studies should try and emulate this procedure to better compare the derived ``widths'' of MS observations. Scatter about the MS is functionally equivalent to a group of galaxies following identical evolutionary tracks with an initial 1$\sigma$ spread in formation times of $\sim 1.4$\,Gyr. Scatter about the MS (i.e. $\sim 0.2$\,dex uncertainty in SFR at a given mass) can thus be directly translated into a scatter in time (i.e. $\sim 10$\,--\,$15\%$ uncertainty in the age of the Universe at a given mass and SFR).
\item $sBzK$-selected samples have systematically smaller scatters than most other studies included here, and are likely substantially biased compared to other selection mechanisms. More generally, we find that selection effects and other systematic effects can have a big impact on the slope of the MS and should be taken into account when conducting future surveys and interpreting results.
\item With our new calibrations, we report possible evidence for $(1+z)^{-0.8}$ evolution in $q_{\textrm{IR}}$ (i.e. $\psi_{1.4}/\psi_{\textrm{other}}$ goes as $(1+z)^{\sim 0.8}$). While the exact meaning of the observed evolution is still uncertain, this at least indicates that possible evolution in pre-existing radio assumptions should be considered when interpreting (stacked) radio SFR data in the future.
\item The SFH of a typical MS galaxy involves a combination of approximately linearly rising, constant, and exponentially declining SFHs. These SFHs can be most easily generated using delayed-$\tau$ models, which should ideally be used in future studies to in order avoid possible biases when deriving physical properties outside of the mass. In addition, the fractional mass growth of a typical MS galaxy is approximately linear for the majority of its lifetime, with deviations at early and late times due to high sSFRs in the early Universe and significant stellar mass loss from older stellar populations, respectively.
\item The evolution of the SFR at fixed mass is well fit by a log-linear evolution in time. Furthermore, fitting MS evolution as a function of time significantly improves the quality of the fits to the MS relative to previous fits as a function of redshift.
\end{enumerate}

Existing studies on star formation represent a strong and consistent constraint on galaxy evolution over the past 12 billion years. The consistency of this constraint, however, is masked by inconsistent calibrations, and future studies should use a standard set of assumptions (or provide conversion factors) in order to make results directly comparable. We propose one such standard in this work; however, as all studies included here have been calibrated to a set of common assumptions, any future study should be able to easily convert all these results to a modified set of assumptions without too much difficulty. Similarly, future studies of SFGs should attempt to select their samples on  more uniform criteria (such as $\textrm{NUV}rJ$ or $UVJ$) and focus increased attention on the effects of various SFHs on SED fitting procedures in order to obtain more robust SED-derived SFRs and better constrain systematics. The methods of extracting more robust parameters from galaxy SED fits (e.g., the median masses used in So14) should be investigated in future studies.

By properly calibrating existing MS studies, we arrive at a consistent picture of star forming galaxies out to $z \gtrsim 5$. However, this picture is one where high-$z$ galaxies have ever-higher masses and SFRs, sharpening a series of puzzles surrounding star formation in the early universe. It is difficult both empirically and theoretically -- from fully empirical tracks \citep{leitner12} or semi-empirical parametrizations \citep{wetzel+13,behroozi+13} to semi-analytic models \citep{somervilleprimack99,somerville+08,mitchell+14} and numerical simulations \citep{finlator+06,katsianis+13} -- to produce massive SFGs in the early Universe. However, there seems to be no signs of deviation from this trend even out to $z\sim 5$\,--\,$6$ (L12; St14). The downsizing paradigm (i.e. ``anti-hierarchical'' growth, with increasingly massive galaxies found at earlier redshifts) has long provided a theoretical challenge for merger-driven evolutionary models \citep{fontanot+09}, and the consistency of this picture out to high redshift seems to further increase tension between theory and observation.

Based on these results, pushing observations to only slightly higher redshifts should move us from the regime of ``downsizing'' to one of ``upsizing'' (with the resulting details likely to yield greatly improved models). If instead astronomers continue to find even more massive SFGs at earlier times, we are likely to arrive in a scenario where there was not enough time for these extremely high-$z$ objects to form given our current understanding of masses, structure formation, and the limits from the CMB on reionization. Regardless of the eventual outcome, future high-redshift observations are poised to provide answers to current unresolved questions surrounding galaxy formation and evolution in the early Universe.

\acknowledgements
The authors would like to thank the anonymous referee for thorough and insightful comments that greatly improved the quality of this work. The authors would also like to thank Martin Elvis, Daniel Masters, David Sanders, and David Sobral for their input and helpful comments, as well as Douglas Finkbeiner for advising the junior thesis course where much of this work was completed. In addition, the authors would  like to thank Herv\'{e} Aussel, Kevin Bundy, Brian Feldstein, Olivier Ilbert, Emil Khabiboulline, Alex Krolewski, John Moustakas, Alvio Renzini, and Michael Strauss for useful discussions and/or comments. In addition, we would like to thank Stephane Arnouts, Rychard Bouwens, Alison Coil, Loretta Dunne, Alex Karim, Mariska Kriek, Kyoungsoo Lee, Sam Leitner, Georgios Magdis, John Moustakas, Eric Murphy, Kai Noeske, Seb Oliver, Samir Salim, Paola Santini, Hyunjin Shim, and David Sobral for providing us with data and/or answering questions about their work. We wish to especially thank Alex Karim, Eric Murphy, and Olivier Ilbert for insightful discussions about stacking and radio SFR observations. JSS would like to thank Rebecca Bleich for all her support. JSS was partially supported by grants from the Harvard Office of Career Services' Weismann International Internship Program and the Harvard College Research Program.

\begin{appendix}
\section{Data}\label{app:data}

We give brief descriptions of each of the data sets included in this study below, and describe our methodology in converting the reported observations to our common calibration. We encourage anyone interested in more details to read the actual paper(s) in question and/or contact the original authors. The main assumptions used to derive the MS relations for each study are summarized in Table~\ref{tab:msfr_info}. MS relations, plus other general information, are listed in Table~\ref{tab:msfr}. All studies listed here do not (explicitly) include AGN in their analysis, and remove them via hard X-ray detection matching, SED fitting, and/or power-law fits to the observed IR emission.

\textbf{{\citet{chen+09} (C09)}}. C09 observes $\sim \sn{5}{5}$ and $\sim 3000$ galaxies in the SDSS Data Release Four (DR4; \citealt{adelman-mccarthy+06}) and a combination of the Deep Evolutionary Exploratory Probe 2 (DEEP2; \citealt{davis+07}) survey and Palomar Observatory Wide Infrared (POWIR) survey at $0.005 < z < 0.22$ and $0.75 < z < 1.0$, respectively. Although C09 do not provide a MS relationship for their fits (which we will term C09(1) and C09(2), respectively) in the paper itself, the fit they \textit{do} provide in their Figure~11 does not seem to include all data points (excluding the ones on the low mass end), and is calculated without assuming extinction. This fit is does not utilize an effective cut to exclusively analyze SFGs.

We derive a MS-esque $M_*$\,--\,$\psi$ relation in a self-consistent manner as the rest of the studies listed here as follows. We first take their derived sSFRs for both their SDSS and DEEP2 datasets, listed in their Table~2, and correct them for extinction based on the PDFs in their Figure~6 and Figure~7. We keep the 1$\sigma$ error bars the same, as the shapes of the dust-corrected PDFs are almost identical to those without dust. We fit all the data points with a simple linear fit using total least squares regression with Scipy's \texttt{ODRpack} in order to account for bin size as well as the SFR PDFs. For the resulting $M_*$\,--\,$\psi$ relationship, we observe that a strong ``levelling out'' of SFRs at higher masses occurs in both datasets, likely due to increasing amounts of quiescent galaxies included in their mass bins, and therefore opt to exclude the 11\,--\,12 mass bins from the C09(1) sample and the 11.5\,--\,12 mass bin from the C09(2) sample.

The SFRs are derived by finding the best fit between model Balmer absorption features and the observed features from the composite stacked spectra in each mass bin. Masses are derived for the SDSS sample using the same procedure as described in B04 and S07 assuming a Kroupa IMF. Masses are derived for the DEEP2 sample using the same procedure but assuming a Chabrier IMF, and are corrected to a Kroupa IMF via a conversion factor of 1.12. Extinctions and other parameters are derived via the Balmer decrement as detailed in Table~\ref{tab:msfr_info}.

\textbf{Coil et al. (2014, in prep.) (C14)}. C14 observe $\sim 165000$ ($\sim 97000$; $z<0.2$) and $\sim 36000$ ($\sim 24000$; $0.2 < z < 1.0$) galaxies (mass-complete sample) from SDSS -- matched with data from GALEX -- and the PRism MUlti-object Survey (PRIMUS; \citealt{coil+11}; \citealt{cool+13}; M13), respectively. All the objects have high quality spec-z's, and thus are not subject to uncertainties and contamination that concern photo-z's. Both masses and SFRs are derived using SED-fitting, as described in Table~\ref{tab:msfr_info}, and are taken from M13.

In order to separate quiescent galaxies from SFGs, C14 divides the sample based on the minimum between the bimodal distribution in the $M_*$\,--\,$\psi$ plane as a function of redshift, which is well fit by $\log\psi = -10.139 + 0.85\log M_* +3.72z - 2.7z^2$. All galaxies above the cut are classified as SFGs, while all those below are classified as quiescent. This is in contrast to most other studies, which have used a color-color diagnostic to separate out star-forming and quiescent systems (e.g., \citealt{ilbert+13}). As we find that C14's sample has much larger scatter than the majority of other published MS studies (with the exception of So14's low-$z$ sample), we hypothesize that this difference in SFG-selection leads to larger differences in the overall sample (ignoring possible systematics from SED-derived SFRs). Color-color diagnostics often seem to lead to ``star-forming''-classified systems having a relatively tight M-SFR relation while ''quiescent''-classified systems occupy an extremely large SFR for fixed mass (see, e.g., Figure~18 from S07 and/or Figure~7 from \citealt{schiminovich+07}). These quiescent-classified systems, even while they occupy mostly the lower range on the M-SFR diagram, have a significant long tail up to higher SFRs that overlaps with the SF-systems \citep{schiminovich+07}. 

If this is true, then this would bias the minimum of an added bimodal distribution towards lower values. This would not only increase the apparent scatter for SF systems on the MS, but bias the slope downwards towards lower values (especially the more strongly/exclusively this effect occurs in higher mass bins). As PRIMUS is mass-complete, not only would the scatter increase because at a given redshift C14 has probed down to lower (s)SFRs (i.e. there is not a built-in (s)SFR-esque cut like one that can occur with many mass-incomplete surveys), but because at all redshifts PRIMUS probes there's a decent fraction of quiescent systems in the sample \citep{brammer+11}.

\textbf{\citet{daddi+07} (D07)}. D07 observe mid-IR (MIR), far-IR (FIR), submillimeter, radio, and UV emission from 1291 $sBzK$ galaxies in the Great Observatories Origins Deep Survey (GOODS; \citealt{giavalisco+04}) North (GOODS-N; 273 galaxies) and South (GOODS-S; 1018 galaxies) fields from $1.4 < z < 2.5$. Multiband photometry in the optical and NIR is taken from \citet{giavalisco+04}; MIR and FIR from \textit{Spitzer}'s Multiband Imaging Photometer for Spitzer (MIPS; \citealt{rieke+04}); submillimeter from the Submillimetre Common-User Bolometer Array (SCUBA; \citealt{holland+99}) maps of \citet{borys+03} and \citet{pope+05}; and radio from Very Large Array (VLA) observations taken from an early \citet{morrison+10} catalog. $L_{\textrm{IR}}$ was derived either by fitting CE01 and DH02 templates or via the IR\,--\,radio correlation from, e.g., \citet{yun+01}, and $L_{\textrm{UV}}$ was derived from $K$-corrected and extinction corrected $B$-band flux following D04. SFRs are obtained via the $L_{\textrm{UV}}$\,--\,$\psi$ relationship outlined in D04. Photo-z's for GOODS-N were determined by fitting to the empirical templates of \citet{coleman+80} as described in D04. Photo-z's for GOODS-S are taken from \citet{grazian+06}, which uses PEGASE.2 models with D-SFHs assuming a \citet{ranabasu92} IMF (slightly steeper than a Salpeter IMF) and a primordial metallicity. Masses are obtained from \cite{fontana+04} as detailed in Table~\ref{tab:msfr_info}.

\textbf{\citet{dunne+09} (D09)}. D09 analyze 1.4\,GHz radio emission from VLA observations \citep{ivison+07,ibar+08} of 23185 galaxies from the Ultra-Deep Survey (UDS) portion of the United Kingdom Infrared Telescope (UKIRT) Infrared Deep Sky Survey (UKIDSS; \citealt{lawrence+07}). Photo-z's were derived as described by \citet{cirasuolo+10} and noted in Table~\ref{tab:msfr_info}. Masses are computed from rest-frame K-magnitudes (calculated using the best-fitting SED template) using the $M_*/L$ relation given by the Millennium simulation \citep{delucia+06} in the same manner as \citet{serjeant+08} using a Salpeter IMF. The parameters which go into this conversion are detailed in Table~\ref{tab:msfr_info}. SFRs are calculated using both the \citet{condon92} and \citet{bell03} conversions ($\sim 0.87$ that of KE12), although only the \citet{bell03} conversion is used in this work so that D09's results are directly to K11's. 

We note that the \citet{bell03} calibration actually has two components, a linear component for most masses that simply scales with the luminosity, and a luminosity-dependent component for luminosities lower than a characteristic value. However, this non-linearity only affects the lowest mass bins in D09's lowest redshift bin. As $M_*$\,--\,$\psi$ fits are not presented in D09, we directly fit the data (L. Dunne, priv. comm.) used in their Figure~13 with Scipy's \texttt{ODRpack}.

\textbf{\citet{elbaz+07} (E07)}. E07 use data from two samples -- SDSS DR4 ($ 0.015 < z < 0.1$) and GOODS (both fields; $0.8 < z < 1.2$) -- which will be referred to as E07(1) and E07(2), respectively. Both data sets are selected via $M_{\textrm{bol}}$ and rest-frame $U$\,--$g$. E07(1) includes H$\alpha$ emission from 19590 galaxies, and derives SFRs using the B04 calibration, scaling results from a Kroupa to a Salpeter IMF dividing by 0.7. Masses are computed based on \citet{kauffmann+03} and detailed in Table~\ref{tab:msfr_info}, with results scaled from a Kroupa to a Salpeter IMF by dividing by 0.7. Extinctions are derived via the Balmer decrement. E07(2) includes UV emission from $\sim 1200$ galaxies. Extinctions and SFRs are derived using the calibrations presented in D04, and masses are derived as detailed in Table~\ref{tab:msfr_info}.

\textbf{\citet{elbaz+11} (E11)}. E11 observe IR emission from 648 LIRGs from the CE01 sample observed with the Infrared Space Observatory (ISO; \citealt{kessler+96}), AKARI \citep{murakami+07}, and the Great Observatories All-Sky LIRG Survey (GOALS; \citealt{armus+09}) from $0 < z < 0.1$. AKARI data was cross-matched with the Infrared Astronomical Satellite (IRAS; \citealt{neugebauer+84}) Faint Sources Catalog ver. 2 \citep{moshir+92} and SDSS DR7, and supplemented with data from the AKARI/Far-Infrared Surveyor (FIS; \citealt{kawada+07}) All-Sky Survey Bright Source Catalogue ver 1.0 and photo-z's from \citet{hwang+10}. Masses are derived as in E07. $L_{\textrm{IR}}$ is calculated from $L_{8 \mu m}$, and SFRs are calibrated on K98 assuming a Salpeter IMF.

\textbf{\citet{karim+11} (K11)}. K11 observe 1.4\,GHz emission from $> 10^5$ $\textrm{NUV}rJ$ galaxies in the Cosmic Evolution Survey (COSMOS) field from $0.2 < z < 3.0$. SFRs are derived using the calibration from \citet{bell03} assuming a Chabrier IMF, which on average gives SFRs 50\% lower than those of KE12 (see Table~\ref{tab:msfr_info}, as well as Appendix\,\ref{app:bell}). As noted previously, the \citet{bell03} calibration has two components, a linear component for most masses that simply scales with the luminosity, and a luminosity-dependent component for luminosities lower than a characteristic value. For the mass bins K11 use to derive their MS relations (since they wanted mostly {\qq}representative{\qq} populations) this lower mass component is unimportant. We calibrate on the higher luminosity component which dominates the fit for all relevant masses, especially at higher redshifts. Photo-z's, masses, and other parameters are derived as detailed in Table~\ref{tab:msfr_info}.

%We wish to note that, in both our uncorrected and corrected initial fits, \textit{SFRs calculated from K11's MS relations display a very notable uptick towards higher redshifts}, similar both qualitatively and quantitatively to $\log(1+z)$ behavior and in disagreement with other SFR indicators (including radio SFRs from \citealt{pannella+09}). Examining the fits listed in their Table~5 of the evolution of specific SFR ($\equiv \psi/M_*$) as a function of redshift, we notice that their best fit is $\propto (1+z)^{\sim 4.3}$, with a power law index $\sim 1$ greater than the evolution commonly seen in other studies ($\propto (1+z)^{\sim 3.5}$; \citealt{rodighiero+11}). After \textbf{confirming???} that K11 has overcorrected their SFRs, we opt to subtract this component from the fit, and find that these $\log(1+z)$-corrected SFRs are in excellent agreement with other measurements from the literature.

\textbf{\citet{kashino+13} (K13)}. K13 observe H$\alpha$ emission from 271 $sBzK$ galaxies in the COSMOS field using the Fiber Multi-Object Spectrograph (FMOS; \citealt{kimura+10}) on the Subaru telescope from $1.4 < z < 1.7$. These are selected from the catalog of \citet{mccracken+10}, based on deep near-IR imaging ($K_s < 23$) from the Canada-France-Hawaii Telescope (CFHT) and optical imaging ($B_J$, $z^+$) from Subaru. SFRs are calibrated on K98 assuming a Salpeter IMF, and extinctions are calculated via the Balmer decrement when available (and averaged by stacking in several mass bins otherwise). Photo-z estimates (and masses) are taken from \citet{ilbert+09} based on photometry as described in \citet{capak+07}.

\textbf{\citet{lee+11} (L11)}. L11 observe 1913 LBGs in the NOAO Deep Wide-Field Survey (NDWFS; \citealt{jannuzidey99}) from $3.3 < z < 4.1$. SFRs are derived from the FUV based on the K98 calibration assuming a Salpeter IMF, while masses (and phot-z's) are derived by fitting a Chabrier IMF as detailed in Table~\ref{tab:msfr_info}. L11 notes that the slope they derive changes from $\alpha = 0.8$\,--\,$1$ depending on the assumed extinction parameters, and fit the MS assuming a slope of unity. We, however, take this variation to be intrinsic error on a slope of $\alpha = 0.9 \pm 0.1$ and recalculate the fit with their normalization scheme, propagating errors accordingly.

\textbf{\citet{lee+12} (L12)}. L12 observes 2952 and 846 LBGs in the GOODS fields from $3.4 < z < 4.4$ to $4.4 < z < 5.6$, respectively. The relationships presented in L12, however, are in terms of $M_{1700}$ rather than $\psi$, and do \textit{not} include extinction corrections. To convert $M_{1700}$ to $\psi$, we take their best fit to the lower-$z$ sample of $\log M=-0.415 \, M_{1700} + 1.2$ (from their equation 13), and the best fit to the higher-$z$ sample of $\log M=-0.442 \, M_{1700} + 0.2$, derived by fitting the median points presented in Figure~3, for the lower and higher redshift bins, respectively, and convert them to luminosities and SFRs using the KE12 relation for the FUV. We correct for extinction using B12's IRX observations at 1600\,\AA assuming the extinction is comparable at 1700\,\AA. This gives us the $M_*$\,--\,$\psi$ relations of $\log SFR = 0.79 \, \log M - 6.34$ and $\log SFR = 0.73 \, \log M - 5.69$ for the two samples, respectively.

We check our assumed extinction correction by examining the data from L11, which exhibits a similar $\log M = -0.413\, M_{1700} + 1.367$ relation (see their Figure~3) but which also includes a derived $M_*$\,--\,$\psi$ relation. We find our results are quite comparable, and agree to within a factor of $\sim 1.5$. We choose to keep the B12 value for consistency with the rest of the papers included here.

As L12 do not provide the actual redshift distribution of the sample included in the fit, we estimate the distribution from spectroscopic follow-up observations by \citet{vanzella+09}. Based on their Figure~6, we find the median redshifts distribution reported in L12 ($\sim 3.7$ and 5.1) are reasonably consistent with the spec-z distribution of observed $B$- and $V$-drops. We therefore take the distribution presented in \citet{vanzella+09}  as representative of the full sample, and assign redshift ranges of 3.4\,--\,4.4 and 4.4\,--\,5.6 (both slightly weighted towards the lower end) for L12's lower and higher redshift sample, respectively. For consistency with other works (where we have picked the midpoint of the distribution), we choose $z_{\textrm{med}}$ to be 3.9 and 5.0 rather than the exact values reported in L12. We find that the difference in time this would lead to is only $\sim$ 100\,Myr and 30\,Myr, respectively -- quite small compared to the observed rate of MS evolution. Masses and photo-z's are derived as detailed in Table~\ref{tab:msfr_info}.

\textbf{\citet{magdis+10} (M10)}. M10 observes UV emission from 106 LBGs from $2.8 < z < 3.2$, taken from \citet{magdis+08} and encompassing a variety of fields. SFRs are determined from $L_{1500}$, using conversions from CB07 models assuming a Chabrier IMF that correspond to $\sim 0.82$ that of the K98 calibration (see Table~\ref{tab:msfr_info}). Masses are derived using CB07 models assuming a Chabrier IMF, which M10 shows on average are lower than those of BC03 models by a factor of $\sim 1.4$. Extinction is modeled as C00, and derived from the IRX\,--\,$\beta$ relation of M99. As M10 provide a slope for their fit ($\sim .91$) but do not provide the normalization, we derive the normalization based on the average mass ($\sim 5\times 10^{10} M_\odot$) and sSFR ($\sim 4.6\,\textrm{Gyr}^{-1}$) of their sample.

\textbf{\citet{noeske+07} (N07)}. N07 observes H$\alpha$ (plus various other emission lines calibrated to H$\alpha$), UV, and IR emission from 2905 galaxies in the All-Wavelength Extended Groth Strip International Survey (AEGIS; \citealt{davis+07}) with $0.2 < z < 0.7$. $L_{\textrm{IR}}$ (and IR SFRs) were determined following \citet{lefloch+05} calibrated on \citet{bell+05}, using CE01 templates and a Kroupa IMF, and SFRs were derived via emission lines, UV+IR, or corrected UV emission when IR measurements were unavailable. Extinction was calculated based on the Balmer decrement. Masses were obtained from SED fits to optical/NIR photometry by \citet{bundy+06} and are detailed in Table~\ref{tab:msfr_info}.

\textbf{\citet{oliver+10} (O10)}. O10 observes FIR emission from $\sim \sn{8}{5}$ galaxies in the Space Infrared Telescope Facility (SIRTF; now \textit{Spitzer}) Wide-Area Infrared Extragalactic Survey (SWIRE; \citealt{lonsdale+03}) from $0 < z < 0.8$. Observations are given in the paper at higher redshifts but are not included here because they do not fulfill our selection criteria (i.e. they only include 2 points in their fit). $L_{\textrm{IR}}$ is derived by fitting a Sc galaxy template from \citet{polletta+06} to 70\,$\mu$m or 160\,$\mu$m emission. Masses (and photo-z's) are determined by SED fitting using the templates of \citet{rowan-robinson+08}, which assumes the same IMF. These templates are empirical, but are regenerated to higher resolution using SPS modelling in order to derive corresponding physical parameters.

SFRs are derived via the calibration presented in \citet{rowan-robinson+08} via a scaling factor and assuming a given fraction of UV energy is absorbed by the dust, which are higher than K98 SFRs by a factor of 1.13. Since the SFR is assuming some UV contribution, which increases the bolometric SFR at a given IR luminosity, we would expect this to indeed be higher. However, by converting back to K98, we can remove the assumed extinction correction and implement our R12a values for consistency with the rest of the results listed here. As this assumes a Salpeter IMF integrated from $.15$\,--$120\,M_\odot$ \citep{babbedge+04}, rather than the usual $0.1$\,--\,$100\,M_\odot$, we use this offset to convert between the standard Salpeter IMF integration range and the one used here. This is most likely an overcorrection, but only leads to an additional change in MS normalization of $-\alpha \times 0.05$\,dex, approximately equivalent to the Chabrier to Kroupa IMF conversion factor assumed here. We fit our MS evolution both with and without this additional correction factor and find that the impact on our results is negligible.

\textbf{\citet{pannella+09} (P09)}. P09 observe 1.4\,GHz emission from 11798 $sBzK$ galaxies in the COSMOS field from $1.0 < z < 3.0$. SFRs are derived using the calibration from \citet{yun+01}, which is calibrated to the K98 relationship assuming a Salpeter IMF. We find that the SFRs derive differ from those offered in KE12 by a factor of $0.93$ (see Table~\ref{tab:msfr_info}), and correct for this accordingly. Masses are taken from \citet{mccracken+10}, which are computed based on a K-band luminosity-to-mass conversion following D04. We correct these based on the results o \citet{arnouts+07}. Photo-z's are calculated from using the same procedure as D07.

\textbf{\cite{rodighiero+11} (R11)}. R11 observe UV and IR emission from 19567 and 698 galaxies, respectively, in the COSMOS and GOODS-S fields from $1.5 < z < 2.5$. The UV sample consists of $sBzK$ galaxies taken from D07 (GOODS-S) and \citet{mccracken+10} (COSMOS), while the flux-limited IR sample was observed with \textit{Herschel}'s Photodetector Array Camera \& Spectrometer (PACS; \citealt{poglitsch+10}). Photo-z's for the $sBzK$ sample were taken from \citet{ilbert+09} and D07 for the COSMOS and GOODS-S fields, respectively, with masses and SFRs have been computed using the same procedure as D07. For the PACS sample, photo-z's (and masses) were derived by cross-matching to the catalog of \citet{ilbert+10} and then fitting via the procedure outlined in \citet{rodighiero+10a}. $L_{\textrm{IR}}$ is derived from PACS fluxes using a P07 and G10 templates as described in \citet{rodighiero+10b}, with SFRs determined from D07.

\textbf{\citet{reddy+12} (R12)}. R12 observe UV and IR emission from 302 LBGs from $1.5 < z < 2.6$, taken from multiple fields detailed in their Table~1. $L_{\textrm{IR}}$ and $L_{\textrm{UV}}$ are derived from $K$-corrected fluxes taken from \textit{Spitzer} MIPS $24\,\mu$m observations per \citet{reddy+10} and broadband photometry, respectively. SFRs are derived via combined UV+IR emission assuming a Salpeter IMF calibrated on the K98 relations. R12 discuss the effects changing the UV-SFR conversion would have on the subsample of galaxies with very young fitted ages in their Section~4.3 and Appendix~A. However, as the majority of their sample consists of older galaxies by design (by imposing restrictions on the SED fitting procedure), this effect should not significantly impact our results. All objects have spec-z's. Extinction is derived via the IRX-$\beta$ relation. Masses are determined as per Table~\ref{tab:msfr_info}.

\textbf{\citet{salim+07} (S07)}. S07 observes 48295 $r$-selected galaxies from the joint SDSS DR4 and GALEX data set from $0.005 < z < 0.22$. SFRs (and masses) are derived from UV\,--\,$z$-band SED-fitting assuming a Chabrier IMF, and are compared against H$\alpha$ SFRs (and masses) taken from B04 and converted from a Kroupa to a Chabrier IMF by dividing by 1.06. The SFR conversion is based on BC03 SPS models which are used to provide an empirical SFR calibration. Extinctions and other parameters are listed in Table~\ref{tab:msfr_info}.

\textbf{\citet{santini+09} (S09)}. S09 observes UV and IR emission from $7909$ galaxies in the GOODS MUltiwavelength Southern Infrared Catalog (GOODS-MUSIC; \citealt{grazian+06}; S09) from $0.3 < z < 2.5$, although only $7877$ are included in their fits (see their Section~2.3). SFRs are derived via combined UV+IR emission assuming a Salpeter IMF and scaled based on the calibration in \citet{bell+05} where IR data is available; otherwise, it is determined via extinction-corrected UV emission derived via SED fitting (wich is used to derive the masses and photo-z's) as detailed in Table~\ref{tab:msfr_info}.

S09 employ a $2$$\sigma$ clipped fitting procedure, and many of the SED-derived SFRs tend to be the galaxies that lie outside this range (i.e., quiescent). We attempt to bypass this problem by assuming that SED-derived SFRs from UV data only have similar dust extinction values to those with observed UV+IR emission (i.e. we assume the properties of UV-only and UV+IR objects are similar) and derive proper UV+IR weights using R12.

\textbf{\citet{shim+11} (S11)}. S11 observes 31 and 41 H$\alpha$ emitters (HAEs) from $3.8 < z < 5.0$ in the GOODS-N and GOODS-S fields, respectively, taken from various spectroscopic observing programs in the field \citep{ando+04,vanzella+05,vanzella+06,vanzella+08}. The sample is fitted using CB07 SPS models as detailed in Table~\ref{tab:msfr_info}. In order to avoid line contamination from $H\alpha$, the IRAC ch1 measurements were excluded from this procedure. Objects with bad fits ($\chi^2 > 5$) were excluded from the analysis, leaving a total of 64 remaining objects. As all these objects have secure spec-z's, the fitting procedure is only used initially to determine the masses and not the photo-z's. 

Using the templates, S11 derive photometric H$\alpha$ fluxes and equivalent widths based on the IRAC ch1 excess (the amount the observed flux exceeds that of the best-fit SED model) and correcting for assumed [NII] contamination. Results are cross-checked with 15 galaxies from \citet{erb+06} and found to be in relatively good agreement ($\sim 0.25$\,dex scatter; see their Fig.~5). Using the K98 $L$\,--\,$\psi$ conversions for UV and H$\alpha$, S11 finds that the average  $\psi_{H\alpha}/\psi_{\textrm{UV}}$ ratio (both uncorrected for extinction) is $\sim 6$, more than a factor of 2 above the median extinction correction in the UV ($\sim 2.5$) that would be implied by B12. Based on these results, we do not choose to apply any additional extinctions here.

\textbf{\citet{salmi+12} (S12)}. S12 observes UV and IR emission from 543 $K$-band selected galaxies in the GOODS-S field from $0.5 < z < 1.3$, taken from the $K$-band selected catalogue of D04. The majority of these objects (70\%) have spec-z's; the rest have photo-z's taken from \citet{grazian+06}. Masses are determined using the method describe in \citet{leborgnerocca-volmerange02} and as used in E07 and E11. $L_{\textrm{IR}}$ is derived using CE01 templates, while $L_{1500}$ is extrapolated from the observed photometry using the best-fitting SED model without extinction, assuming a Chabrier IMF. The extinction is derived by comparing the latter to the total SFR, and used to correct for rest frame colors based on C00. Although there are multiple MS relationships derived in the paper, we only include the one that is most comparable to those from the literature, notably the one with masses derived via SED fitting (rather than their ``empirical'' mass) that does not take into account color or morphology.

\textbf{\citet{sobral+14} (So14)}. So14 observes H$\alpha$ emission from 1742, 637, 515, and 807 galaxies at $z \sim 0.40, 0.84, 1.47$ and $2.23$, respectively, in the COSMOS and UDS fields as part of the High Redshift Emission Line Survey (HiZELS; \citealt{sobral+13}). The data used to select the samples are taken from deep and wide narrow-band surveys using UKIRT, Subaru, and the Very Large Telescope (VLT) designed to select HAEs (and thus cut on SFR; no other strict cuts are applied except to attempt to remove contaminating line emitters from the sample). Redshifts are determined photometrically and are taken from the catalogs of \citet{ilbert+09} and \citet{cirasuolo+10}. Masses are derived as outlined in Table~\ref{tab:msfr_info} following \citet{sobral+11}, which use CB07 models. As mentioned above, unlike for M10 and other LBG samples, the correction factor here is closer to $\sim 1.6$ ($C_{M_*,S} = +0.20$\,dex; D. Sobral, priv. comm.) due to the increased prevalence of the TP-AGB phase on the integrated light.

masses have been derived using both the best-fitted SED as well as the median mass across all solutions in the entire multi-dimensional parameter space for each source that lie within 1$\sigma$ of the best-fit. In order to maintain consistency with other studies included here, the best-fitting masses are used in this analysis; however, we note that the median masses seem more robust, as we discuss in Appendix~\ref{app:mass_disc}. SFRs are derived using the relation from K98 corrected to a Chabrier IMF, which gives values $\sim 0.56$ and $0.82$ those of K98 and KE12, respectively. Dust corrections are applied to the data using the empirical relations of \citet{garnbest10}.

As $M_*$\,--\,$\psi$ relations are not presented in So14 directly, we obtained the data from D. Sobral (priv. comm.) and fit $M_*$\,--\,$\psi$ relations directly using a procedure analogous to that used in St14. We exclude any galaxies with masses below $10^{8.5} M_\odot$ ($z \sim 0.4$) and $10^{9.5} M_\odot$ (higher $z$ samples) in order to avoid incompleteness issues (D. Sobral, private comm.), leaving a final sample of 305, 392, 376, and 605 galaxies in each redshift bin. Each of the fixed-redshift relations is fit using running medians in bins of 0.1\,dex rather than the individual points in order to avoid biases from outlying galaxies. We impose a minimum of 10 objects per bin order for a median to be included in the fit to avoid biases from the edges where there are a small number of galaxies, and derive errors using resampling. The results are presented in Table~\ref{tab:msfr}. 

As So14 does not introduce a color-color cut or some other selection criteria to separate SFGs from quiescent galaxies, the fits have shallower slopes and lower SFRs at higher masses (due to the existence of larger quenched populations as you go upwards in mass), as well as with much higher scatter at lower masses (due to both better detection limits and a larger quenched population). This is also discussed in C14's data description.

\textbf{Steinhardt et al. (2014, subm.) (St14)}. St14 observe 3398 galaxies from the flux-limited Spitzer Large Area Survey with Hyper-Suprime Cam (SPLASH; Capak et al. 2014, in prep.) survey using Epoch 2 data at redshifts $4 < z < 6$. The data is multiwavelength, with observations from the Ultra Deep Survey with the Visible and Infrared Survey Telescope for Astronomy (VISTA) telescope (UltraVISTA; \citealt{mccracken+12}), \textit{Spitzer}, and Hyper-Suprime-Cam (HSC). Epoch 2 observations, however, do not contain HSC data. Photo-z's, masses, luminosities, and extinctions are derived via SED fitting as described in Table~\ref{tab:msfr_info}. SFRs are calibrated on K98 but assuming a Chabrier IMF, with $\psi_{tot} [M_\odot \textrm{yr}^{-1}] = (L_{\textrm{TIR}} + 2.3 \, L_{\textrm{NUV}} ) \, \sn{8.6}{-11} [L_\odot]$, where the total IR luminosity is defined as KE12 and NUV is taken at 2300\,{\AA} (S. Arnouts, priv. comm.). Weighting according to B12, we find this conversion gives SFRs approximately a factor of $\sim 0.65$ those of KE12 (-0.19\,dex) and correct for it accordingly. Note that the fit and results included here were taken from a preliminary analysis; see St14 for the final MS fits.

\textbf{\citet{whitaker+12} (W12)}. W12 observe UV and IR emission from 22816 $U-V$ vs. $V-J$ ($UVJ$) selected galaxies from $0 < z < 2.5$ in the NOAO Extremely Wide-Field Infrared Imager (NEWFIRM) Medium-Band Survey (NMBS; \citealt{whitaker+11}), which encompasses two fields within COmassOS and AEGIS. $L_{\textrm{IR}}$ is derived based on a single template that is the log average of DH02 templates with $1 < \alpha < 2.5$ following \citet{wuyts+08}, \citet{franx+08}, and \citet{muzzin+10}, while $L_{\textrm{UV}}$ is derived based on best-fitting B11 models. SFRs are derived via the calibration presented in \citet{franx+08}, itself based on the calibration presented in \citet{bell+05}. Photo-z's are derived using EAZY with PEGASE.2 and M05 templates assuming a Kroupa IMF. Masses are derived using the Fitting and Assessment of Synthetic Templates code (FAST; \citealt{kriek+09}) assuming a Chabrier IMF.

\textbf{\citet{zahid+12} (Z12)}. Z12 use data from three separate $z$-selected star-forming samples:  SDSS DR7, from $0.04 < z < 0.1$; DEEP2, from $0.75 < z < 0.82$; and the sample from \citet{erb+06}, from $1.41 < z < 2.57$ -- which will be referred to as Z12(1), Z12(2), and Z12(3), respectively. Z12(1) includes H$\alpha$ emission from $\sim \sn{2}{5}$ galaxies, and derives SFRs based on \citet{brinchmann+04} [B04] (with additional improvements given by S07, including aperture corrections), and scale results from a Kroupa to a Chabrier IMF dividing by 1.06. The strong emission lines of each galaxy are fit using the nebular emission models of CL01. Z12(2) includes H$\beta$ emission from 1348 galaxies, with SFRs derived based on K98 (assuming $L_{H\alpha} = 2.86 \, L_{H\beta}$), and scale results from a Salpeter to a Chabrier IMF dividing by 1.7. Z12(3) includes H$\beta$ emission from 87 galaxies, and derives SFRs as Z12(2). For Z12(1), extinctions are determined from the Balmer decrement (assuming a C89 extinction curve). These are then used to parametrize extinction as a function of mass and metallicity, calibrated on \citet{kobulnickykewley04}, using a similar formulation to \citet{xiao+12}. This parametrization is then applied to Z12(2) and Z12(3) galaxies. Photo-z's and masses are derived consistently for all samples, as detailed in Table~\ref{tab:msfr_info}.

\textbf{{\citet{brinchmann+04} (B04)}}. While not utilized directly in our analysis, several papers here rely on results taken from and/or based on B04's findings as well as the online joint Max Planck Institute for Astrophysics (MPA)/Johns Hopkins University (JHU) SDSS catalog. For clarity, we briefly summarize features of the paper and online catalogs here. B04's original sample is based on SDSS DR1, with spectroscopically derived masses taken from \citet{kauffmann+03}. Fiber-based SFRs are derived using a combination of H$\alpha$ emission and the emission line-calibrated 4000{\AA} break, with total SFRs calculated using an emission line-calibrated photometric aperture correction. The older version of the MPA/JHU catalog is based on DR4, but otherwise calculates masses and SFRs identically to B04. The current online MPA/JHU catalog, by contrast, is based on DR7, with masses derived photometrically via SED fitting without GALEX data. Aperture-corrected total SFRs are likewise derived using SED fitting.

\section{Testing the Effect of Catastrophic Errors on Main Sequence Relations}\label{app:eta_test}

Because most photo-z codes first determine a photometric redshift and use that as the basis for further analysis of each galaxy, it is essential to understand the uncertainties in photometric redshifts (photo-z's) and how they might impact the MS. Most directly, uncertainty in the determined redshift will introduce additional uncertainty into any redshift-dependent relationship as they scatter objects into and out of each redshift bin. A more substantial problem is that redshift determination is highly degenerate with the other properties inferred for each galaxy. This might affect classification, so that a particular class of galaxies might be mistakenly flagged as star-forming or excluded from the sample, as well as more directly the mass and SFR.

We investigate the effect of errors in photo-z's on the inferred $M_*-SFR$ relation at lower redshift using the GOODS-MUSIC catalog \citep{fontana+06}, which had spec-z's for 1858 out of 18409 objects in the GOODS field (1484 high quality spectra with flags 3 and 4) and the Le PHARE photo-z code. This spectroscopic subset (all flags) was fit with LE\_PHARE using 27 BC03 models (no empirical models to account for non-SF galaxies), with $\sim$\,100 ages from 1\,Myr to 13\,Gyr, and a very rough grid of $\Delta z = 0.05$ ($z_{max} = 6$) with no extinction values. In order to bound the effects of poor photo-z determination, we chose the worst possible performance conditions and simulated blind (and sloppy) use of both the code and the subsequent input: we did not include possible contributions from emission lines, we did not refine our result using the spec-z's as a prior, use systematic shifts to correct for systematic offsets in each band, imposed no cut on our spec-z sample based on quality flags, applied no quality cuts to any of the photometric input, and substituted several filters in the input file with similar ones from other surveys. As expected, the resulting photometric fits had a high rate of catastrophic failures ($\sim 50\%$).

\begin{figure*}[!ht]
\plotone{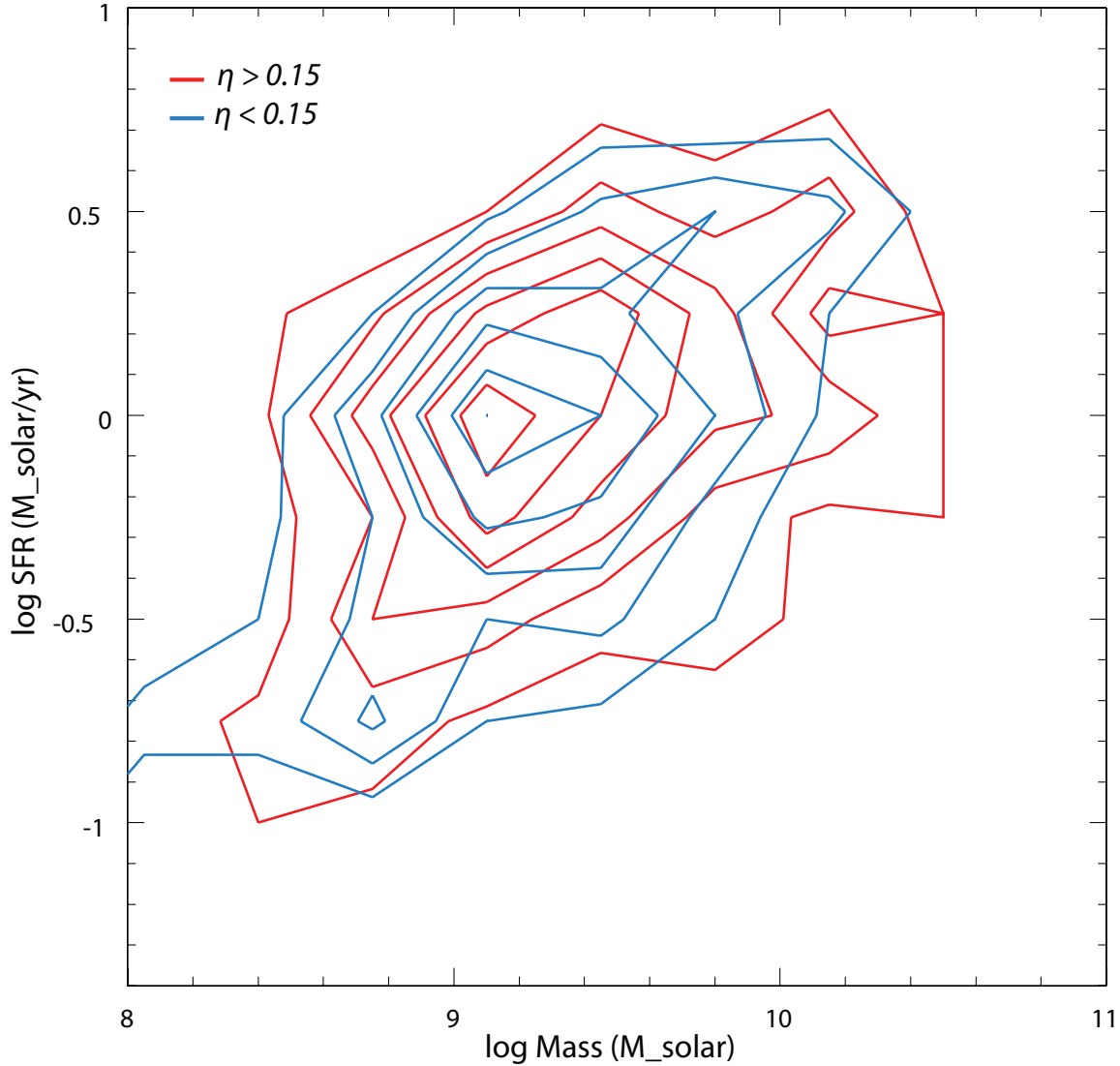}
\vspace{-10pt}
\caption{A comparison of the derived $M_*$\,--\,$\psi$ relation using (a) well-fit photometric redshifts ($\eta < 0.15$; light blue) and (b) poorly-fit photometric redshifts ($\eta > 0.15$; red) from the GOODS-MUSIC catalog. The best-fit star-forming main sequence has a similar slope of $\alpha \sim 0.6$ in both cases, although the ``bad'' photo-z sample displays slightly more scatter.}
\label{fig:eta_test}
\end{figure*}

The sample was then divided into subsamples of {\qq}good{\qq} ($\eta < .15$) and {\qq}bad{\qq} ($\eta > .15$) photo-z's in order to example the effect of poor photo-z determination on the MS.  We find similar $M_*$\,--\,$\psi$ correlations in both samples (Figure~\ref{fig:eta_test}), with $\alpha \sim 0.6$ for both, along with similar mass distributions (although the {\qq}bad{\qq} photo-z sample displayed noticeably more scatter). Based on these results, even large errors in LE PHARE's derived photo-z's do not induce any noticeable $M_*$\,--\,--$\psi$ correlations in the data (compared to the well-matched sample), although it does seem to increase the scatter (at least at lower redshifts). While this might not be true at higher redshifts or for data with very specific types of catastrophic failures (e.g., St14), it at least seems to indicate that the relatively small percentage of catastrophic errors seen in previously published samples should have little impact on derived MS relations.

\section{On Calibrations and Disagreements Among Radio Star Formation Rates}\label{app:bell} 

At higher redshifts, all of the studies included here have assumed that $q_{\textrm{IR}}$ is unchanging with time. However, this assumption might be incorrect, as radio samples the SFR in two different ways. The first is through non-thermal (i.e. synchrotron) radiation, which dominates at longer wavelengths (i.e. 1.4\,GHz) and underpins the tight IR\,--\,radio correlation. While the mechanism connecting the two is not well understood, it is thought to be due to cosmic ray electrons from supernovae being trapped in the magnetic field of the galaxy. This ratio is potentially is sensitive to the CMB temperature as the CMB can cool (suppress) these cosmic ray electrons through inverse Compton scattering if there isn't a high gas density that can shield the magnetic field \citep{murphy09}.

The second is through Bremsstrahlung (free-free) radiation from HII regions, which dominates at shorter wavelengths ($\lambda < \sim 1$\,cm in local galaxies). This should be independent of the CMB temperature. As a result, the conversion from radio flux at a fixed wavelength may vary from a single power law as radio SFR studies have assumed (and hence evolution in $q_{\textrm{IR}}$). This variation would likely be redshift-dependent and also sensitive to the gas content/geometry of the galaxy. Such an evolution in $q_{\textrm{IR}}$, which we would expect to go as $(1+z)$ from a simple CMB temperature-suppression scenario, seems to match the observed systematic difference. A similar line of reasoning is taken in \citet{carilli+08}, who find that for the range of fields considered typical for spiral arms (a few $\mu$G), and for starburst galaxy nuclei ($\sim 100\,\mu$G), inverse Compton losses off the CMB dominate synchrotron losses in a typical ISM at $z > 0.5$ and in starburst nuclei at $z > 4$.

There are multiple issues that might cast doubt on this line of reasoning. First and foremost, such logic leads to a \textit{suppression} in radio luminosity, rather than the systematic \textit{increase} that is observed. It also does not strictly scale as $(1+z)$ \citep{murphy09}. In addition, this finding conflicts with those of \citet{sargent+10}, who investigated the IR\,--\,radio properties of galaxies in the COSMOS field and did not see any evidence for an evolution of $q_{\textrm{IR}}$. Although they are naturally limited by the data available (IR observations with \textit{Spitzer} only), their analysis argues convincingly against such trends in the data. Additionally, the evolution in $q_{\textrm{IR}}$ should only really be a major influencing factor at high redshift where direct detections are scarce; it shouldn't affect low-redshift observations. Our blanket $(1+z)$ correction might then be incorrect at lower redshifts.

Furthermore, as mentioned above, for CMB cooling to work efficiently at a wide range of higher redshifts there must some energy density in the magnetic fields across the star forming regions that is similar to what is observed in spiral arms locally. Any evolution (enhancement) of typical magnetic field strengths will tend to directly counteract cooling losses. As galaxies at high redshift are more gas rich and highly star-forming (although quite different from local ULIRGS), such an enhancement in the typical magnetic field strengths seems quite plausible. These higher gas densities might also serve to shield the magnetic field from CMB cooling losses, again leading to a deviation from this $(1+z)$ behavior. Deeper JVLA observations might be able to untangle the two effects, as short wavelengths will have a larger free-free emission component that is not affected by possible redshift evolution. Given the uncertainties, however, this line of reasoning remains an open possibility.

Most basically, such a disagreement might arise be due to the new calibration from KE12 (taken from \citealt{murphy+11}) we adopt, which might inadvertantly boost the radio SFRs relative to the other calibrations. As the majority of the calibrations from KE12 seem consistent with each other, such an error would have to occur in the $q_{\textrm{IR}}$ conversion from $L_{\textrm{IR}}$ to $L_{1.4}$. At first glance, such a miscalibration seems plausible: the KE12 radio calibration gives SFRs approximately \textit{twice} that of K11's \citet{bell03} calibration (converted to a Chabrier IMF) even though both use similar IMFs and \citet{murphy+11} uses the \citet{bell03}'s reported $q_{\textrm{IR}}$. 

However, \citet{bell03} makes several differing key assumptions that increase his SFRs by $\sim 50$\,--\,$60\%$. First, by using an energy budget argument to account for the IR emission generated by older stars, \citet{bell03} reduces his calibration by $\sim 10\%$ of its original value. In addition, by using a slightly different $q_{\textrm{IR}}$ factor (2.52, after he limits his sample to only more luminous IR galaxies) for his conversion than reported for the full sample (2.64), his calibration is further reduced by $\sim 30\%$.

Taking these factors into account (plus the $\sim 7\%$ Chabrier-to-Kroupa IMF conversion assumed here), we end up with a $\sim 50$\,--\,$60\%$ increase in the reported K11 coefficient. This brings it well within the observed scatter (0.26\,dex) in the IR\,--\,radio correlation itself, and is acceptable given the differences in modeling assumptions between the two studies. It also is relatively close to the empirically calculated $L_{\textrm{IR}}$\,--\,$\psi$ conversion reported in \citet{murphy+12}, which is $\sim 23\%$ lower than that of the \citet{murphy+11} calibration used here. Furthermore, using the direct supernovae-to-1.4\,GHz conversion reported in \citet{murphy+11}, we get values favoring those of the KE12 relations, an independent piece of evidence that seems to suggest that our chosen $L$\,--\,$\psi$ calibration is more accurate than the one reported in K11.

In addition to this good agreement with other calibrations, this line of reasoning also does not seem likely because our $C_R$ correction (Table~\ref{tab:corr}) only adjusted the normalizations of each of the radio samples (D09, P09, and K11) by the same factor, not their redshift evolution. Indeed, we find that the unaltered radio SFRs and our pre-$C_R$ corrected radio SFRs display similar evolutions in both redshift and time (see the last few fits in Tables~\ref{tab:fits} and~\ref{tab:fits_z}). So the main effect of our new calibration was mainly to bring these pre-existing differences in the data to light. Note that the differences between the KE12 and \citet{yun+01} calibration mainly arise from the IMF conversion (Salpeter to Kroupa), the altered definition of $q_{\textrm{IR}}$ being defined by the TIR instead of FIR, and the $\sim 15\%$ difference in IR calibration between that of K98 and KE12.

Another possibility could involve evolution in either $q_{\textrm{IR}}$ or the radio spectral index $\alpha$. To derive radio SFRs at higher redshifts, all of the studies included here have assumed that $q_{\textrm{IR}}$ and $\alpha$ are redshift-independent and that a constant radio spectral index can be universally applied to all galaxies. While \citet{sargent+10} indicates these assumptions seem to hold to $z \sim 2$, they might not necessarily be true at all redshifts.

An additional explanation might be that the radio stacking procedure itself leads to a systematic overestimate in the observed 1.4\,GHz luminosity. In their stacking analysis, \citet{condon+12} find that their estimates of source counts from lower angular resolution data are much lower than those of \citet{owenmorrison08}, which are derived from much higher angular resolution data. They conclude that the disagreement is likely due to count corrections made for partial resolution of extended sources in the high-resolution 1.4\,GHz beam. Survey catalogs are complete to a fixed brightness cutoff (flux per beam), so extended sources with lower brightnesses but higher integrated flux densities will be missed/under-represented in the beam.

The corrections needed to convert source brightnesses to source flux densities (as well to account for missing sources) become quite large near the brightness cutoff as the angular resolution approaches the median angular size of faint sources. As flux goes as $(1+z)^{-1}$, if these corrections happened to be too large, they might end up contributing to much larger apparent source luminosities at higher redshifts with a similar evolution to what we observe here. 

It is important to note, however, that the conclusions of \citet{condon+12} are largely based on the comparison of the source counts derived in \citet{owenmorrison08}; no other study (including K11) has found source counts at the low end that are as high (A. Karim, priv. comm.). In addition, the radio disagreements appear to be systematic, rather than isolated to any particular study. This would imply that the wide variety of different stacking procedures used in D09, P09, and K11 (see also \citealt{roseboombest14}) -- which are in excellent agreement with each other -- are all overestimating source counts, which seems unlikely. As a result, ultimately we remain unsure which (if any) of the above explanations are the likely cause of the observed $(1+z)^{\sim 0.8}$ systematic offset between radio-based SFRs compared to SFRs derived from other indicators.

\section{Interpreting Cross-Correlational Scatters Among SFR Indicators}\label{app:sfr_disc}

There are several possible interpretations of the relatively constant $\sigma_{cc}$ values indicated by cross-correlational SFR-indicator studies. The first would be that one of the SFR indicators (e.g., H$\alpha$) is intrinsically more reliable than the others (taking the differences in the relative timescales probed into account), and thus should be taken as having very little internal scatter compared to the rest. However, the fact that all indicators seem to share similar $\sigma_{cc}$ values seems to imply that the intrinsic scatter for most is around equivalent ($\sigma_{int} \sim .2$\,dex); otherwise, they should display smaller scatter when being compared with H$\alpha$ versus how much they display relative to each other. This would then imply that the data measured with said SFR indicator should have intrinsically less scatter than those measured via other means, and should be more representative of the population. 

If we ignore the discrepant smaller $sBzK$ scatters (see \citet{wuyts+11} and K13 for discussions on extinction-corrected UV SFRs often used by $sBzK$ studies; see also \S\,\ref{subsubsec:bzk}), we find that on average the H$\alpha$ indicators  have less scatter, with $\sigma_d\sim 0.25$ compared to $\sim$\, 0.3-0.35 for other sources. However, the fact that these observations, which not only span different times but also differently sized time bins, show similar scatter is concerning, since as noted above there shouldn't be effects like this unless they apply to all bins equally. A casual inspection shows that this isn't true, and that scatters range both relative to $t$ and $\Delta t$, in seemingly an uncorrelated way, which tends to rule out this idea.

For instance, we can assume that, due to better observational constraints, on average H$\alpha$-based studies have less scatter than UV or IR ones (as we observe in So14's data), with $\sigma_d\sim.25$. We can then assume that $H\alpha$ has an intrinsic scatter (e.g., from internal galaxy properties and observational errors) that is $\lesssim .2$\,dex. We pick fiducial values ranging from $\sigma_{\textrm{int,H}\alpha} \sim 0$\,--\,$.15$\,dex ($\sigma_t \sim 0.2$\,--\,$0.25$\,dex). This implies that the intrinsic scatter among the other observables ranges should range from  $\sigma_{\textrm{int}} \sim .25$\,--\,$.3$\,dex, in order to satisfy to $\sigma_{cc} \sim 0.3$\,dex constraint we observe. Adding these in quadrature, we get that the \textit{minimum} $\sigma_d$ must range between 0.32 and 0.39\,dex. The former (the generous case) is already in tension with several data points, and the latter contradicts most others. Since this is in tension with both the observations listed here and cross-correlation conclusions, we reject this hypothesis. We also construct this same argument for all the other SFR indicators (when possible), and reach similar conclusions. Thus, there does not seem to be one superior SFR indicator which displays smaller internal scatter among the data collected here.

Alternately, we can assume that the observed scatter in each study (i.e. MS observation) is almost entirely the result of the scatter among the SFR indicator (i.e. that the SFR indicator scatter is completely unrelated to internal galaxy properties). This then implies almost perfectly synchronous evolution among objects in a given mass bin, which seems unreal given that range of ages that these studies are covering. Simple differences in formation times due to, e.g., \citet{pressschechter74}, should generate an intrinsic dispersion among SFRs unless evolution was rapidly convergent before/during the era of highest SFR activity ($1 < z < 3$; \citealt{hopkinsbeacom06}) to within $\sim .05$ dex. This would imply that galaxies are synchronous enough to be considered (with proper calibration) standard candles! It would also imply that the processes that trigger and drive star formation are completely deterministic, and dominate at all times over stochastic processes like mergers (at least while galaxies are on the MS). 

We reject this line of reasoning due to the concerns mentioned earlier in this section, namely that if $\sigma_{cc} \sim 0.3$\,dex  among all indicators, there must be intrinsic scatter among each indicator, and so this situation cannot be viable. We could also assume the other extreme, that the scatter among SFR indicators is in fact completely due to internal galaxy properties (e.g., metallicity) that modify the observed emission, rather than the indicators themselves. Cross-correlational studies also seem to rule the extreme version of this hypothesis out, although the results from S12 (See their Table~1) seem to imply that this must be true at least to some extent.

\section{Intrinsic Scatter Among SED-Fitted Stellar Masses}\label{app:mass_disc}
While empirical indicators that can be used to cross-check SED-fitted masses have not been established, it is nonetheless possible to evaluate the intrinsic scatter by comparing different mass determinations that have been derived using the same data and fitting procedures (so that the only difference is interpreting the final output). We opt to do this using the two masses taken from So14 (D. Sobral, priv. comm.), who derive their masses (see Appendix~\ref{app:data}) using two different procedures. The first uses the most common method of taking the mass from the best-fitting SED, used in all studies included in this work ($M_{*,B}$). The second involves taking the median of all the masses that have been derived from fits that lie within 1$\sigma$ parameter space of the best-fitting SED ($M_{*,M}$). 

As expected, while the ``best-fit'' mass and the ``median'' mass tend to be tightly correlated, the best-fit mass is very sensitive to small changes in the parameter space and/or error estimations, while the median mass tends to be robust against such variations (So14). We find that the correlation between the two mass estimates for all the galaxies included in the HiZELS dataset (3004) is well parameterized by a linear fit in log-log space, where
\begin{equation}
\log M_{*,M} = (1.05 \pm 0.01) \, \log M_{*,B} - (0.76 \pm 0.28),
\end{equation}
and $\sigma = .32$\,dex. This fit covers $\log M_{*,M} = 6.8$\,--\,$11.4$ and $\log M_{*,B} = 6.1$\,--\,$11.1$, and has been fit using the both same procedures outlined in St14 and Appendix~\ref{app:data} as well as just a standard fit to all the data points (the differences are negligible). This relationship is not 1:1, likely due to the sensitivity of the best-fit mass to the grid space and fitting procedures leading to nonlinear dependencies. In principle, the relation between the two could vary even more if different (and more) models are used in the SED fitting process. We find that the average median mass is $\sim 0.2$\,dex smaller than the best-fit mass, which suggests that while larger masses tend to be favored by the best fit, somewhat smaller masses are in fact more common among most of the fits that are only marginally worse than the best fit.

There is also evidence for possible variations in the $M_{*,B}$\,--\,$M_{*,M}$ relation over time. Splitting up our sample into the four subsamples used in So14, the slopes for the $z \sim 0.40$ $(N=1108)$, $0.84$ $(N=635)$, $1.47$ $(N=511)$, and $2.23$ $(N=750)$ samples are $1.10 \pm 0.02$, $1.06 \pm 0.02$, $0.99 \pm 0.03$, and $0.80 \pm 0.02$, respectively. At fixed mass ($\log M_{*,B} = 10.5$), the differences in mass, $\Delta M \equiv \log\left(M_{*,M}/M_{*,B}\right)$, are $-0.13 \pm 0.03, -0.09 \pm 0.01, -0.27 \pm 0.01,$ and $-0.33 \pm 0.01$, respectively. At lower redshifts, the relationship is generally steeper (i.e., they agree more at higher masses) and the differences between the two mass estimates smaller, while at higher redshifts the relationship is shallower (i.e., agrees more at lower masses) and the disagreements somewhat larger ($\sim 0.3$\,dex vs. $\sim 0.1$\,dex).

Directly comparing the MS relations from these two mass estimates, in all cases the scatter is lower among studies which use the median masses. For the $z \sim 0.40, 0.84, 1.47,$ and $2.23$ samples, the scatter (in dex) around the best fit decreases from 0.49, 0.25, 0.23, and 0.24 to 0.43, 0.21, 0.21, and 0.20. If the median masses are more accurate indicators of the true mass, then these decreases give some indication of the intrinsic scatter (or at least, some sense of an upper bound) present in SED-fitted masses. Assuming that this intrinsic scatter has been subtracted in quadrature from the original scatter in much the same way that this work has done for the SFRs (see \ref{app:sfr_disc}), these correspond to intrinsic scatters (dex) of 0.23, 0.14, 0.09, and 0.13. In the worst case, this indicates that the intrinsic scatter in mass is comparable to that in SFR; in the best case, it's slightly lower, at around $\sim 0.1$\,--\,$0.15$\,dex.

Given our findings, while the median masses do indeed seem to be more robust than the best-fit masses, the exact relationship between the two (and its time dependence) is uncertain. The methodology also looks promising to derive more robust estimates of other SED parameters such as SFRs, stellar ages, etc., and might also serve as an alternate way to estimate the errors on the output parameters. All of these should be further explored by future studies.

\section{Effects of Various Fitting Assumptions}
\label{app:details}

As can be seen both from the median $\sigma_i$'s and the errors on the fit, our method as a whole is robust both in time and in mass. The small $\sigma_i$'s, in most cases $\lesssim \sigma_t$, provides additional support that most MS observations are consistent with each other. No matter how accurate our results seem to be, however, there always is the risk that we might ignore important systematics present within our fitting procedure. In this section, we investigate the effects various changes in our fitting parameters affect our result.

As can be seen in Tables~\ref{tab:fits} and~\ref{tab:fits_z}, a simple average over MS observations while ignoring mass information tends to hide possible differential mass evolution when fitting to all the data, as selection effects mainly cancel each other out. We find that for the majority of our fits, there is $> 3\sigma$ evidence for differential mass evolution (i.e. different rates of time evolution at fixed stellar mass for different masses, or $\alpha(t) = \alpha_t t + \alpha_c$ has a nonzero $\alpha_t$). These are sometimes as high as $\alpha_t \sim -0.04$\,dex per Gyr, but more typically are $\alpha_t\sim -0.02$\,dex per Gyr. These negative $\alpha_t$'s suggest more rapid mass evolution occurring for higher masses and seem to match the observed changes in slope. By contrast, some extrapolated fits provide evidence for \textit{positive} $\alpha_t$'s, which would imply more rapid $\psi(t)$ evolution of \textit{lower}-mass objects (at fixed mass) rather than higher-mass ones, in contrast to the lower slopes we seem to observe.

In general, our calibrated SFRs show smaller interpublication scatters ($\sigma_i$) (typically $\sim 0.1$\,--\,$0.2$\,dex) than their uncalibrated counterparts, and are $\gtrsim 0.1$\,dex smaller than the interpublication scatters between sSFR observations included in \citet{behroozi+13} (around $\sim 0.3$\,dex). They also display less time-dependencies in the MS slopes $\alpha(t)$: while both sets of data display similar slopes at $z \sim 0$, uncorrected data shows steeper slopes (and slightly smaller absolute SFRs) at higher redshifts. Given the robustness of our fitting procedure and the improved results after adjusting results to our common calibration, we choose to exclude the uncorrected and mass-independent fits from further analysis.

We now shift our discussion to our varying minimum threshold for the number of objects we opt to include for each mass bin ($N_{\textrm{bin}}$), and how altering the minimum threshold can affect our results. As can be seen from Table~\ref{tab:fits}, within each cut variations in N tend to only shift the $\alpha_c$ by a maximum of $\sim 0.15$, and more typically $\sim 0.1$\,dex. We do not discuss the other parameters extensively here because they all tend to be degenerate. Changes in $\alpha_c$ are compensated by changes in $\alpha_t$, and usually in $\beta_t$ and $\beta_c$ ($\beta(t) = \beta_t t + \beta_c$) as well since they're all fitting the same data (and tend to fit them about equally well, if the median $\sigma_i$ values are taken at face value). The overall evolution of the MS ultimately remains about the same.

The main reasons why these changes are nonzero have to due with biases towards overweighting individual studies. As we include several studies with multiple data points and large mass ranges (S09; O10), having a low $N_{\textrm{bin}}$ tends to lead to biases towards these data points near the edges of the mass ranges, pushing them downwards/upwards on the low/high mass side (such as with the mixed fits). This effect more generally also is more prominent in cuts with less points overall, where a small number of data points can change some of the specifics of the fit more than they would otherwise (such as the FIR, stacked fits or the UV, non-stacked fits).

For very high $N_{\textrm{bin}}$, this change comes about because of the reduced mass ranges that meet our criteria. Since we have restricted much of the dynamical range by design, fewer studies are consequently moving in and out of our mass bins. This makes our mass-dependent fit more similar to our extrapolated fits as we lose some of our mass-dependence from the slope/normalization averaging process. As can be seen in Table~\ref{tab:fits}, our extrapolated fits tend to find less time evolution ($< \alpha_t$) in the MS slope. Due to the degeneracy of $\alpha_t$ with $\alpha_c$, this leads to an overall decrease in $\alpha_c$ as well. This effect is best approximated when there are lots of data points included in the fit; due to the sensitive nature of fits with a small number of data points ($\lesssim 15$ observations), this effect might be quite different and sensitive to the individual data points (and their parent studies) included in the fit. In order to balance robustness while keeping a large enough dynamical range, we advocate only using fits where $N_{\textrm{bin}} \gtrsim 10$ and the total number of data points is $\gtrsim 15$. All other fits should be used cautiously.

%In all cases, the reason why MS fits change as a function of the minimum number of data points required for the fit, $N_{min}$, is due to how the decreasing mass ranges allowed for the fits exclude some data at high and low redshifts. Due to downsizing, data at higher redshifts are more likely to contain higher mass galaxies, and thus tend to dominate the fit at higher $z$. At lower $z$ (i.e. in the nearby Universe), lower mass galaxies tend to be more common due to detection limits that prevent us from observing them at higher redshifts. MS evolution is in fact differential in mass, i.e. at fixed mass, galaxies with larger masses experience a stronger decline in SFR over time compared to galaxies with smaller mass (see Figures~\ref{fig:sfr_t_relations},~\ref{fig:m10i}, and~\ref{fig:m11i}). As the observed amount is not one long continuous linear function (groups of data entering and leaving in mass bins introduce discontinuities; see Figures~\ref{fig:a_fit} and~\ref{fig:b_fit}), excluding the edges tends to decrease the observed amount of differential mass evolution, and hence the extent to which the MS's mass-dependence (MS slope) evolves over time. Although this slightly alters the change in slopes, the fits compensate by increasing the change in normalization, leading to almost identical MS evolution among these samples over time.

So far, we have limited this discussion to our MS$(t)$ fits from Table~\ref{tab:fits}. However, they also hold for our MS$(z)$ fits from Table~\ref{tab:fits_z}. In order to decide which functional form of the MS provides a superior fit to the data, we turn our attention towards functional robustness, interpublication scatters, and a direct comparison. In terms of robustness, we find that the MS($z$) functional form is about as robust as the MS($t$) one in terms of the quality of the fits themselves, with both methods having similar orders of errors and variances. In most cases, however, the errors on the MS($z$) fit (especially for the power law index) are fractionally larger than the MS($t$) one. In addition, we note that the MS($z$) fit is not stable with respect to $N_{\textrm{bin}}$, with the power-law index changing drastically with variations in $N_{\textrm{bin}}$ for the same cuts.

For interpublication scatters, we find that on average $\sigma_{i,z}$ is $\sim 0.02$\,--\,$0.05$\,dex greater than $\sigma_{i,t}$. As both functions are fitting the same set of data, higher $\sigma_{i,z}$ values seem to indicate that the chosen MS($z$) parametrization is less effective than the MS($t$) one at fitting the data. We also find that it serves as a worse predictor for high-$z$ MS observations compared to the MS$(t)$ forms plotted in Figures~\ref{fig:sfr_t_relations_selection},~\ref{fig:sfr_t_relations},  and~\ref{fig:m105ri}, leading to differences in SFR at fixed mass ($\Delta\psi = \psi_z - \psi_t$) of several tenths of a dex at $z \gtrsim 3$, and of $\sim 0.1$\,dex at $z \sim 0$. While it does provide reasonable fits to within the range to which it has been applied, the failures to extrapolate MS($z$) to higher redshifts where it severely overpredicts available data leads us to favor the MS($t$) case. We thus deem the MS($z$) fits inferior to the MS($t$) ones (i.e., time-dependent parametrizations of the MS are superior to redshift-dependent ones), and will focus our discussion in this discussion on the latter.

As can be seen from the fits in \S\,\ref{subsec:ms_evol} and in Tables~\ref{tab:fits} and~\ref{tab:fits_z}, excluding the radio data ($\sim 15$ observations; 3 studies) from the analysis shifts the fit towards larger amounts of time evolution in the slope. However, after removing all stacked data (which also removes the stacked IR observations of O10), we find that the fit does not change substantially, indicating that the radio data are not biased relative to their other stacked counterparts. Relative to the mixed data, the fit which includes all available data points also tends to favor increased time evolution. This is because of the dichotomy in slopes between mixed and bluer data illustrated in Figure~\ref{fig:slope_selection}. As most UV observations are centered at high redshift (as FIR data are unavailable), this biases the average slopes (and SFRs) at these redshifts and increases the slope evolution towards low redshift where these observations are less prevalent. 

This behavior becomes especially apparent looking at the fit for only the bluer data, whose best fit indicates a general slope \textit{increase} over time. The subsets for observations for specific SFR indicators display much of the same trends. For UV (SFR) data, the problem is worse due to the decreased sample size, and gives a best fit that implies a slope of significantly greater than unity at late times. For combined UV+IR and IR (SFR) data, the strong evolution in the slope is due to a combination of R12's LBGs at higher redshift biasing the slopes upwards and the strong redshift evolution present in observations such as W12. These are the same problems that characterize the fit that included all observations. Finally, looking at the $(1+z)^{\sim 0.8}$-corrected radio observations relative to their uncorrected counterparts, we find that the main differences between the two data sets is their time evolution, as expected.

\section{Impacts of Assumptions on Main Sequence Evolutionary Tracks}\label{app:leitner}

To estimate possible errors in our MSI tracks due to our assumption of an initial seed mass of $10^7$, we examine MSI tracks generated by smaller seed masses of $10^5$ and $0$\,$M_\odot$. In order to arrive at the same final masses as our original seed mass, these new seed masses require earlier formation times, with average offsets of $\sim 0.25$ and $\sim 0.5$\,Gyr, respectively. These indicate that it takes $\sim 250$\,--\,$500$\,Myr to grow $10^7\,M_\odot$ assuming continuous MS-like star formation, or an average SFR of $0.02$\,--\,$0.04\,M_\odot$\,yr$^{-1}$. As galaxies as high as $z \sim 6$ ($t \sim 900$\,Myr) are observed to have masses of $\sim 10^{9-11}\,M_\odot$ (L12; \citealt{stark+13}; St14), this level of growth seems insufficient to generate some of the massive SFGs in the extremely early Universe (and also seems to indicate significant amounts of mass assembly $<10^7 M_\odot$ must place in bursts).

Indeed, even assuming $t_{form} = 0$ and $\log M_{*,0} = 7$, we are just barely able to generate galaxies on the order of $10^9\,(10^{10})\,M_\odot$ by $z \sim 7\,(6)$. In order to account for possible (albeit unlikely) MS-like mass growth, we investigate adding $\sim 500$\,Myr age offsets to our stellar mass loss prescriptions. These lead to SFRs on the order of $10$\,--\,$500\,M_\odot$\,yr$^{-1}$, which seem reasonable given the gas densities and merger rates in the very early Universe. We find this offset leads to variations in the final mass on the order of a few percent, the same level of uncertainty inherent in the approximation itself \citep{leitner12}. The results of reported here do not change if we include this additional age/growth component.

Besides initial seed mass, other results from MSI (e.g., SFHs) can also sensitive to the assumed evolution of the MS. In order to investigate the impact different parametrized MS evolutions might play in our MSI procedure, we calculate tracks for several of our best fits presented in Table~\ref{tab:best_fits}. We find that for most measurements, the variations in formation times (assuming a fixed initial seed mass) vary between 5\,--\,10\% while SFRs at any given time/mass vary by $\sim 0.1$\,dex. This is due to the fact that the actual SFR of a MS galaxy at any given mass and time is very similar across most of the studies compiled here -- most apparent differences in evolution (e.g., the strong evolution in slope from W12 compared to the relatively steep slopes found other studies at similar redshifts) are almost entirely offset by the derived evolution of other parameters (see Figure~\ref{fig:sfr_t_relations}). For radio-based measurements, however, the differences can be more significant (formation times varying by $\sim 15\%$, SFHs by several tenths of a dex), although the exact magnitude depends on the final mass of the galaxy. Based on these considerations, we conclude our results are relatively robust to the exact form of the MS. This also implies that while specific quantities reported in other MSI-based analyses (e.g., \citealt{munozpeeples14}) might have larger systematic errors than reported, their main conclusions should be unaffected.

\section{Extrapolations of Main Sequence Fits to Low Redshift}
\label{app:low_z}

Since we have excluded the first and last 2\,Gyr of data from our fits, at lower redshifts we have the ability to examine what differing fitting techniques, sample selection, and other systematic effects have on the determination of MS fits. We find  different parameters from E07, Z12, C09, and S07 (in order of increasing median redshifts), all of which have used SDSS data from DR4 \citep{adelman-mccarthy+06} with the exception of Z12 and C14, who use DR7 \citep{abazajian+09} and their own individual methods to determine MS relations for $> 10^5$ galaxies (see Appendix~\ref{app:data}). As a separate check, we also compare these results to those of O10 (SWIRE). All listed parameters will be for E07, Z12, C14, C09, S07, and O10, respectively, unless indicated otherwise.

We first examine the slopes of the individual MS determinations. For these six studies, we find $\alpha = 0.77, 0.71 \pm 0.01, 0.477 \pm 0.004, 0.35 \pm 0.09, 0.65$, and $0.77 \pm 0.02$. Excluding C09 and C14 as before, we see that MS slopes for (essentially) the same sample range from 0.77 to 0.65. This wide range in the $M_*$\,--\,$\psi$ behavior is much larger than the single quoted error (the only ones provided are $0.01$), and so we can easily conclude that (excluding selection differences) systematic errors on the order of $\sim 0.1$\,dex or larger dominate the MS slope error budget at these low redshifts. Even when comparing results which used the same SFR indicators (E07 and Z12 use H$\alpha$, while S07 and C14 use SED-fitted values) or SPS models (S07, Z12, and C14 use BC03), the variation in slope is still $\sim 0.06$\,--\,$0.2$. As an independent check on the validity of any of the slopes listed here, O10's data gives a slope on the higher end of the other low-$z$ estimates (0.77). We note that our extrapolated slope at $z \sim 0.1$ is $\sim 0.5$, in better agreement with C14's observed slope rather than those of E07 and Z12.

We next examine the normalizations (at $M_* = 10^{9}$ and $10^{10}$) for the samples. For these six studies, $\log\psi(9) = -0.49, -0.40, -0.26, -0.26, -0.59,$ and $-0.78$, and $\log\psi(10) = 0.28, 0.31, .22, 0.09, 0.06,$ and $-0.01$. At lower masses, the agreement among the different studies is good, with $\log\psi(9)$ ranging from -0.59 to -0.40 (excluding C09 and C14, who both display much higher values). Again, the absolute SFR among studies with the same SFR indicator (E07 and Z12) differ by $\sim 0.1$\,dex. The variation is larger, however, among the studies with the same SPS models ($\sim 0.2$\,dex), which seems to indicate that the SFR indicator is a more important driver for differences rather than SPS model. We also find that the SFRs derived from O10 are $\sim 0.2$\,dex lower than even S07's. So while the slopes are identical between O10 and E07, the absolute normalizations differ by $\sim 0.3$\,dex. Thus a comparison in slope is not enough to determine the robustness of any MS measurement -- normalizations (especially relative normalizations) should be taken into account.

At $\log\psi(10)$, however, we get a slightly diferent picture. Here, while SFRs derived from H$\alpha$ (E07 and Z12) and those derived from other methods (absorption lines, SED fitting, and IR for C09, S07/C14, and O10, respectively) exhibit good agreement within each group ($\Delta\log\psi(10)\lesssim 0.1$\,dex; C14, which straddles both groups, is the exception), they are offset by $\sim 0.2$\,dex from each other. In other words, SFRs derived from H$\alpha$ tend to be systematically higher than those derived through other means. While this tendency is true over the entire fitted mass range, it becomes most severe (due to the steeper slopes) at higher mass.

We find that the SFRs provided by our best fits tend to favor the H$\alpha$ SFRs for $M_* \lesssim 10^{9.8}$ or so, and prefer the other studies for higher masses. This trend is similar to that seen at high redshift, and probably indicates a similar problem either with extinction corrections or selection effects at high and low masses. Alternately, it could indicate systematic variations in the SED fitting procedure (perhaps due to SFH or other parameters) at higher and/or lower masses (see also St14). The wide range in fitted MS slopes and normalizations suggest that many of the systematics involved in determining MS parameters have been severely underestimated. We estimate the magnitude of these effects on the MS slope to be of order $\sim 0.2$ or larger using just the data included here, in good agreement with \citet{abramson+14}, although we note that if we include other slopes from the literature (e.g., B04 measures a slope of $\sim 0.9$) the differences might be as large as $\sim 0.4$.

\section{The Non-Unity Slope of the Main Sequence and the Star-Forming Stellar Mass Function}
\label{app:mass_function}

As noted by \citet{peng+10}, \citet{lilly+13}, and \citet{abramson+14}, a MS slope of less than unity implies that the slope $\alpha_s$ of the SF mass function below the characteristic mass $M_*$ should steepen with time. More explicitly, \citet{peng+10} argue that the observed constancy of $M^*$ and $\alpha_s$ for SFGs implies that the quenching of galaxies around and above $M^*$ must be proportional to their SFRs. As the shape of the mass function appears to be relatively unchanged since $z \sim 2$ \citep{marchesini+09,ilbert+10,ilbert+13,muzzin+13,sobral+14}, the fact that the majority of our results give MS slopes significantly less than unity and display time-dependent evolution even further away from unity could be seen as somewhat concerning. As we have spent a significant portion of this work arguing the robustness of our fitting procedure and the widespread agreement among MS observations, we would like to spend some time discussing this possible problem.

As argued by \citet{abramson+14}, the below-unity slope of the MS may in large part due to studies failing to separate out bulge and disk components of their constituent galaxies. By analyzing only the disk components of SFGs at $z \sim 0$, they find the slope of this ``disk-limited'' MS (MS$_\textrm{disk}$) is approximately unity. They then propose that this new approximately unity MS$_\textrm{disk}$ slope solves the steepening mass function problem if one makes the reasonable assumption that mass-growth is dominated by \textit{in situ} disk SF.

As noted in Appendix~\ref{app:low_z}, the systematics in deriving a MS slope are extremely large (as high as $\sim 0.4$), even when using the same dataset (i.e. SDSS). Consequently, one might assume that these errors overwhelm all MS slopes measured in this literature. If this is the case, all MS observations (excluding non-selective ones) included in this work could technically be seen as consistent with slopes of $\sim 0.75$ (indeed, many seem to cluster around $\sim 0.6$\,--\,$0.8$ at a wide range of redshifts), and our results as consistent with a scenario where the MS slope out to $z \sim 2$ is approximately constant and the MS$_\textrm{disk}$ slope is approximately unity. As a result, $\alpha_s$ would remain unchanged over the same redshift range, in agreement with observations. While we cannot refute such a view as we are ultimately uncertain how large a role systematics play in MS observations, we  find this position unfavorable given the good results from, e.g., So14 at reproducing the cSFR out to $z \sim 2.5$.

Alternately, we note that while most of our MS fits imply some form of time-dependent MS slope, some of our MS fits (e.g. those which include only stacked, mixed MS observations) give time-independent parametrizations with slopes similar to those seen in \citep{abramson+14}. Although many of these are derived from smaller, more selective subsets of the MS observations included here, it is fully possible that they are a more accurate parametrization of MS evolution. Once future observations at low-$z$ can rigorously demonstrate that their MS slopes are robust and agree on appropriate selection criteria for differentiating between SF and quiescent galaxies, their results should be able to serve as additional constraints on the fits we provide here and hopefully provide a concrete answer to the apparent tension in the data.

\section{List of Acronyms}\label{app:acronyms}

Many acronyms are used throughout the paper in our discussions of concepts and issues surrounding the SFG MS, some used only once or twice, and others much more frequently. For convenience, we have compiled them all into Table~\ref{tab:acronyms} for easy reference.

\newpage

\clearpage
\LongTables
\begin{landscape}
\begin{center}
\scriptsize
\begin{deluxetable*}{l c c c c c c c c c c c c c}
\tabletypesize{\scriptsize}
\tablewidth{0pt}
\setlength{\tabcolsep}{0.5pt}
\tablecaption{Main Sequence Derivations and their Assumptions
\label{tab:msfr_info}}
\tablehead{
\colhead{Paper} &
\colhead{IMF} &
\colhead{SFR indicator} &
\colhead{$\psi_{\textrm{K98}}/\psi_{\textrm{KE12}}$} &
\colhead{$(h,\Omega_m,\Omega_\Lambda)$} &
\colhead{Selection} &
\colhead{Type} &
\colhead{SPS Model} &
\colhead{SFH} & 
\colhead{$Z (Z_\odot)$} &
\colhead{Extinction curve} &
\colhead{Lines} &
\colhead{Stacked} &
\colhead{N}
}
\startdata
B12 & Salpeter & FUV & 1.00/0.63 & (.7,.3,.7) & UV/LBG & BL & BC03\,\tablenotemark{n} & C\,\tablenotemark{n} & 0.2\,\tablenotemark{n} & ?\,\tablenotemark{n} & No\,\tablenotemark{n} & Yes & 1 \\
C09(1) & Kroupa & H$\alpha$/H$\beta$\,\tablenotemark{w} & 0.48/0.71 & (.7,.3,.7) & OPT/None & Non\,\tablenotemark{v} & BC03 & D & $0.2$\,--\,$1$ & CF00 & No & Yes & 1 \\
C09(2) & Kroupa & H$\alpha$/H$\beta$\,\tablenotemark{w} & 0.48/0.71 & (.7,.3,.7) & OPT\,\tablenotemark{1}/None & Non\,\tablenotemark{v} & BC03 & D & $0.2$\,--\,$1$ & CF00 & No & Yes & 1 \\
C14 & Chabrier & UV SED & --/1.0 & (.7,.3,.7) & OPT\,\tablenotemark{1}/$M_*$-$\psi$\,\tablenotemark{t} & Non\,\tablenotemark{v} & BC03 & DRB & $0.2$\,--\,$1.5$ & CF00 & No & No & 7 \\
D07 & Salpeter & FUV\,\tablenotemark{p} & 0.81/1.29 & (.73,.26,.74) & OPT/$sBzK$ & BL & BC03 & D & $0.02$\,--\,$2.5$\, & P84 & No & No & 1 \\
D09 & Salpeter & 1.4\,GHz & --/0.87 & (.71,.27,.73) & OPT/None & MI & BC03\,\tablenotemark{q} & D & ?\,\tablenotemark{r} & C00/M95 & No & Yes & 6 \\
E07(1) & Salpeter & H$\alpha$ & 0.94/1.38 & (.7,.3,.7) & OPT/blue & BL & BC03 & DRB & $0.005$\,--\,$2.5$\, & CF00 & No & No & 1 \\
E07(2) & Salpeter & FUV & 0.81/1.29 & (.7,.3,.7) & OPT/blue & BL & PEGASE.2 & ?\,\tablenotemark{a} & ?\,\tablenotemark{a} & ?\,\tablenotemark{a} & No & No & 1 \\
\enddata
\end{deluxetable*}
\end{center}
\clearpage
\end{landscape}

\newpage

\clearpage
\LongTables
\begin{landscape}
\begin{center}
\scriptsize
\begin{deluxetable*}{l c c c c c c c c c c c c c c c c}
\tabletypesize{\scriptsize}
\tablewidth{0pt}
\setlength{\tabcolsep}{0.5pt}
\tablecaption{Main Sequence Relationships \label{tab:msfr}}
\tablehead{
\colhead{Paper} &
\colhead{${z}_{\textrm{med}}$} & 
\colhead{${z}_{\textrm{range}}$} & 
\colhead{$\Delta t$} &
\colhead{$\alpha$} & 
\colhead{$\beta$} &
\colhead{$\psi(10.5)$} &
\colhead{$\sigma$} &
\colhead{$\sigma_d$} &
\colhead{$\sigma_t$ } &
\colhead{$\log M_*$ range} &
\colhead{Sample} &
\colhead{Survey} &
\colhead{Area/Vol}
}
\startdata
E11 & 0.05 & 0.0-0.1\,\tablenotemark{a} & 1.301 & 1 & -9.60 & 0.9 & 0.26 & 0.26 & 0.16 & 9.6-11.3 & 648 & various\,\tablenotemark{a} & 38960\,\tablenotemark{a}/289.86 \\
E07(1) & 0.06 & $0.015$-$0.1$ & 1.094 & $0.77$ & $-7.44$ & 0.645 & $0.25$ & $0.25$ & $0.15$ & 9.1-11.2 & 19590 & SDSS & 705/5.23 \\
Z12(1) & 0.07 & $0.04$-$0.1$ & 0.758 & $0.71 \pm 0.01$ & $-6.78 \pm 0.1$ & 0.675 & $0.25$ & $0.25$ & $0.15$ & 8.5-10.4 & $\sim$\,$\sn{2}{5}$ & SDSS & 8200/56.93 \\
C14(1) & 0.1 & 0.0-0.2 & 2.432 & $0.477 \pm 0.004$ & $ -4.57 \pm 0.04$ & 0.43 & 0.65 & 0.65 & 0.61 & 9.0-11.3 & 101973 & GALEX+SDSS & 2505/138.56 \\
O10 & 0.1 & 0.0-0.2 & 2.432 & $0.77 \pm 0.02$ & $-7.88 \pm .22$ & 0.205 & 0\,\tablenotemark{b} & 0\,\tablenotemark{b} & 0\,\tablenotemark{b} & 9.1-11.6 & 8400 & SWIRE & 11.33/0.63 \\
C09(1) & 0.11 & 0.005-0.22 & 2.570 & $0.35 \pm 0.09$ & $-3.56 \pm .87$ & 0.115 & 0\,\tablenotemark{b} & 0\,\tablenotemark{b} & 0\,\tablenotemark{b} & 9.0-12.0 & $\sim$\,$\sn{5}{5}$ & SDSS & 4783/346.88 \\
S07 & 0.11 & $0.005$-$0.22$ & 2.570 & $0.65$ & $-6.33$ & 0.495 & $0.3$ & $0.29$ & $0.21$ & 9-11.1 & 48295 & GALEX+SDSS & 645/46.78 \\
C14(2) & 0.25 & 0.2-0.3 & 0.986 & $0.48 \pm 0.01$ & $ -4.36 \pm 0.14$ & 0.66 & 0.53 & 0.53 & 0.49 & 9.0-10.9 & 3423 & PRIMUS & 9/1.06 \\
\enddata
\end{deluxetable*}
\end{center}
\clearpage
\end{landscape}

\newpage

\clearpage
\begin{center}
\scriptsize
\begin{deluxetable*}{l c c c c c c c c c}
\tabletypesize{\scriptsize}
\tablewidth{0pt}
\setlength{\tabcolsep}{2pt}
\tablecaption{Main Sequence Calibration Offsets \label{tab:corr}}
\tablehead{
\colhead{Paper} &
\colhead{Redshift} &
\colhead{$C_M$} &
\colhead{$C_\psi$\,\tablenotemark{f}} &
\colhead{$C_C$} &
\colhead{$C_S$} &
\colhead{$C_E$} &
\colhead{$C_L$} &
\colhead{$C_R$} &
\colhead{${C}_{\textrm{tot}}$}
}
\startdata
\citet{elbaz+11} & 0.05 & +0.21 & -0.06 & +0.00 & +0.00 & +0.00\,\tablenotemark{a} & +0.00 & +0.00 & +0.15 \\
\citet{elbaz+07}(1) & 0.06 & +0.16 & -0.14 & +0.00 & +0.00 & +0.00 & +0.00 & +0.00 & +0.02 \\
\citet{zahid+12}(1) & 0.07 & -0.02 & +0.01 & +0.00 & +0.00 & +0.00 & +0.00 & +0.00 & -0.01 \\
Coil et al. (2014, in prep.)(1) & 0.1 & -0.01 & +0.03 & +0.00 & +0.00 & +0.00 & +0.00 & +0.00 & +0.02 \\
\citet{oliver+10} & 0.1 & +0.20 & -0.12 & +0.00 & +0.00 & +0.09 & +0.00 & +0.00 & +0.17 \\
\citet{chen+09}(1) & 0.11 & +0.00 & +0.15 & +0.00 & +0.00 & +0.00 & +0.00 & +0.00 & +0.15 \\
\citet{salim+07} & 0.11 & -0.02 & -0.09 & +0.00 & +0.00 & +0.00 & +0.00 & +0.00 & -0.11 \\
Coil et al. (2014, in prep.)(2) & 0.25 & -0.01 & +0.03 & +0.00 & +0.00 & +0.00 & +0.00 & +0.00 & +0.02 \\
\enddata
\tablecomments{Calibration offsets to MS normalizations based on Table~\ref{tab:calibration} and outlined in \S\,\ref{sec:technicalities}, in dex. Subscripts are as follows: $M = \textrm{mass}$, $\psi = \textrm{SFR}$, $C = \textrm{cosmology}$, $S = \textrm{SPS Model}$, $E = \textrm{Extinction}$, $L = \textrm{Emission Lines}$, $R = \textrm{Radio}$ (possible evolution in $q_{\textrm{IR}}$), and $\textrm{tot} = \textrm{total}$.
%\\
%\\
\tablenotemark{a}\,{An extinction correction (for missing/unobscured UV emission) is not taken into account for E11 because they've selected a sample of LIRGS, which are display much higher dust attenuation (and hence IR emission) than the typical MS galaxy.}
\tablenotemark{b}\,{As mentioned earlier, So14's mass adjustment for CB07 models result in a shift upwards of $\sim 0.2$\,dex, rather than the slightly lower $\sim 0.15$\,dex found by M10.}
\tablenotemark{c}\,{Includes both an adjustment for IMF (Salpeter to Kroupa) and for systematic biases present in using K-band luminosity-to-mass conversions \citep{ilbert+10}.}
\tablenotemark{d}\,{We use the R12a-derived extinction correction here as the median redshifts of each of the respective mass bins used in the fit tends to be closer to $z \sim 3.25$\,--\,3.5 than to 4.}
\tablenotemark{e}\,{Given that the $\psi_{H\alpha}$-to-$\psi_{\textrm{UV}}$ ratio reported in S11 ($\sim 6$) is larger than the extinction correction that would otherwise be applied from B12 ($\sim 2.5$), we decide not to include any additional extinction corrections.}
\tablenotemark{f}\,{Includes a -0.1\,dex adjustment (see \S\,\ref{subsec:radio}) for stacked radio data to account for the fact that the true mean (derived through median stacks) is biased above the median for a log-normal distribution.}
}
\end{deluxetable*}
\end{center}
\clearpage

\newpage

\clearpage
\begin{center}
\scriptsize
\begin{deluxetable*}{l c c c c c c c c c}
\tablewidth{0pt}
\setlength{\tabcolsep}{2pt}
\tabletypesize{\scriptsize}
\tablecaption{Calibrated Main Sequence relationships \label{tab:msfr_corr}}
\tablehead{
\colhead{Paper} &
\colhead{$\rom{z}{\textrm{avg}}$} & 
\colhead{$\rom{z}{\textrm{min}}$} &
\colhead{$\rom{z}{\textrm{max}}$} &
\colhead{$\Delta t$ (Gyr)} &
\colhead{$\alpha$} & 
\colhead{$\beta$} &
\colhead{$\log\psi(10.5)$} &
\colhead{$\sigma_t$ (dex)} &
\colhead{$\log M$ range} 
}
\startdata
\citet{elbaz+11} & 0.05 & 0.0 & 0.1 & 1.301 & 1 & -9.45 & 1.05 & 0.16 & 9.4-11.1 \\
\citet{elbaz+07}(1) & $0.06$ & 0.015 & 0.1 & 1.094 & $0.77$ & $-7.42$ & 0.665 & $0.14$ & 8.9-11.0 \\
\citet{zahid+12}(1) & $0.07$ & 0.04 & 0.1 & 0.758 & $0.71 \pm .01$ & $-6.79 \pm 0.1$ & 0.665 & $0.15$ & 8.5-10.4 \\
Coil et al. (2014, in prep.)(1) & 0.1 & 0.0 & 0.2 & 2.432 & $0.477 \pm 0.004$ & $-4.55 \pm 0.04$ & 0.45 & 0.61 & 9.0-11.3 \\
\citet{oliver+10} & 0.1 & 0.0 & 0.2 & 2.432 & $0.77 \pm .02$ & $-7.71 \pm .22$ & 0.375 & 0 & 8.9-11.4 \\
\citet{chen+09}(1) & 0.11 & 0.005 & 0.22 & 2.570 & $0.35 \pm .09$ & $-3.41 \pm .87$ & 0.265 & 0 & 9.0-12.0 \\
\citet{salim+07} & $0.11$ & 0.005 & 0.22 & 2.570 & $0.65$ & $-6.44$ & 0.385 & $0.21$ & 9.0-11.1 \\
Coil et al. (2014, in prep.)(2) & 0.25 & 0.2 & 0.3 & 0.986 & $0.48 \pm 0.01$ & $-4.34 \pm 0.14$ & 0.68 & 0.49 & 9.0-10.9 \\
\enddata
\tablecomments{Same as Table~\ref{tab:msfr}, but with the adjustments from Table~\ref{tab:corr} applied. Mass ranges have been adjusted for IMF, cosmology, SPS model, and emission line corrections, plus any additional offsets noted in Table~\ref{tab:corr} and/or in \S\,\ref{sec:technicalities}.}
\end{deluxetable*}
\end{center}
\clearpage

\newpage

\clearpage
\LongTables
\begin{landscape}
\begin{center}
\scriptsize
\begin{deluxetable*}{l c c c c c c c c c c c c c c c}
\tablewidth{0pt}
\setlength{\tabcolsep}{1pt}
\tabletypesize{\scriptsize}
\tablecaption{Parametrizations of Main Sequence Evolution as a Function of Time \label{tab:fits}}
\tablehead{
\colhead{Fit} &
\colhead{Data} &
\colhead{Sel} &
\colhead{Stk} &
\colhead{SFR} &
\colhead{$\rom{N}{\textrm{min}}$} &
\colhead{$\rom{N}{\textrm{tot}}$} &
\colhead{$\log M_*$} &
\colhead{$z$} &
\colhead{$\alpha_c$} &
\colhead{$\alpha_t$} &
\colhead{$\beta_c$} &
\colhead{$\beta_t$} &
\colhead{$a(10.5)$} &
\colhead{$b(10.5)$} &
\colhead{$\sigma_i$}
}
\startdata
1 & C(E) & All & All & All & N/A & 51(39) & 8.5-11.5 & 0.25-3.0 & $0.77$ & $-0.014$ & $-5.76$ & $-0.02$ & $-0.166$ & $2.346$ & 0.15 \\
2 & U(E) & All & All & All & N/A & 51(39) & 8.5-11.5 & 0.25-3.0 & $0.77$ & $-0.014$ & $-5.62$ & $-0.05$ & $-0.197$ & $2.487$ & 0.15 \\
3 & C(M) & All & All & All & 5 & 51(39) & 8.5-11.5 & 0.25-3.0 & $0.98 \pm 0.01$ & $-0.044 \pm 0.002$ & $-7.94 \pm 0.14$ & $0.3 \pm 0.02$ & $-0.166 \pm 0.002$ & $2.317 \pm 0.01$ & 0.1 \\
4 & U(M) & All & All & All & 5 & 51(39) & 8.5-11.5 & 0.25-3.0 & $1.05 \pm 0.02$ & $-0.054 \pm 0.003$ & $-8.57 \pm 0.16$ & $0.38 \pm 0.03$ & $-0.195 \pm 0.002$ & $2.444 \pm 0.012$ & 0.11 \\
5 & C(M) & All & All & All & 10 & 51(39) & 9.2-11.3 & 0.25-3.0 & $0.92 \pm 0.02$ & $-0.037 \pm 0.003$ & $-7.4 \pm 0.2$ & $0.22 \pm 0.03$ & $-0.165 \pm 0.001$ & $2.313 \pm 0.01$ & 0.1 \\
6 & U(M) & All & All & All & 10 & 51(39) & 9.2-11.3 & 0.25-3.0 & $1.0 \pm 0.02$ & $-0.047 \pm 0.002$ & $-8.03 \pm 0.18$ & $0.3 \pm 0.02$ & $-0.193 \pm 0.001$ & $2.434 \pm 0.01$ & 0.11 \\
7 & C(M) & All & All & All & 15 & 51(39) & 9.4-11.2 & 0.25-3.0 & $0.89 \pm 0.02$ & $-0.033 \pm 0.002$ & $-7.08 \pm 0.16$ & $0.18 \pm 0.03$ & $-0.165 \pm 0.001$ & $2.315 \pm 0.008$ & 0.09 \\
8 & U(M) & All & All & All & 15 & 51(39) & 9.4-11.2 & 0.25-3.0 & $0.97 \pm 0.02$ & $-0.044 \pm 0.003$ & $-7.73 \pm 0.2$ & $0.27 \pm 0.03$ & $-0.194 \pm 0.001$ & $2.443 \pm 0.01$ & 0.11 \\
\enddata
\tablecomments{Col. 1: Fit number. Col. 2: Type of data used for fit, where C = Calibrated data, U = uncalibrated data, (M) = fitted in bins of mass, (E) = extrapolated/simple average, (R) = \textit{excluding} radio data, (Q) = \textit{excluding} our radio-specific $C_R$ offsets, and (z) = \textit{including} high-$z$ data. Col. 3: Selection type (bluer (BL)/ mixed (MI)/non-selective (NON)/BL+MI (all)); see \S\,\ref{subsubsec:selection}. Col. 4: Stacked? (yes/no/all). Col. 5: SFR indicators included in the fit. Col. 6: Minimum number of data points in a given mass bin required to include the bin in the fit. Col. 7: Total number of data points included in our selection criteria before (and after) excluding the first and last 2\,Gyr of data. Col. 8: Range of mass bins included in the fit, in units of 0.1\,dex. Col. 9: Total span of the data in redshift (using median redshifts of each sample). Cols. 10\,--\,13: Best-fit MS parameters, with $\alpha_c$, $\alpha_t$, $\beta_c$, and $\beta_t$ defined via $\psi(M_*,t) = 10^{\beta(t)} \, M_*^{\alpha(t)}$, where $\alpha(t) = \alpha_t \, t + \alpha_c$ and $\beta(t) = \beta_t \, t + \beta_c$. Cols. 14 \& 15: Best-fit parameters at fixed mass, where $a(10.5)$ and $b(10.5)$ are defined via $\log\psi = a(\log M_*) \, t + b(\log M_*)$ for $\log M_* = 10.5$. Col. 16: The median interpublication scatter at fixed mass between studies included in the fit. Errors on the parameters are derived by calculating the standard deviation of their distributions after 100 runs, with mass ranges for each study randomly adjusted by up to $\pm 0.2$\,dex in each run. These are $\sim 50\%$ greater than the formal fitted errors in most cases. Errors are not listed for the extrapolated fits since not MS parameters used in the fit  included errors (i.e. some studies did not report errors on their fit); they are likely similar to the errors seen in our mass-dependent fits.}
\end{deluxetable*}
\end{center}
\clearpage
\end{landscape}

\newpage

\clearpage
\LongTables
\begin{landscape}
\begin{center}
\scriptsize
\begin{deluxetable*}{l c c c c c c c c c c c c c c c}
\tablewidth{0pt}
\setlength{\tabcolsep}{1pt}
\tabletypesize{\scriptsize}
\tablecaption{Parametrization of Main Sequence Evolution as a Function of Redshift\label{tab:fits_z}}
\tablehead{
\colhead{Fit} &
\colhead{Data} &
\colhead{Sel} &
\colhead{Stk} &
\colhead{SFR} &
\colhead{$\rom{N}{\textrm{min}}$} &
\colhead{$\rom{N}{\textrm{tot}}$} &
\colhead{$\log M_*$} &
\colhead{$z$} &
\colhead{$a_m$} &
\colhead{$a_0$} &
\colhead{$b_m$} &
\colhead{$b_0$} &
\colhead{$a(10.5)$} &
\colhead{$b(10.5)$} &
\colhead{$\sigma_i$}
}
\startdata
1 & C(E) & All & All & All & N/A & 51(39) & 8.5-11.5 & 0.25-3.0 & $0.25$ & $0.26$ & $0.61$ & $-5.95$ & $2.84$ & $0.46$ & 0.17 \\
2 & U(E) & All & All & All & N/A & 51(39) & 8.5-11.5 & 0.25-3.0 & $0.25$ & $0.83$ & $0.61$ & $-6.17$ & $3.4$ & $0.24$ & 0.15 \\
3 & C(M) & All & All & All & 5 & 51(39) & 8.5-11.5 & 0.25-3.0 & $0.69 \pm 0.05$ & $-4.41 \pm 0.47$ & $0.48 \pm 0.02$ & $-4.59 \pm 0.16$ & $2.85 \pm 0.03$ & $0.42 \pm 0.01$ & 0.11 \\
4 & U(M) & All & All & All & 5 & 51(39) & 8.5-11.5 & 0.25-3.0 & $0.88 \pm 0.05$ & $-5.85 \pm 0.47$ & $0.42 \pm 0.02$ & $-4.26 \pm 0.16$ & $3.37 \pm 0.03$ & $0.2 \pm 0.01$ & 0.11 \\
5 & C(M) & All & All & All & 10 & 51(39) & 9.2-11.3 & 0.25-3.0 & $0.52 \pm 0.05$ & $-2.54 \pm 0.52$ & $0.53 \pm 0.02$ & $-5.09 \pm 0.16$ & $2.85 \pm 0.03$ & $0.43 \pm 0.01$ & 0.11 \\
6 & U(M) & All & All & All & 10 & 51(39) & 9.2-11.3 & 0.25-3.0 & $0.69 \pm 0.05$ & $-3.93 \pm 0.52$ & $0.48 \pm 0.02$ & $-4.81 \pm 0.16$ & $3.36 \pm 0.02$ & $0.21 \pm 0.01$ & 0.11 \\
7 & C(M) & All & All & All & 15 & 51(39) & 9.4-11.2 & 0.25-3.0 & $0.4 \pm 0.06$ & $-1.41 \pm 0.58$ & $0.56 \pm 0.02$ & $-5.4 \pm 0.2$ & $2.85 \pm 0.03$ & $0.43 \pm 0.01$ & 0.11 \\
8 & U(M) & All & All & All & 15 & 51(39) & 9.4-11.2 & 0.25-3.0 & $0.61 \pm 0.05$ & $-3.03 \pm 0.52$ & $0.5 \pm 0.02$ & $-4.99 \pm 0.17$ & $3.37 \pm 0.02$ & $0.22 \pm 0.01$ & 0.11 \\
\enddata
\tablecomments{Data are classified as Table~\ref{tab:fits}, except the parameters listed are for a fitted function of the form $\psi(M_*,z) = 10^{b(M_*)} \, (1+z)^{a(M_*)}$, where $a(\log M_*) = a_m \, \log M_* + a_0$ and $b(M_*) = b_m \, \log M + b_0$ are the power-law index and the normalization of the MS. $a_m$ contains information about the steepness of the mass dependence (i.e. the amount of differential mass evolution), while the other three parameters mostly provide normalizations. $a(10.5)$ and $b(10.5)$ are defined at $\log M_* = 10.5$, and track the evolution of the SFR from the present day ($z = 0$) to higher redshifts. As in Table~\ref{tab:fits}, errors on the parameters are derived by calculating the standard deviation of their distributions after 100 runs, with mass ranges for each study randomly adjusted by up to $\pm 0.2$\,dex in each run. These are $\sim 50\%$ greater than the formal fitted errors in most cases. Errors are not listed for the extrapolated fits since not MS parameters used in the fit included errors (i.e. some studies did not report errors on their fit); they are likely similar to the errors seen in our mass-dependent fits.}
\end{deluxetable*}
\end{center}
\clearpage
\end{landscape}

\newpage

\clearpage
\LongTables
\begin{landscape}
\begin{center}
\begin{deluxetable*}{l l l c}
\tabletypesize{\scriptsize}
\tablewidth{0pt}
\setlength{\tabcolsep}{1pt}
\tablecaption{Best-fit Main Sequence Evolution
\label{tab:best_fits}}
\tablehead{
\colhead{Data} &
\colhead{Best MS Fit} &
\colhead{Best $\log\psi(t,10.5)$ fit} &
\colhead{Number}
}
\startdata
``Mixed'' (preferred fit) & $\log\psi(M_*,t) = \left(0.84 \pm 0.02 - 0.026 \pm 0.003 \times t\right) \log M_* \nonumber - \left(6.51 \pm 0.24 - 0.11 \pm 0.03 \times t\right)$ & $\log\psi(10.5,t) = (-0.16 \pm 0.02)\,t + (2.28 \pm 0.20)$ & 49 \\
``Mixed'' w/o $C_R$ & $\log\psi(M_*,t) = \left(0.90 \pm 0.05 - 0.037 \pm 0.006 \times t\right) \log M_* - \left(6.84 \pm 0.51 - 0.19 \pm 0.06 \times t\right)$ & $\log\psi(10.5,t) = (-0.20 \pm 0.02)\,t + (2.65 \pm 0.20)$ & 151 \\
``Mixed'' w/ high-$z$ obs & $\log\psi(M_*,t) = \left(0.80 \pm 0.02 - 0.022 \pm 0.003 \times t\right) \log M_* - \left(6.09 \pm 0.23 - 0.07 \pm 0.03 \times t\right)$ & $\log\psi(10.5,t) = (-0.16 \pm 0.02)\,t + (2.30 \pm 0.20)$ & 159 \\
``Mixed'' w/ high-$z$ obs w/o $C_R$ & $\log\psi(M_*,t) = \left(0.96 \pm 0.05 - 0.045 \pm 0.006 \times t\right) \log M_* - \left(7.41 \pm 0.48 - 0.27 \pm 0.06 \times t\right)$ & $\log\psi(10.5,t) = (-0.20 \pm 0.02)\,t + (2.64 \pm 0.20)$ & 167 \\
All & $\log\psi(M_*,t) = \left(0.89 \pm 0.02 - 0.033 \pm 0.002 \times t\right) \log M_* - \left(7.08 \pm 0.16 - 0.18 \pm 0.03 \times t\right)$ & $\log\psi(10.5,t) = (-0.16 \pm 0.02)\,t + (2.31 \pm 0.20)$ & 7 \\
All w/o $C_R$ & $\log\psi(M_*,t) = \left(0.95 \pm 0.04 - 0.042 \pm 0.005 \times t\right) \log M_* - \left(7.38 \pm 0.43 - 0.25 \pm 0.06 \times t\right)$ & $\log\psi(10.5,t) = (-0.19 \pm 0.02)\,t + (2.56 \pm 0.20)$ & 127 \\
All w/ high-$z$ obs & $\log\psi(M_*,t) = \left(0.85 \pm 0.01 - 0.027 \pm 0.002 \times t\right) \log M_* - \left(6.68 \pm 0.14 - 0.12 \pm 0.02 \times t\right)$ & $\log\psi(10.5,t) = (-0.16 \pm 0.02)\,t + (2.28 \pm 0.20)$ & 135 \\
All w/ high-$z$ obs w/o $C_R$ & $\log\psi(M_*,t) = \left(1.02 \pm 0.02 - 0.051 \pm 0.003 \times t\right) \log M_* - \left(8.23 \pm 0.21 - 0.35 \pm 0.03 \times t\right)$ & $\log\psi(10.5,t) = (-0.18 \pm 0.02)\,t + (2.48 \pm 0.20)$ & 143 \\
\enddata
\tablecomments{Our best fits to available MS data taken from Table~\ref{tab:fits}. Col. 1: Subset of data that was fit. Col. 2: Best-fit parametrization of the time evolution of the $\log\psi$\,--\,$\log M_*$ relation. Col. 3: Best-fit parametrization of $\log\psi$ over time at $\log M_* = 10.5$. Errors on the slope and normalization from Table~\ref{tab:fits} have bee added in quadrature with $0.2$\,dex and $0.02$\,dex, respectively, in order to better represent intrinsic uncertainties and MS scatter. Col. 4: Numbers corresponding to the relevant row in Table~\ref{tab:fits} that the best fits have been drawn from. All of these fits do not include non-selective data (see \S\,\ref{subsubsec:selection}) unless specifically stated otherwise (e.g., the ``all'' fits does not include non-selective data). To convert these results to a Salpeter (Chabrier) IMF, just add (subtract) 0.21 (0.03) dex from $\log M_*$ and $\log \psi$ and propagate changes accordingly.}
\end{deluxetable*}
\end{center}
\clearpage
\end{landscape}

\newpage

\begin{center}
\begin{deluxetable*}{l l}
\tabletypesize{\scriptsize}
\tablewidth{0pt}
\setlength{\tabcolsep}{2pt}
\tablecaption{Acronyms and Shorthand Notation
\label{tab:acronyms}}
\tablehead{
\colhead{Term} &
\colhead{Usage}
}
\startdata
\textbf{TERMINOLOGY} & \textbf{SHORTHAND} \\
Blue Straggles & BS \\
Cosmic Microwave Background & CMB \\
Color-Magnitude Diagram & CMD \\
Elmegreen-Silk (Relation) & ES \\
``Extended'' KS (Relation) & EKS \\
Equivalent Width & EW \\
Extinction: $B - V$ & $E$($B$\,--\,$V$) \\
\enddata
\tablecomments{A list of acronyms used throughout the paper. Organized in alphabetical order and separated by terms and surveys.}
\end{deluxetable*}
\end{center}
\clearpage

\clearpage

\end{appendix}
\clearpage

%\nocite{*}\bibliography{ms}

\end{document}